\documentclass[aps,pra,twocolumn,notitlepage,superscriptaddress,showpacs,nofootinbib]{revtex4-1}
\usepackage{amsmath}
\usepackage{amssymb}
\usepackage{amsfonts}
\usepackage{enumerate}
\usepackage{mathrsfs}
\usepackage{graphicx}
\usepackage{epstopdf}
\usepackage{enumerate}
\usepackage{changepage}
\usepackage{bm}
\usepackage{multirow}
\usepackage{lipsum}
\usepackage{booktabs}
\usepackage{tikz}
\usepackage{extarrows}
\usepackage{multirow}
\usepackage{makecell}
\usepackage{tabularx}
\usepackage{graphicx}
\usepackage{float}
\usepackage{afterpage}
\usepackage{placeins}

\newtheorem{definition}{Definition}

\newtheorem{lemma}{Lemma}
\newtheorem{proposition}{Proposition}
\newtheorem{conjecture}{Conjecture}
\newtheorem{example}{Example}
\newtheorem{corollary}{Corollary}

\def\bcj{\begin{conjecture}}
	\def\ecj{\end{conjecture}}
\def\bcr{\begin{corollary}}
	\def\ecr{\end{corollary}}
\def\bd{\begin{definition}}
	\def\ed{\end{definition}}
\def\bea{\begin{eqnarray}}
	\def\eea{\end{eqnarray}}
\def\bem{\begin{enumerate}}
	\def\eem{\end{enumerate}}
\def\bex{\begin{example}}
	\def\eex{\end{example}}
\def\bim{\begin{itemize}}
	\def\eim{\end{itemize}}
\def\bl{\begin{lemma}}
	\def\el{\end{lemma}}
\def\bma{\begin{bmatrix}}
	\def\ema{\end{bmatrix}}
\def\bpf{\begin{proof}}
	\def\epf{\end{proof}}
\def\bpp{\begin{proposition}}
	\def\epp{\end{proposition}}
\def\bqu{\begin{question}}
	\def\equ{\end{question}}
\def\br{\begin{remark}}
	\def\er{\end{remark}}

\def\squareforqed{\hbox{\rlap{$\sqcap$}$\sqcup$}}
\def\qed{\ifmmode\squareforqed\else{\unskip\nobreak\hfil
		\penalty50\hskip1em\null\nobreak\hfil\squareforqed
		\parfillskip=0pt\finalhyphendemerits=0\endgraf}\fi}
\def\endenv{\ifmmode\;\else{\unskip\nobreak\hfil
		\penalty50\hskip1em\null\nobreak\hfil\;
		\parfillskip=0pt\finalhyphendemerits=0\endgraf}\fi}
\newenvironment{proof}{\noindent \textbf{{Proof.~} }}{\qed}
\def\Dbar{\leavevmode\lower.6ex\hbox to 0pt
	{\hskip-.23ex\accent"16\hss}D}
\makeatletter
\def\url@leostyle{%
	\@ifundefined{selectfont}{\def\UrlFont{\sf}}{\def\UrlFont{\small\ttfamily}}}
\makeatother
\def\bcj{\begin{conjecture}}
	\def\ecj{\end{conjecture}}
\def\bcr{\begin{corollary}}
	\def\ecr{\end{corollary}}
\def\bd{\begin{definition}}
	\def\ed{\end{definition}}
\def\bea{\begin{eqnarray}}
	\def\eea{\end{eqnarray}}
\def\bem{\begin{enumerate}}
	\def\eem{\end{enumerate}}
\def\bex{\begin{example}}
	\def\eex{\end{example}}
\def\bim{\begin{itemize}}
	\def\eim{\end{itemize}}
\def\bl{\begin{lemma}}
	\def\el{\end{lemma}}
\def\bpf{\begin{proof}}
	\def\epf{\end{proof}}
\def\bpp{\begin{proposition}}
	\def\epp{\end{proposition}}
\def\bqu{\begin{question}}
	\def\equ{\end{question}}
\def\br{\begin{remark}}
	\def\er{\end{remark}}

\def\btb{\begin{tabular}}
	\def\etb{\end{tabular}}

	\newcommand{\nc}{\newcommand}
	
	\nc{\bbA}{\mathbb{A}} \nc{\bbB}{\mathbb{B}} \nc{\bbC}{\mathbb{C}}
	\nc{\bbD}{\mathbb{D}} \nc{\bbE}{\mathbb{E}} \nc{\bbF}{\mathbb{F}}
	\nc{\bbG}{\mathbb{G}} \nc{\bbH}{\mathbb{H}} \nc{\bbI}{\mathbb{I}}
	\nc{\bbJ}{\mathbb{J}} \nc{\bbK}{\mathbb{K}} \nc{\bbL}{\mathbb{L}}
	\nc{\bbM}{\mathbb{M}} \nc{\bbN}{\mathbb{N}} \nc{\bbO}{\mathbb{O}}
	\nc{\bbP}{\mathbb{P}} \nc{\bbQ}{\mathbb{Q}} \nc{\bbR}{\mathbb{R}}
	\nc{\bbS}{\mathbb{S}} \nc{\bbT}{\mathbb{T}} \nc{\bbU}{\mathbb{U}}
	\nc{\bbV}{\mathbb{V}} \nc{\bbW}{\mathbb{W}} \nc{\bbX}{\mathbb{X}}
    \nc{\bbY}{\mathbb{Y}} \nc{\bbZ}{\mathbb{Z}}
	
	\nc{\bA}{{\bf A}} \nc{\bB}{{\bf B}} \nc{\bC}{{\bf C}}
	\nc{\bD}{{\bf D}} \nc{\bE}{{\bf E}} \nc{\bF}{{\bf F}}
	\nc{\bG}{{\bf G}} \nc{\bH}{{\bf H}} \nc{\bI}{{\bf I}}
	\nc{\bJ}{{\bf J}} \nc{\bK}{{\bf K}} \nc{\bL}{{\bf L}}
	\nc{\bM}{{\bf M}} \nc{\bN}{{\bf N}} \nc{\bO}{{\bf O}}
	\nc{\bP}{{\bf P}} \nc{\bQ}{{\bf Q}} \nc{\bR}{{\bf R}}
	\nc{\bS}{{\bf S}} \nc{\bT}{{\bf T}} \nc{\bU}{{\bf U}}
	\nc{\bV}{{\bf V}} \nc{\bW}{{\bf W}} \nc{\bX}{{\bf X}}
	\nc{\ba}{{\bf a}} \nc{\be}{{\bf e}} \nc{\bu}{{\bf u}}
	\nc{\brr}{{\bf r}} \nc{\bx}{{\bf x}}  \nc{\bz}{{\bf z}} \nc{\bv}{{\bf v}} \nc{\bt}{{\bf t}} \nc{\bs}{{\bf s}}

	\nc{\cA}{{\cal A}} \nc{\cB}{{\cal B}} \nc{\cC}{{\cal C}}
	\nc{\cD}{{\cal D}} \nc{\cE}{{\cal E}} \nc{\cF}{{\cal F}}
	\nc{\cG}{{\cal G}} \nc{\cH}{{\cal H}} \nc{\cI}{{\cal I}}
	\nc{\cJ}{{\cal J}} \nc{\cK}{{\cal K}} \nc{\cL}{{\cal L}}
	\nc{\cM}{{\cal M}} \nc{\cN}{{\cal N}} \nc{\cO}{{\cal O}}
	\nc{\cP}{{\cal P}} \nc{\cQ}{{\cal Q}} \nc{\cR}{{\cal R}}
	\nc{\cS}{{\cal S}} \nc{\cT}{{\cal T}} \nc{\cU}{{\cal U}}
	\nc{\cV}{{\cal V}} \nc{\cW}{{\cal W}} \nc{\cX}{{\cal X}}
	\nc{\cZ}{{\cal Z}}
	
	\nc{\hA}{{\hat{A}}} \nc{\hB}{{\hat{B}}} \nc{\hC}{{\hat{C}}}
	\nc{\hD}{{\hat{D}}} \nc{\hE}{{\hat{E}}} \nc{\hF}{{\hat{F}}}
	\nc{\hG}{{\hat{G}}} \nc{\hH}{{\hat{H}}} \nc{\hI}{{\hat{I}}}
	\nc{\hJ}{{\hat{J}}} \nc{\hK}{{\hat{K}}} \nc{\hL}{{\hat{L}}}
	\nc{\hM}{{\hat{M}}} \nc{\hN}{{\hat{N}}} \nc{\hO}{{\hat{O}}}
	\nc{\hP}{{\hat{P}}} \nc{\hR}{{\hat{R}}} \nc{\hS}{{\hat{S}}}
	\nc{\hT}{{\hat{T}}} \nc{\hU}{{\hat{U}}} \nc{\hV}{{\hat{V}}}
	\nc{\hW}{{\hat{W}}} \nc{\hX}{{\hat{X}}} \nc{\hZ}{{\hat{Z}}}
	
	\nc{\hn}{{\hat{n}}}
	
	\def\min{\mathop{\rm min}}

	\def\tr{\mathop{\rm Tr}}

	\newcommand{\ket}[1]{|#1\rangle}
	
	\newcommand{\ketbra}[2]{|#1\rangle\!\langle#2|}

	\usepackage[
	colorlinks,
	linkcolor = blue,
	citecolor = blue,
	urlcolor = blue]{hyperref}
	\def \qed {\hfill \vrule height7pt width 7pt depth 0pt}
	
	\newcounter{lastnote}

\begin{document}
		\title{Experimental Multipartite Entanglement Detection With Minimal-Size Correlations}
\author{Dian Wu}
\email[]{dwu@lps.ecnu.edu.cn}
\thanks{These two authors contributed equally to this work.}
\affiliation{Advanced Institute of Precision Spectroscopy, East China Normal University, Shanghai 200241, China}
 \affiliation{Shanghai Branch, Hefei National Laboratory, Shanghai 201315, China}

\author{Fei Shi}
\email[]{shifei@hku.hk}
\thanks{These two authors contributed equally to this work.}
 \affiliation{QICI Quantum Information and Computation Initiative, School of Computing and Data Science,
The University of Hong Kong, Pokfulam Road, Hong Kong SAR, China}

\author{Jia-Cheng Sun}
\affiliation{Advanced Institute of Precision Spectroscopy, East China Normal University, Shanghai 200241, China}

\author{Bo-Wen Wang}
\affiliation{Advanced Institute of Precision Spectroscopy, East China Normal University, Shanghai 200241, China}

\author{Xue-Mei Gu}
\affiliation{Institut für Festkörpertheorie und Optik, Friedrich-Schiller-Universität Jena, Max-Wien-Platz 1, 07743 Jena, Germany}

\author{Giulio Chiribella}
\email[]{giulio@cs.hku.hk}
\affiliation{QICI Quantum Information and Computation Initiative, School of Computing and Data Science,
The University of Hong Kong, Pokfulam Road, Hong Kong SAR, China}
\affiliation{Department of Computer Science, Parks Road, Oxford, OX1 3QD, United Kingdom}
 \affiliation{Perimeter Institute for Theoretical Physics, Waterloo, Ontario N2L 2Y5, Canada}

\author{Qi Zhao}
\email[]{zhaoqi@cs.hku.hk}
 \affiliation{QICI Quantum Information and Computation Initiative, School of Computing and Data Science,
The University of Hong Kong, Pokfulam Road, Hong Kong SAR, China}

\author{Jian Wu}
 \affiliation{Advanced Institute of Precision Spectroscopy, East China Normal University, Shanghai 200241, China}

 \affiliation{Chongqing Key Laboratory of Precision Optics, Chongqing Institute of East China Normal University,\\
 Chongqing 401121, China}
 
\affiliation{\mbox{Collaborative Innovation Center of Extreme Optics,\! Shanxi University,\! Taiyuan,\! Shanxi~030006,\! China}}


\begin{abstract}

Multiparticle entanglement is a valuable resource for quantum technologies, including measurement based quantum computing, quantum secret sharing, and  a variety of quantum sensing applications.
The direct way to detect
this resource 
is to observe correlations arising from  local measurements performed simultaneously on all particles. However, this approach is increasingly vulnerable to measurement imperfections when the number of particles grows, and becomes unfeasible for large-scale entangled states.  It is  therefore crucial to devise detection methods that minimize the number of simultaneously measured particles.  
Here we provide the first experimental demonstration of   multipartite entanglement detection with minimal-size correlations, showing that our setup is robust to misalignment of the local measurement bases and enables the certification of genuine multipartite entanglement in a regime where the direct approach fails.  Overall, our results indicate a promising route to the experimental detection of genuine multipartite entanglement in large-scale entangled states. 

\end{abstract}

	\maketitle

\textit{\textbf{Introduction.}}\textbf{---} Entanglement is one of the cornerstones of quantum information science, and is as one of the key resources in quantum computation \cite{raussendorf2003measurement,2020Deterministic,zhao2025entanglement},  communication \cite{karlsson1998quantum,yin2020entanglement,proietti2021experimental}, and  metrology \cite{chin2012quantum,demkowicz2014using,pezze2018quantum}.    With the advancement of quantum technologies, the size and complexity of the entangled states accessible in the laboratory is rapidly increasing; for example, globally entangled states of  51-qubit and 95-qubit    have been recently generated in superconducting qubit systems \cite{cao2023generation,jiang2025generation}.

As the size of the accessible entangled states scales up, certifying the presence of genuine multipartite entanglement (GME)—the strongest form of nonclassical correlation—becomes increasingly important.  This task, however, presents  major challenges.  The direct approach to GME detection requires either performing global measurements  ({\em e.g.,}  measurements on a basis that include the entangled state of interest)  or estimating correlations among local measurements performed simultaneously on all particles  \cite{eibl2004experimental,toth2005entanglement,toth2005detecting,kiesel2005experimental,kiesel2007experimental,lu2007experimental,wieczorek2009experimental,prevedel2009experimental}.  When the number of particles becomes large, however, this approach becomes problematic: on the one hand, large-scale entangled measurements are hard to implement, and on the other hand correlations among a large number of particles are very sensitive to experimental imperfections.     For example, it has been found that small deviations between the single-particle measurements required by the theory and the measurements actually performed in the laboratory can result in a high rate of false positives \cite{seevinck2007local,rosset2012imperfect,morelli2022entanglement,cao2024genuine}.   
Therefore, it is essential to reduce the number of particles and to develop detection methods that involve only a small number of particles. 

Over the past two decades, significant efforts have been devoted to detection methods that require measurements on fewer particles \cite{toth2005entanglement,toth2007optimal,toth2009spin,gittsovich2010multiparticle,liang2015family,miklin2016multiparticle,paraschiv2018proving,frerot2021optimal,Chen_2025}, and to the study of related problems, such as the entanglement marginal problem \cite{navascues2021entanglement} and the entanglement transitivity problem \cite{tabia2022entanglement}.
Very recently, Ref. \cite{shi2025entanglement} introduced the notion of {\em entanglement detection length (EDL)}, defined as the minimum number of particles that must be simultaneously measured in order to detect genuine multipartite entanglement.      
The EDL sets the ultimate in-principle limit to the size of the correlations required for entanglement detection,  and provides useful guidance in designing robust, scalable detection schemes using only few-particle observables. 
However, no systematic experimental demonstrations of entanglement detection have reached the EDL limit.

In this letter, we present a proof-of-concept demonstration of entanglement detection at the EDL limit.  Using a photonic platform,   we detect the multipartite entanglement of states in  three important classes: W states, Dicke states, and cluster states.  Our experiment includes a high-fidelity preparation of  prepare three- and four-qubit states      and the verification of their GME by performing few-particle measurements at the EDL limit. For the four-qubit Dicke state, we show that the EDL witness is more robust to measurement errors than the projector witness that relies on global measurements, representing the first experimental demonstration of the advantage of EDL-sized measurements for entanglement detection. 


\textit{\textbf{EDL witnesses.}}\textbf{---}
A density matrix of  $n$-particles  
is called {\em biseparable}  if it is a convex combination of  separable states over different bipartitions, and is called  \emph{genuinely entangled} if it is not biseparable \cite{guhne2009entanglement}. 
In the following, we will label the $n$ particles by integers from 1 to $n$, and we will represent subsystems of the $n$-particle system by subsets of the set $ \{1,\dots,  n\}$.   
For a genuinely entangled state $\rho$, we say that a collection of subsets $\cS:=\{S_1,  \dots,  S_t\}$ \emph{ detect $\rho$'s GME} if the GME of the state $\rho$ can be certified by performing measurements on the subsystems corresponding to the subsets in $\cS$  \cite{shi2025entanglement}.  

\begin{figure}[t]
   \raggedright
   \includegraphics[scale=0.38]{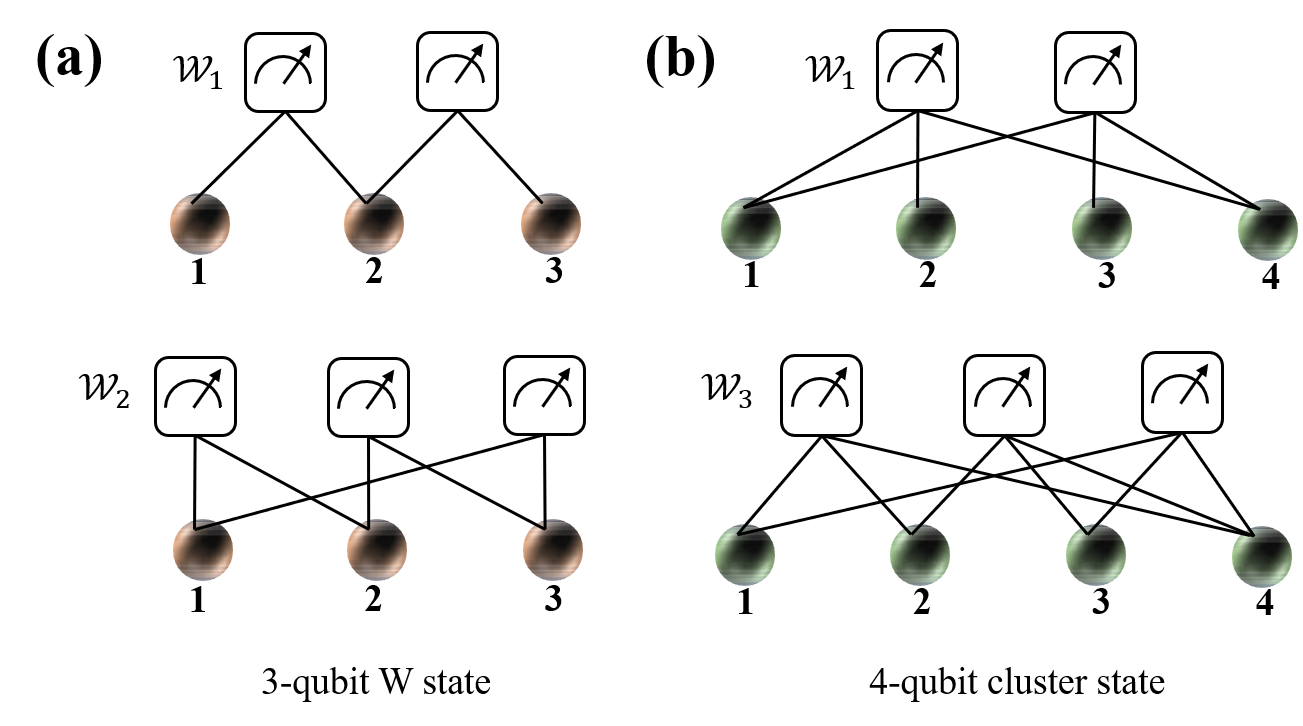}
  \caption{\footnotesize Entanglement detection length (EDL) is the minimum number of parties that must be jointly measured to certify a state’s genuine multipartite entanglement (GME).  For example, the EDL for a W state is $2$, whereas the EDL for a cluster state is $3$. (a) For the $3$-qubit W state, we consider two EDL witnesses, $\cW_1$ and $\cW_2$, constructed from two‑body joint measurements over the pairs $\{12, 23\}$
and $\{12, 23, 13\}$, respectively. (b) For the $4$-qubit  cluster state, we consider two EDL witnesses, $\cW_1$ and $\cW_3$, constructed from three‑body joint measurements over the triplets $\{124, 134\}$
and $\{124, 234, 134\}$, respectively. } \label{figure:edl}
 \end{figure}

For a collection of subsets that detect $\rho$'s GME,  an important question is how many particles are contained in each subset, that is, how many particles have to be jointly measured  in order to detect $\rho$'s GME.   The minimum such number is called the {\em EDL of the state $\rho$}  \cite{shi2025entanglement}, and will hereafter be denoted by $l(\rho)$.  
An illustration of the notion of EDL  in a few special cases is provided  in Fig.~\ref{figure:edl}.

By definition, any experimental scheme for detecting the GME of the state $\rho$ must involve  at least one $k$-particle measurement with $k\geq l(\rho)$.
 Since measurements on larger numbers of particles are generally more sensitive to noise, it is important to design entanglement detection schemes that reach the EDL limit, that is, entanglement detection schemes using measurements on exactly $l(\rho)$ particles.  
 
  To enable detection of GME at the EDL limit, we use a suitable variant of the method of entanglement witnesses \cite{guhne2009entanglement}.  An $n$-particle observable $W$  is a witness for $\rho$'s GME  if $\cW$ has  nonnegative expectation values for all biseparable states, and a negative expectation value on $\rho$.   Using semidefinite programming (SDP), we show that, if a collection of subsets $\cS$ detects $\rho$'s GME, then it is possible to construct a witness $\cW$ of $\rho$'s GME and is supported on $\cS$   (see End Matter), that is, a witness of the form  
    $\cW:=\sum_{S\in \cS}B_{S}\otimes \bbI_{S^c}$,
where $B_{S}$ is a Hermitian operator on the Hilbert space associated to particles in the set $S$, and $\bbI_{S^{\rm c}}$ is an identity operator on the Hilbert space associated to particles in the complementary set  $S^{\rm c} :  =  \{1,\dots , n\}\setminus S$.    Then, we define an {\em EDL witness} as a witness associated to a collection of EDL-sized subsets $\cS$, that is, a collection $\cS$ such that $|S|  =  l(\rho)$ for every subset $S \in \cS$.

  \begin{figure*}[t]
    \centering
    \includegraphics[scale=0.35]{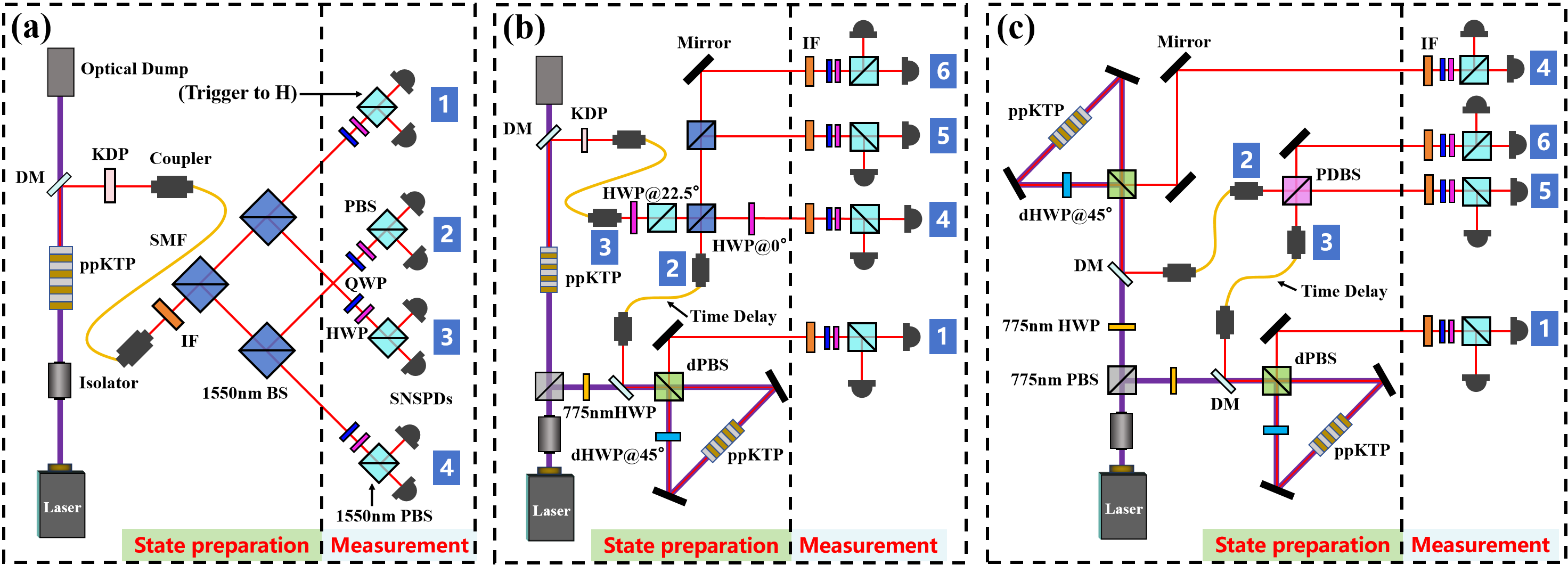}
    \caption{\footnotesize Experimental setup. A mode-locked Ti:sapphire laser (775\,nm, 80\,MHz) pumps a periodically poled KTiOPO$_4$ (ppKTP) crystal, producing photon pairs via type-II spontaneous parametric down-conversion (SPDC). An optical isolator directly after the laser suppresses back-reflections. Residual pump light is removed with a dichroic mirror (DM), and temporal walk-off is compensated using a potassium dihydrogen phosphate (KDP) crystal. After spectral filtering with interference filters (IFs), two-photon interference is implemented on beam-splitter / polarization-dependent beam-splitter (BS/PDBS) networks; the interphoton delay is tuned by a motorized translation stage to ensure temporal overlap. Each output mode is analyzed by a standard quarter-wave-plate–half-wave-plate–polarizing-beam-splitter (QWP–HWP–PBS) module implementing the required local projective settings. Photons are coupled into single-mode fibers (SMFs) and detected with superconducting nanowire single-photon detectors (SNSPDs, $\approx 90\%$ efficiency); fourfold coincidences are recorded with a multichannel coincidence counter. Here, dHWP and dPBS denote a dual-wavelength half-wave plate (775/1550\,nm) and a dual-wavelength polarizing beam splitter (775/1550\,nm), respectively. (a) Four-photon Dicke and heralded three-photon $W$ states. (b) Four-photon $W$ states. (c) Four-photon cluster states.
 }
    \label{figure:D4W4C4}
 \end{figure*}

We now provide EDL witnesses for   four paradigmatic examples of states:  W states of 3 and 4 qubits, 4-qubit Dicke state, and 4-qubit cluster state, explicitly given by
\begin{equation}
\begin{aligned}
    \ket{\text{W}_3}=(&\ket{001}+\ket{010}+\ket{100})/\sqrt{3}, \\
    \ket{\text{W}_4}=(&\ket{0001}+\ket{0010}+\ket{0100}+\ket{1000})/2,\\
    \ket{{\text{D}_4}}=(&\ket{0011}+\ket{0101}+\ket{0110}\\
    &+\ket{1001}+\ket{1010}+\ket{1100})/\sqrt{6},\\
    \ket{\text{C}_4}=(&\ket{0000}+\ket{0011}+\ket{1100}-\ket{1111})/2.
\end{aligned}
\label{eq:5}
\end{equation}
The EDL of these states is   
$l(\ket{\text{W}_3})=l(\ket{\text{W}_4})=l(\ket{\text{D}_4})=2$, and $l(\ket{\text{C}_4})=3$, respectively~\cite{shi2025entanglement},  and EDL witnesses using only 2-particle and 3-particle correlations are constructed in the End Matter (see also Table~\ref{table:summary} for a quick summary.)

\begin{figure}[t]
    \raggedright
    \includegraphics[scale=0.08]{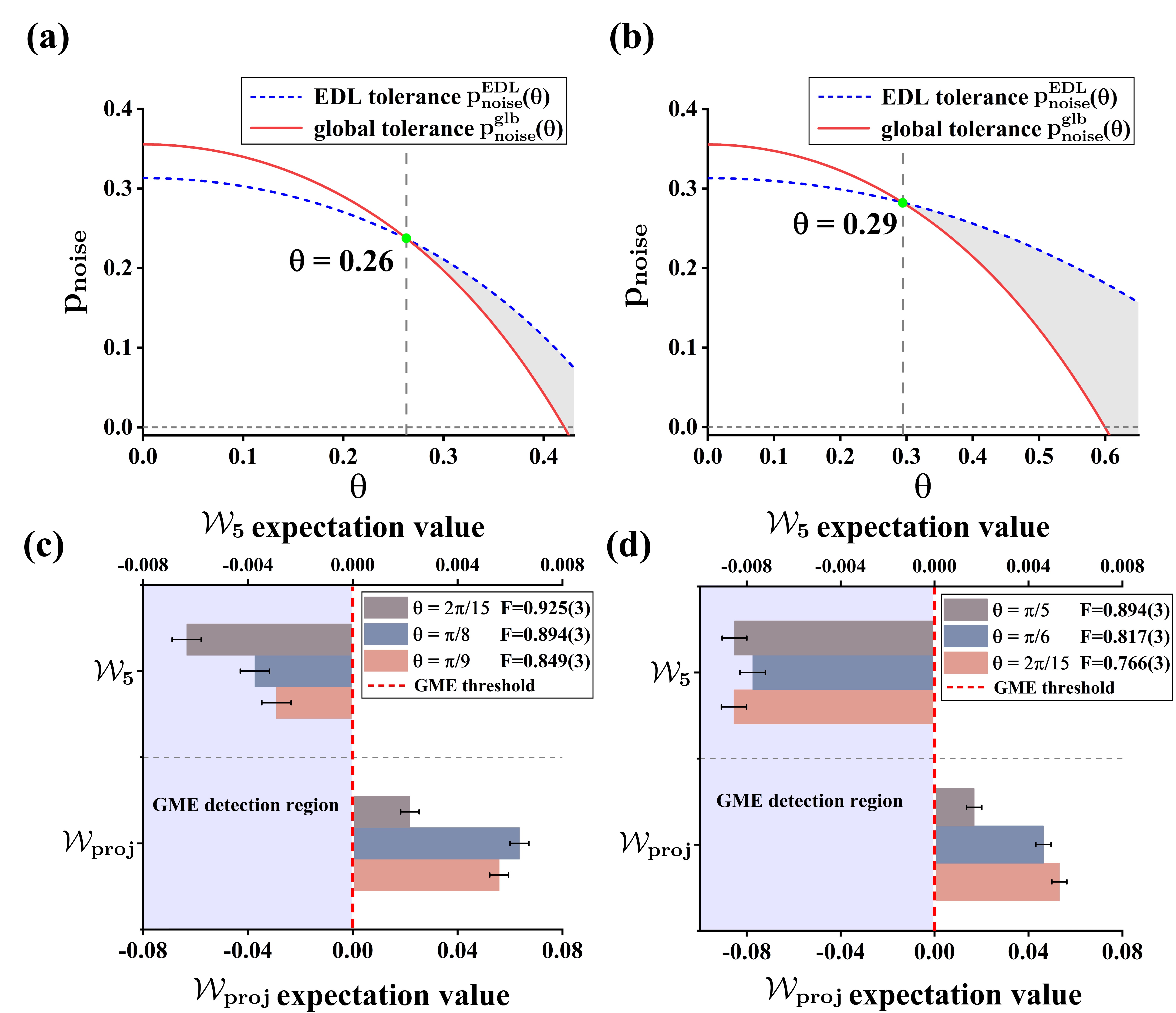}
    \caption{\footnotesize Theoretical predictions of white noise tolerance and experimental comparison results. (a) White noise tolerances of the EDL witness and the projector/global witness versus measurement bases misalignment $\theta$. When $\theta>0.26$, the EDL witness exhibits greater white noise tolerances than the projector witness. (b) White noise tolerances when the misalignment $\theta$ is only introduced on the basis $Y$. When $\theta>0.29$, the EDL witness exhibits greater white noise tolerances than the projector witness. (c) Experimental results of measurement-noise robustness for the four-photon Dicke state: comparison between the global-measure (fidelity) witness $\cW_\text{proj}$ and the EDL two-body witness $W_5$ (designed for $p_{\mathrm{noise}}=0.313$), evaluated under the global measure with measurement bases misalignment $\theta=2\pi/15$, $\theta=\pi/8$, and $\theta=\pi/9$. (d) Comparison results of measurement noise introduced only on the basis $Y$ with measurement bases misalignment$\theta=\pi/5$, $\theta=\pi/6$, and $\theta=2\pi/15$ at all four measurement stations. Error bars indicate one standard deviation obtained by propagating Poissonian counting statistics of the raw counts.}
\label{figure:compare}
\end{figure}

We now analyze the robustness of EDL witnesses to noise and measurement imperfections. 
In general,  entanglement witnesses are robust to small amounts of white noise. If an $n$-qubit state $\rho$ can be detected by the witness $\cW$, then its noisy version
$\rho(p)=(1-p)\rho+p(\bbI/{2^n})$
can be also detected by $\cW$, provided that the amount of white noise $p$ is below  the  maximum value   $p_{\text{noise}}:=\tr(\cW\rho)/[\tr(\cW\rho)-\tr(\cW)/2^n]$. 
For example, let us consider two witnesses for the 4-qubit Dicke state. The first witness  is an  EDL witness based on the collection of subsets  $\{12,23,34,14,13,24\}$. This witness, denoted by  $\cW_5$ in the End Matter, has  noise robustness $p_{\text{noise}}=0.3131$.    The second witness  is the projector witness,  $\cW_{\text{proj}}=(2/3)\bbI-\ketbra{\text{D}_4}{\text{D}_4}$, which requires measurements on all qubits and
  has noise robustness   $p_{\text{noise}}=0.3556$ \cite{toth2009practical}. 
While the projector witness appears to have higher robustness than the EDL witness, it is important to note that the robustness was defined under the assumption that the setup in the laboratory performs exactly the right measurements appearing in the decomposition of the witness.  We now show that, in the presence of measurement imperfections, the EDL witness can offer higher robustness. For instance, 
assume that the measurement bases for each particle are misaligned as follows:
\begin{equation}
\begin{aligned}
      X&\rightarrow X(\theta)=\cos(\theta)X+\sin(\theta)Y,\\
      Y&\rightarrow Y(\theta)=\cos(\theta)Y+\sin(\theta)Z,\\
      Z&\rightarrow Z(\theta)=\cos(\theta)Z+\sin(\theta)X,
\end{aligned}
\label{eq:6}
\end{equation}
we compare the white noise tolerances $p_{\text{noise}}$  of these two witnesses in Fig.~\ref{figure:compare}(a).  When $\theta>0.26$,  the  EDL witness exhibits greater white noise tolerances than the projector witness. Notably, this advantage of the EDL witness persists even when the misalignment angle 
 is introduced solely on $Y$, as shown in Fig.~\ref{figure:compare}(b).

It is also interesting to anlyze different EDL witnesses for the same state, and to observe how the noise robustness depends on the subsets of measured particles.  In the cases under consideration, we find that increasing the  number of  subsets typically increases the robustness. 
For example, we compare two witnesses $\cW_1$ and $\cW_2$ for the three-qubit W state. $\cW_1$ uses measurements on the subsets $\{1,2\}$ and $\{2,3\}$, and tolerates $p_{\text{noise}}=0.1859$ noise, whereas $\cW_2$ uses measurements on all 2-qubit  subsets  and tolerates a higher level of noise $p_{\text{noise}}=0.3039$. 
It is also worth mentioning that another  witness for the 3-qubit W state  was previously known \cite{toth2005entanglement}. This witness is also an EDL witness, as it  uses measurements on all two-qubits subsets.   Compared to the witness $\cW_2$ constructed here, it  appears to have less noise tolerance ($p_{\text{noise}}=0.19$.)

\textit{\textbf{Experimental realization.}}\textbf{---}  Our experimental setup for the demonstration of entanglement detection at the EDL limit is depicted in Fig.\ref{figure:D4W4C4}(a), four spatially indistinguishable photons in the polarization state $\vert HVHV \rangle$, with $H$ and $V$ denoting horizontal and vertical polarizations encoding logical 0 and 1, are generated via second-order emission in collinear type-II SPDC. The photons are then distributed into four spatial modes using three 50:50 beam splitters (BSs). Detection of one photon in each mode indicates the successful preparation of the four-photon Dicke state defined in Eq.\eqref{eq:5} \cite{2007Experimental}. In Fig.~\ref{figure:compare}, we compare our setup with the global measurement method commonly used in entanglement detection experiments~\cite{2010Permutationally,2012Permutationally,2019Efficient,2014Experimental},  showing that the EDL offers increased robustness to imperfection in the choice of measurement bases.  

Using a standard fidelity-measurement protocol~\cite{toth2009practical}, we prepare four-photon Dicke states with fidelities of 0.925(3), 0.894(3), and 0.849(3). To emulate calibration drifts that differ across settings, we applied Eq.~(\ref{eq:6}) with a fixed magnitude $\theta=2\pi/15, \pi/8$ and $\pi/9$ to construct \emph{per-setting} offset axes $X(\theta), Y(\theta), Z(\theta)$. The same three offset measurement directions were used identically at all four measurement stations to evaluate both the global projector/fidelity witness and the EDL witnesses. 
Under these conditions, the fidelity drop to 0.611(4), 0.603(4) and 0.645(4), giving $\cW_{\text{proj}}(2\pi/15) = 0.056(4) > 0, \cW_{\text{proj}}(\pi/8) = 0.064(4) > 0, \cW_{\text{proj}}(\pi/9) = 0.022(4) > 0$ and thus no certification of GME, whereas the EDL witness yield $\cW_5(2\pi/15) = -0.0029(6) < 0, \cW_5(\pi/8) = -0.0037(6) < 0, \cW_5(\pi/9) = -0.0063(6) < 0$, certifying GME. A  comparison of the expectation values is shown in Fig.~\ref{figure:compare}(c). Then, we prepare four-photon Dicke state with fidelity 0.817(3), 0.894(3) and 0.766(3) to perform $\theta=\pi/5, \pi/6$ and $2\pi/15$ only on basis $Y(\theta)$ at all four measurement stations. The experimental results still demonstrate the robustness advantage of the EDL method as shown in Fig.~\ref{figure:compare}(d).

\begin{figure}[t]
    \raggedright
    \includegraphics[scale=0.08]{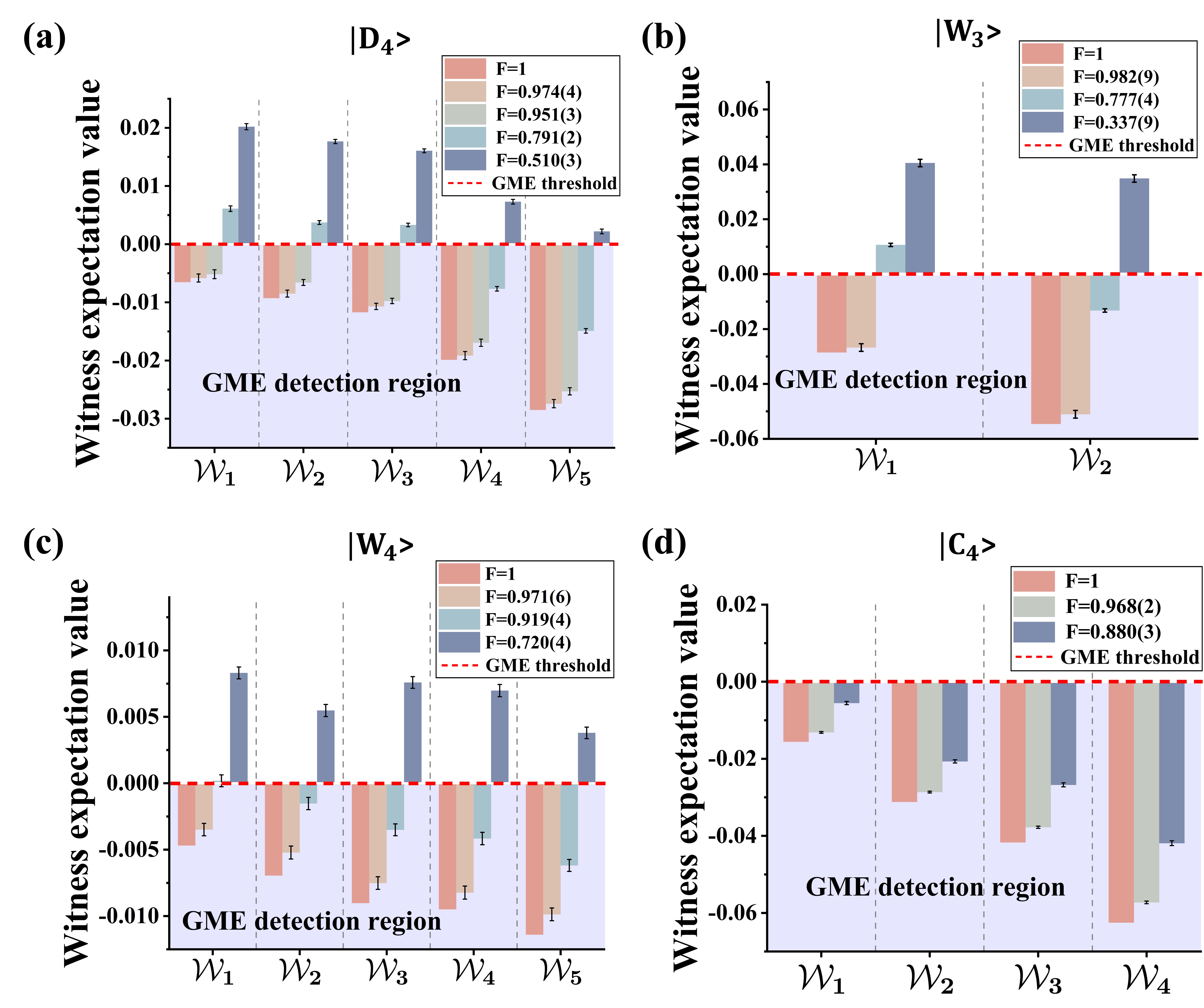}
    \caption{\footnotesize Experimental results. (a) Four-photon Dicke states; (b) three-photon W states; (c) four-photon W states; (d) four-photon cluster states. Bars show measured witness values, with bar colors denoting the experimentally prepared entangled-state fidelities; an ideal $F=1$ reference is included for comparison.
 The red dashed zero line marks the GME threshold (negative values certify genuine multipartite entanglement). EDL-constructed witnesses $\cW_i$ are shown together with their predicted maximum tolerable white-noise fractions $p_i$. Error bars indicate one standard deviation obtained by propagating Poissonian counting statistics of the raw counts.}
\label{figure:all}
\end{figure}

 In addition, we verify the correctness of the witness' predictions within its noise range and at the noise boundary under various noise coefficients. By tuning the experimental parameters, we prepare $\vert \text{D}_4 \rangle$ states with fidelities of 0.974(4), 0.951(3), 0.791(2), and 0.510(3)(see Appendix), corresponding to different $p_\text{noise}$ . This set spans a wide range of noise, allowing a systematic assessment of the robustness of the EDL witnesses and a quantitative comparison between theory and experiment. The corresponding results are shown in Fig.~\ref{figure:all}(a). 

For W states, we prepare states of both 3-qubit and 4-qubit systems. Using the setup in Fig.\ref{figure:D4W4C4}(a), we generate a four-photon Dicke state and herald an $|H\rangle$ photon in mode 1 to prepare the  
three-photon W state. To benchmark the noise resilience of EDL witnesses, we prepare $|\text{W}_{3}\rangle$ with three distinct fidelities—0.982(9), 0.777(4), and 0.337(9) (see Appendix). The two EDL witnesses remain robustly negative across this range, as shown in Fig.\ref{figure:all}(b).


As shown in Fig.\ref{figure:D4W4C4}(b), a pulsed laser is split by a polarizing beam splitter into two paths, generating photon pairs in the states $\vert HV \rangle$ and $(\vert HV \rangle + \vert VH \rangle)/\sqrt{2}$ via SPDC. To ensure temporal overlap of photons in spatial modes 2 and 3 at the beam splitter, a motorized translation stage is used to adjust their relative delay. A Hong–Ou–Mandel interference measurement yields a visibility of 0.989(2) (see Appendix), confirming near-ideal indistinguishability. A four-photon W state is prepared by detecting one photon in mode 1 and one in each of modes 4, 5, and 6 after the beam splitter \cite{2008An}.
The EDL framework provides five EDL witnesses with distinct noise tolerances for the four-photon $\vert \text{W}_4 \rangle$ state. We 
prepare $\vert \text{W}_4 \rangle$ states with fidelities of 0.971(6), 0.919(4), and 0.720(4), and test each witness accordingly (see Appendix). 
As expected from theory, the measurements in Fig.\ref{figure:all}(c) confirm that the state of fidelity 0.919(4) is detected by $\cW_2$ ($p_{\mathrm{noise}} = 0.100$) but fails to be detected by $\cW_1$ ($p_{\mathrm{noise}} = 0.070$).

Within the EDL framework, the EDL for a W state or a Dicke state is $2$, whereas the EDL for a cluster state is $3$.
Guided by this structure, we experimentally implement four EDL witnesses to detect GME in a four-photon cluster state $\vert \text{C}_4 \rangle$. According to the Fig.\ref{figure:D4W4C4}(c), a pulsed laser pumps the ppKTP crystal in a Sagnac loop, generating two entangled photon pairs $|\psi\rangle = \frac{1}{2}|HH\rangle + \frac{\sqrt{3}}{2}|VV\rangle$ through SPDC.  Photons in spatial modes 2 and 3 interfere at a PDBS characterized by transmittance 
$T_V = \frac{1}{3}$, $R_V = \frac{2}{3}$, $T_H = 1$, and $R_H = 0$. A four-photon cluster state \cite{2021Robust} is obtained by detecting one photon in each of output modes 5 and 6, along with one photon in each of local modes 1 and 4. 
Experimental implementations under two distinct conditions (detailed in Appendix), achieving $\vert \text{C}_4 \rangle$ with fidelities of 0.968(2) and 0.880(3). The corresponding experimental results are presented in Fig.\ref{figure:all}(d).

\textit{\textbf{Conclusions.}}\textbf{---}
We reported  the first experimental demonstration of  the entanglement detection length  framework for certifying genuine multipartite entanglement. By preparing a photonic multipartite states with varied configurations and fidelities, we showed that EDL witnesses can achieve accurate certification using only few-body observables supported on the minimum number of particles.  Compared with conventional global-projector methods, the EDL approach appears to have   increased robustness to measurement imperfections,  as measuring less particles generally reduces the impact of basis misalignment.  These results constitute a proof-of-principle that EDL provides a scalable, tomography-free, and noise-robust route to the verification of multipartite entanglement.


\medskip 

{\bf Acknowledgments.} This work was supported by  the Chinese Ministry of Science and Technology (MOST) through grant 2023ZD0300600,  by the National Natural Science Foundation of China under the General Program, Grant No. 12474487, Shanghai Pilot Program for Basic Research
(TQ20240204), and the Science and Technology Commission of Shanghai Municipality (231c1402900).
FS and QZ acknowledge funding from Innovation Program for Quantum Science and Technology via Project 2024ZD0301900, National Natural Science Foundation of China (NSFC) via Project No. 12347104 and No. 12305030, Guangdong Basic and Applied Basic Research Foundation via Project 2023A1515012185, Hong Kong Research Grant Council (RGC) via No. 27300823, N\_HKU718/23, and R6010-23, Guangdong Provincial Quantum Science Strategic Initiative No. GDZX2303007, HKU Seed Fund for Basic Research for New Staff via Project 2201100596.  GC acknowledges Hong Kong Research Grant Council (RGC) through grants  SRFS2021-7S02  and R7035-21F, and the State Key Laboratory of Quantum Information Technologies and Materials.

\begin{center}
    \textbf{End Matter}
\end{center}

\begin{table*}[t]
\centering
\caption{Comparative analysis of multiphoton entanglement state fidelities.} \label{table:comparisons}
\renewcommand\arraystretch{1.4}	
	\renewcommand\tabcolsep{2pt}
\begin{tabular}{|c|c|c|c|}
\hline
\multicolumn{1}{|c|}{{} } & \multicolumn{2}{c|}{{History works} } & \multicolumn{1}{c|}{{Our work} } \\
\hline
\multirow{4}{*}{{$|\text{D}_{4}\rangle$} } & {Kiesel et al. (2007)~\cite{2007Experimental}} & {$0.844\pm0.008$} & \multirow{4}{*}{{$0.974\pm0.004$} } \\
\cline{2-3}
                  & {Chiuri et al. (2012)~\cite{2012Tomographic}} & {$0.870\pm0.010$} &                   \\
\cline{2-3}
                  & {Zhao et al. (2015)~\cite{2015Experimental}} & {$0.904\pm0.004$} &                   \\
\cline{2-3}
                  & {Chen et al. (2023)~\cite{2023On}} & {$0.817\pm0.003$} &                   \\
\hline
\multirow{4}{*}{{$|\text{W}_{3}\rangle$} } & {Lanyon et al. (2009)~\cite{2008Experimentally}} & {$0.900\pm0.030$} & \multirow{4}{*}{{$0.982\pm0.009$} } \\
\cline{2-3}
                  & {Fang et al. (2019)~\cite{2019Three}} & {$0.920\pm0.060$} &                   \\
\cline{2-3}
                  & {Li et al. (2020)~\cite{2020Experimental}} & {$0.866\pm0.007$} &                   \\
\cline{2-3}
                  & {Kumar et al. (2023)~\cite{2023Experimental}} & {$0.873\pm0.011$} &                   \\
\hline
\multirow{4}{*}{{$|\text{C}_{4}\rangle$} } & {Kiesel et al. (2005)~\cite{2005Experimental}} & {$0.741\pm0.013$} & \multirow{4}{*}{{$0.968\pm0.002$} } \\
\cline{2-3}
                  & {Tokunaga et al. (2008)~\cite{2008Generation}} & {$0.860\pm0.015$} &                   \\
\cline{2-3}
                  & {Zhang et al. (2016)~\cite{2016Experimental}} & {$0.952\pm0.003$} &                   \\
\cline{2-3}
                  & {Wu et al. (2021)~\cite{2021Robust}} & {$0.945\pm0.002$} &                   \\
\hline
\end{tabular}
\end{table*}

\begin{table*}[t]
\renewcommand\arraystretch{1.4}	
	\renewcommand\tabcolsep{2pt}
\caption{Summary of EDL wintesses for the 3,4-qubit W states, the 4-qubit Dicke state, and the 4-qubit cluster state}\label{table:summary}
\begin{tabular}{|c|c|c|c|c|}
\hline
States& Marginals & Witnesses & Expectation values &  $p_{\text{noise}}=$\\
\hline
\multirow{2}{*}{$\ket{\text{W}_3}$} &$\cS_2^{[1]}=\{12,  23\}$  &$\cW_1$
 &-0.0285 &0.1859\\
\cline{2-5}
&$\cS_2^{[2]}=\{12, 23, 13\}$ &$\cW_2$
 &-0.0546 &0.3039\\
\hline
\multirow{5}{*}{$\ket{\text{W}_4}$}
&$\cS_2^{[1]}=\{12, 23, 34\}$ &$\cW_1$ &-0.0047   &0.0696\\
\cline{2-5}
&$\cS_2^{[2]}=\{12,23,34,24\}$ &$\cW_2$ &-0.0070 &0.1001 \\
\cline{2-5}
&$\cS_2^{[3]}=\{12, 23, 34, 14\}$ &$\cW_3$ &-0.0090   &0.1261 \\
\cline{2-5}
&$\cS_2^{[4]}=\{12, 23, 34, 24, 14\}$ &$\cW_4$ &-0.0095 
 &0.1319\\
\cline{2-5}
&$\cS_2^{[5]}=\{12, 23, 34, 14, 13, 24\}$ &$\cW_5$ &-0.0114  &0.1541\\
\hline

\multirow{5}{*}{$\ket{\text{D}_4}$}
&$\cS_2^{[1]}=\{12, 23, 34\}$ &$\cW_1$ &-0.0065   &0.0946\\
\cline{2-5}
&$\cS_2^{[2]}=\{12,13,14\}$  &$\cW_2$ &-0.0093 &0.1293\\
\cline{2-5}
&$\cS_2^{[3]}=\{12, 23, 34, 14\}$ &$\cW_3$ &-0.0117  &0.1577\\
\cline{2-5}
&$\cS_2^{[4]}=\{12, 23, 34, 24, 14\}$ &$\cW_4$ &-0.0199 &0.2413\\
\cline{2-5}
&$\cS_2^{[5]}=\{12, 23, 34, 14, 13, 24\}$ &$\cW_5$ &-0.0285  &0.3131\\
\hline

\multirow{4}{*}{$\ket{\text{C}_4}$}
&$\cS_3^{[1]}=\{123, 134\}$ &$\cW_1$ &-0.0156  &$1/5$ \\
\cline{2-5}
&$\cS_3^{[2]}=\{123, 134\}$  &$\cW_2$ &-0.0312  &$1/3$ \\
\cline{2-5}
&$\cS_3^{[3]}=\{123, 134, 234\}$  &$\cW_3$  &-0.0417 
 &$2/5$\\
\cline{2-5}
&$\cS_3^{[4]}=\{123, 124, 134, 234\}$  &$\cW_4$ &-0.0625 
 &$1/2$\\
\hline
\end{tabular}
\end{table*}

The construction of  witnesses using measurement   supported on  a collection of subsets $\cS$ can be formulated as the following semidefinite program (SDP) \cite{jungnitsch2011taming,shi2025entanglement}; specifically, the  SDP for a given entangled state $\rho$ is
\begin{equation}
    \begin{aligned}
        &\alpha(\rho,\cS):=\min \tr(\cW\rho),\\
        &\text{s.t.} \quad \tr(\cW)=1,\\
         &\cW=\sum_{S\in \cS}B_{S}\otimes \bbI_{S^c},\\
         &\text{for all bipartitions $A|A^c$},\\
         &\cW=P_A+Q_A^{T_A}, \quad P_A\geq 0, \quad Q_A\geq 0,
    \end{aligned}
\end{equation}
where, for every subset $S$, $B_{S}$ is a Hermitian operator on $\cH_{S}:  =  \bigotimes_{i\in S} \cH_i$, and $\bbI_{S^c}$ is an identity operator on $\cH_{S^c}$, ${T_A}$  is the partial transpose
with respect to $A$. If $\alpha(\rho,\cS)<0$, then $\cS$ detects $\rho$'s GME,  and the witness is $\cW$, with $p_{\text{noise}}= 2^n\alpha(\rho,\cS)/[2^n\alpha(\rho,\cS)-1]$.

By using the above SDP, we can construct several EDL witnesses for the 3,4-qubit W states, the 4-qubit Dicke state, and the 4-qubit cluster state.
Note that for the EDL witness of $3$-qubit $W$ state,   $Z_1:=Z\otimes I\otimes I$, $X_1X_2:=X\otimes X\otimes I$, $Y_1Y_3:=Y\otimes I\otimes Y$, etc, where $X, Y, Z, I$ are the Pauli matrices.

EDL witnesses for the 3-qubit W state:
 
\begin{enumerate}[1)]
        \item $\cS_2^{[1]}=\{12,23\}$, $\cW_1=[\bbI-0.1748(Z_1+Z_3)-0.415Z_2-0.3426(X_1X_2+Y_1Y_2+X_2X_3+Y_2Y_3)+0.0898(Z_1Z_2+Z_2Z_3)]/8$, $\tr\left(\cW_1\ketbra{\text{W}_3}{\text{W}_3}\right)=-0.0285$, and $p_{\text{noise}}=0.1859$;

\item $\cS_2^{[2]}=\{12,23,13\}$,  $\cW_2=[\bbI-0.2516(Z_1+Z_2+Z_3)-0.2377(
X_1X_2+Y_1Y_2+X_2X_3+Y_2Y_3+X_1X_3+Y_1Y_3)+0.2342(Z_1Z_2+Z_2Z_3+Z_1Z_3)]/8$, $\tr\left(\cW_2\ketbra{\text{W}_3}{\text{W}_3}\right)=-0.0546$, and $p_{\text{noise}}=0.3039$.
    \end{enumerate}

 EDL witnesses for the 4-qubit W state:

\begin{enumerate}[1)]
\item $ \cS_2^{[1]}=\{12,23,34\}$,  $\cW_1=[\bbI-0.2515(Z_2+Z_3)-0.1341(Z_1+Z_4)-0.2176(X_1X_2+Y_1Y_2+X_3X_4+Y_3Y_4)-0.2542(X_2X_3+Y_2Y_3)-0.0052(Z_1Z_2+Z_3Z_4)-0.0885Z_2Z_3]/16$, $\tr\left(\cW_1\ketbra{\text{W}_4}{\text{W}_4}\right)=-0.0047$, and $p_{\text{noise}}=0.0696$;

      \item $\cS_2^{[2]}=\{12,23,34,24\}$,  $\cW_2=[\bbI-0.4166Z_2-0.1428Z_1-0.1782(Z_3+Z_4)-0.2788(X_1X_2+Y_1Y_2)-0.0104Z_1Z_2-0.1561(X_2X_3+Y_2Y_3+X_2X_4+Y_2Y_4)+0.0429(Z_2Z_3+Z_2Z_4)-0.0623(X_3X_4+Y_3Y_4)+0.061Z_3Z_4]/16$, $\tr\left(\cW_2\ketbra{\text{W}_4}{\text{W}_4}\right)=-0.0070$, and $p_{\text{noise}}=0.1001$;

 \item $ \cS_2^{[3]}=\{12,23,34,14\}$, $\cW_3=[\bbI-0.2626(Z_1+Z_2+Z_3+Z_4)-0.1548(X_1X_2+Y_1Y_2+X_2X_3+Y_2Y_3+X_3X_4+Y_3Y_4+X_1X_4+Y_1Y_4)+0.078(Z_1Z_2+Z_2Z_3+Z_3Z_4+Z_1Z_4)]/16$, $\tr\left(\cW_3\ketbra{\text{W}_4}{\text{W}_4}\right)=-0.0090$,  $p_{\text{noise}}=0.1261$,

    \item $\cS_2^{[4]}=\{12,23,34,24,14\}$, 
$\cW_4=[\bbI-0.303(Z_2+Z_4)-0.219(Z_1+Z_3)-0.1634(X_1X_2+Y_1Y_2+X_2X_3+Y_2Y_3+X_3X_4+Y_3Y_4+X_1X_4+Y_1Y_4)+0.0503(Z_1Z_2+Z_2Z_3+Z_3Z_4+Z_1Z_4)+0.0238(X_2X_4+Y_2Y_4)+0.1544Z_2Z_4]/16$, $\tr\left(\cW_4\ketbra{\text{W}_4}{\text{W}_4}\right)=-0.0095$, and $p_{\text{noise}}=0.1319$; 

      \item $\cS_2^{[5]}=\{12,23,34,14,13,24\}$, 
$\cW_5=[\bbI-0.2801(Z_1+Z_2+Z_3+Z_4)-0.1037(X_1X_2+Y_1Y_2+X_2X_3+Y_2Y_3+X_3X_4+Y_3Y_4+X_1X_4+Y_1Y_4+X_1X_3+Y_1Y_3+X_2X_4+Y_2Y_4)+0.0919(Z_1Z_2+Z_2Z_3+Z_3Z_4+Z_1Z_4+Z_1Z_3+Z_2Z_4)]/16$, $\tr\left(\cW_5\ketbra{\text{W}_4}{\text{W}_4}\right)=-0.0114$, and $p_{\text{noise}}=0.1541$.
    \end{enumerate}

 EDL witnesses for the 4-qubit Dicke state:

\begin{enumerate}[1)]
        \item $\cS_2^{[1]}=\{12,23,34\}$,  $\cW_1=[\bbI-0.13(X_1X_2+Y_1Y_2+X_3X_4+Y_3Y_4)-0.5659(X_2X_3+Y_2Y_3)+0.1838(Z_1Z_2+Z_3Z_4)-0.3579Z_2Z_3]/16$, $\tr\left(\cW_1\ketbra{\text{D}_4}{\text{D}_4}\right)=-0.0065$, and $p_{\text{noise}}=0.0946$;

\item $\cS_2^{[2]}=\{12,13,14\}$,  
$\cW_2=[\bbI-0.2711(X_1X_2+Y_1Y_2+X_1X_3+Y_1Y_3+X_1X_4+Y_1Y_4)+0.0641(Z_1Z_2+Z_1Z_3+Z_1Z_4)]/16$, $\tr\left(\cW_2\ketbra{\text{D}_4}{\text{D}_4}\right)=-0.0093$, and $p_{\text{noise}}<0.1293$;

\item $\cS_2^{[3]}=\{12,23,34,14\}$, $\cW_3=[\bbI-0.216(X_1X_2+Y_1Y_2+X_2X_3+Y_2Y_3+X_3X_4+Y_3Y_4+X_1X_4+Y_1Y_4)+0.0266(Z_1Z_2+Z_2Z_3+Z_3Z_4+Z_1Z_4]/16$, $\tr\left(\cW_3\ketbra{\text{D}_4}{\text{D}_4}\right)=-0.0117$, and 
 $p_{\text{noise}}<0.1577$;

      \item $\cS_2^{[4]}=\{12,23,34,24,14\}$, 
$\cW_4=[\bbI-0.1287(X_1X_2+Y_1Y_2+X_2X_3+Y_2Y_3+X_3X_4+Y_3Y_4+X_1X_4+Y_1Y_4)+0.2403(Z_1Z_2+Z_2Z_3+Z_3Z_4+Z_1Z_4)-0.1551(X_2X_4+Y_2Y_4)+0.3132Z_2Z_4]/16$, $\tr\left(\cW_4\ketbra{\text{D}_4}{\text{D}_4}\right)=-0.0199$, and $p_{\text{noise}}=0.2413$;

      \item $\cS_2^{[5]}=\{12,23,34,14,13,24\}$, $\cW_5= [\bbI-0.1228(X_1X_2+Y_1Y_2+X_2X_3+Y_2Y_3+X_3X_4+Y_3Y_4+X_1X_4+Y_1Y_4+X_1X_3+Y_1Y_3+X_2X_4+Y_2Y_4)+0.2368(Z_1Z_2+Z_2Z_3+Z_3Z_4+Z_1Z_4+Z_1Z_3+Z_2Z_4)]/16$, $\tr\left(\cW_5\ketbra{\text{D}_4}{\text{D}_4}\right)=-0.0285$, and 
$p_{\text{noise}}=0.3131$.
    \end{enumerate}

  EDL witnesses for the 4-qubit Cluster state: 

\begin{enumerate}[1)]
      \item $\cS_3^{[1]}=\{124,134\}$,   $\cW_1=[\bbI-0.25(X_1X_2Z_4-Y_1Y_2Z_4+Z_3Z_4+Z_1X_3X_4-Z_1Y_3Y_4)]/16$, $\tr(\cW_1\ketbra{\text{C}_4}{\text{C}_4})=-0.0156$, and $p_{\text{noise}}=1/5$;
\item $\cS_3^{[2]}=\{124,134\}$,  $\cW_2=[\bbI-0.25(X_1X_2Z_4-Y_1Y_2Z_4+Z_1Z_2+Z_3Z_4+Z_1X_3X_4-Z_1Y_3Y_4)]/16$,\\
 $\tr(\cW_2\ketbra{\text{C}_4}{\text{C}_4})=-0.0312$, and $p_{\text{noise}}=1/3$;

\item $\cS_3^{[3]}=\{124,134,234\}$,   $\cW_3=[\bbI+0.1153Z_1Z_2-0.218Z_3Z_4+0.2243(Z_1Y_3Y_4-Z_1X_3X_4+Z_2Y_3Y_4-Z_2X_3X_4)+0.3333(Y_1Y_2Z_4-X_1X_2Z_4)]/16$, $\tr(\cW_3\ketbra{\text{C}_4}{\text{C}_4})=-0.0417 $, and
$p_{\text{noise}}=2/5$;

\item $\cS_3^{[4]}=\{123,124,134,234\}$, $\cW_4=[\bbI+0.0942(Z_1Z_2+Z_3Z_4)+0.2736(Y_1Y_2Z_3-X_1X_2Z_3+Y_1Y_2Z_4-X_1X_2Z_4+Z_1Y_3Y_4-Z_1X_3X_4+Z_2Y_3Y_4-Z_2X_3X_4)]/16$, $\tr(\cW_4\ketbra{\text{C}_4}{\text{C}_4})=-0.0625 $, and $p_{\text{noise}}=1/2$.
  \end{enumerate}

\newpage
\appendix
\begin{widetext}
\section{Determination of the state fidelity and projector witness. }

To describe how  the state $\ket{\psi}$ is prepared experimentally, we usually use the fidelity
\begin{equation}
F_{\ket{\psi}}{(\rho)}:=\tr(\ketbra{\psi}{\psi}\rho). 
\end{equation}
So it is important to decompose $\ketbra{\psi}{\psi}$
using a small number of measurement settings.
The projector witness is give by
\begin{equation}
\cW_{\text{proj}}:=\lambda_{\ket{\psi}}\bbI-\ketbra{\psi}{\psi},
\end{equation}
where $\lambda_{\ket{\psi}}$ is the maximal squared Schmidt coefficient over all bipartitions \cite{bourennane2004experimental}. If  $F_{\ket{\psi}}{(\rho)}> \lambda_{\ket{\psi}}$, then $\rho$ must be genuinely entangled.

\begin{enumerate}[1)]
	\item The 3-qubit W state can be decomposed using five measurement settings \cite{guhne2003investigating}: $Z^{\otimes 3}$, $(Z\pm X)^{\otimes 3}$, $(Z\pm Y)^{\otimes 3}$, i.e.,
	\begin{equation}
	\begin{aligned}
	\ketbra{\text{W}_3}{\text{W}_3}=\frac{1}{24}[&-\bbI-3(ZII+IZI+IIZ)-5(ZZI+ZIZ+IZZ)-7Z^{\otimes 3}\\&
	+(I+Z+X)^{\otimes 3}+(I+Z-X)^{\otimes 3}
	+(I+Z+Y)^{\otimes 3}+(I+Z-Y)^{\otimes 3}]. 
	\end{aligned}   
	\end{equation}
	The projector witness is 
	$\cW_{\text{proj}}=\frac{2}{3}\bbI- \ketbra{\text{W}_3}{\text{W}_3}$.

	\item The 4-qubit W state can be decomposed using seven measurement settings \cite{guhne2007toolbox}: $X^{\otimes 4}$, $Y^{\otimes 4}$, $Z^{\otimes 4}$, $(Z\pm X)^{\otimes 4}$, $(Z\pm Y)^{\otimes 4}$, i.e.,
	\begin{equation}
	\begin{aligned}
	\ketbra{\text{W}_4}{\text{W}_4}=\frac{1}{64}[&-2(ZIII+IZII+IIZI+IIIZ)-4(ZZII+ZIZI+ZIIZ+IZZI+IZIIZ+IIZZ)\\
	&-6(ZZZI+ZZIZ+ZIZZ+IZZZ)-8Z^{\otimes 4}-2X^{\otimes 4}-2Y^{\otimes 4}\\
	&+(I+Z+X)^{\otimes 4}+(I+Z-X)^{\otimes 4}+(I+Z+Y)^{\otimes 4}+(I+Z-Y)^{\otimes 4}];
	\end{aligned}   
	\end{equation}
	The projector witness is $\cW_{\text{proj}}=\frac{3}{4}\bbI- \ketbra{\text{W}_4}{\text{W}_4}$.

	\item The 4-qubit Dicke state can be decomposed using nine measurement settings \cite{toth2009practical}: $X^{\otimes 4}$, $Y^{\otimes 4}$, $Z^{\otimes 4}$, $(X\pm Y)^{\otimes 4}$, $(X\pm Z)^{\otimes 4}$, $(Y\pm Z)^{\otimes 4}$, i.e.,
	\begin{equation}
	\begin{aligned}
	\ketbra{\text{D}_4}{\text{D}_4}=\frac{1}{96}[&4X^{\otimes 4}+2(X+I)^{\otimes 4}+2(X-I)^{\otimes 4}+4Y^{\otimes 4}+2(Y+I)^{\otimes 4}+2(Y-I)^{\otimes 4}+16Z^{\otimes 4}-(Z+I)^{\otimes 4}\\&-(Z-I)^{\otimes 4}-2(X+Z)^{\otimes 4}-2(X-Z)^{\otimes 4}-2(Y+Z)^{\otimes 4}-2(Y-Z)^{\otimes 4}+(X+Y)^{\otimes 4}+(X-Y)^{\otimes 4}]. \\
	\end{aligned}   
	\end{equation}
	The projector witness is $\cW_{\text{proj}}=\frac{2}{3}\bbI- \ketbra{\text{D}_4}{\text{D}_4}$

	\item The 4-qubit  cluster state can be decomposed using nine measurement settings \cite{tokunaga2006fidelity}: $Z^{\otimes 4}$, ZZXX, XXZZ, YYZZ, ZZYY, XYYX, YXXY, XYXY, YXYX i.e.,
	\begin{equation}
	\begin{aligned}
	\ketbra{\text{C}_4}{\text{C}_4}=\frac{1}{16}[&\bbI+ZZII+XXZI+IZXX+IIZZ-YYZI+ZIXX+Z^{\otimes 4}+XYYX\\&+XXIZ-IZYY+YXYX-YYIZ-ZIYY+XYXY+YXXY]. 
	\end{aligned}   
	\end{equation}
	The projector witness is $\cW_{\text{proj}}=\frac{1}{2}\bbI- \ketbra{\text{C}_4}{\text{C}_4}$.
\end{enumerate}

\section{Experimental Details}
\subsection{Preparation of Entanglement Sources}
To further validate the correctness of the EDL method, this work accomplished the preparation of multiple high-fidelity multiphoton entangled states, demonstrating significant improvements over previous studies. Table II presents a comparative analysis between representative works in multiphoton entanglement and our current achievements.

\begin{figure}[t]
	\centering
	\includegraphics[scale=0.3]{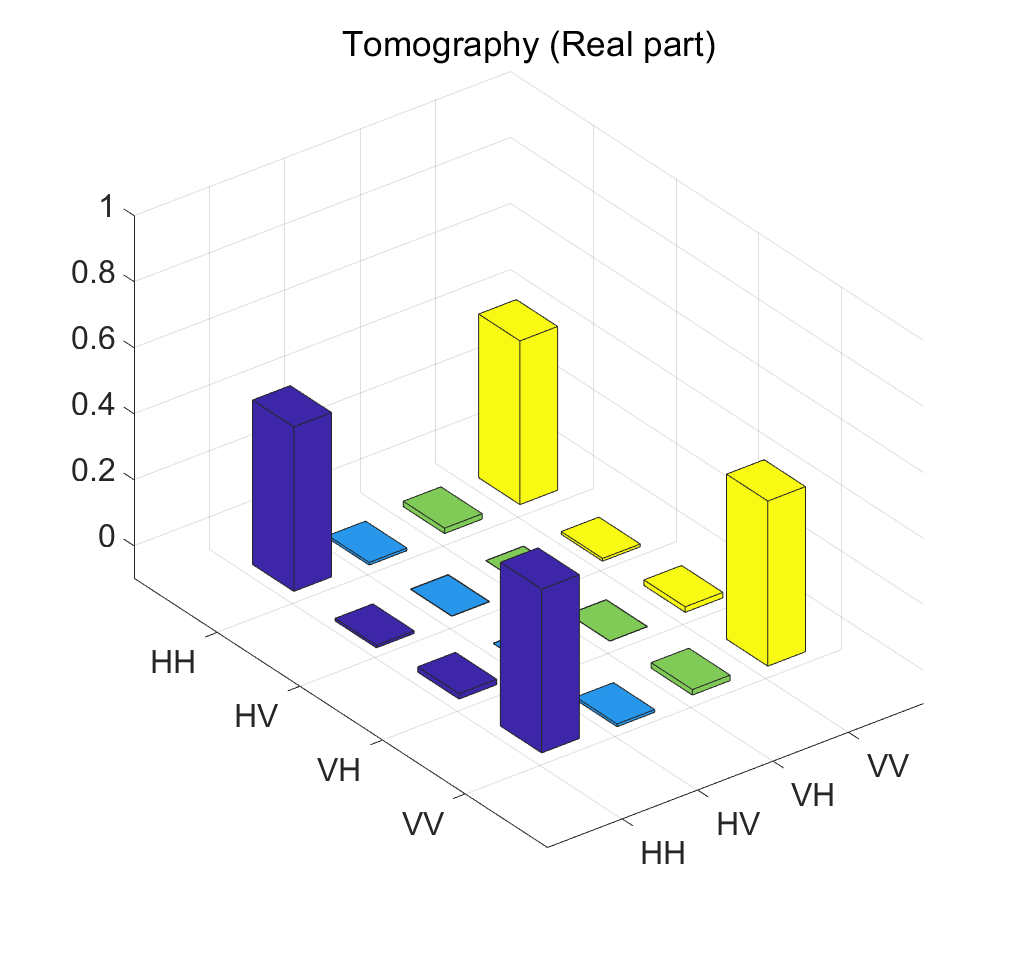}
	\includegraphics[scale=0.3]{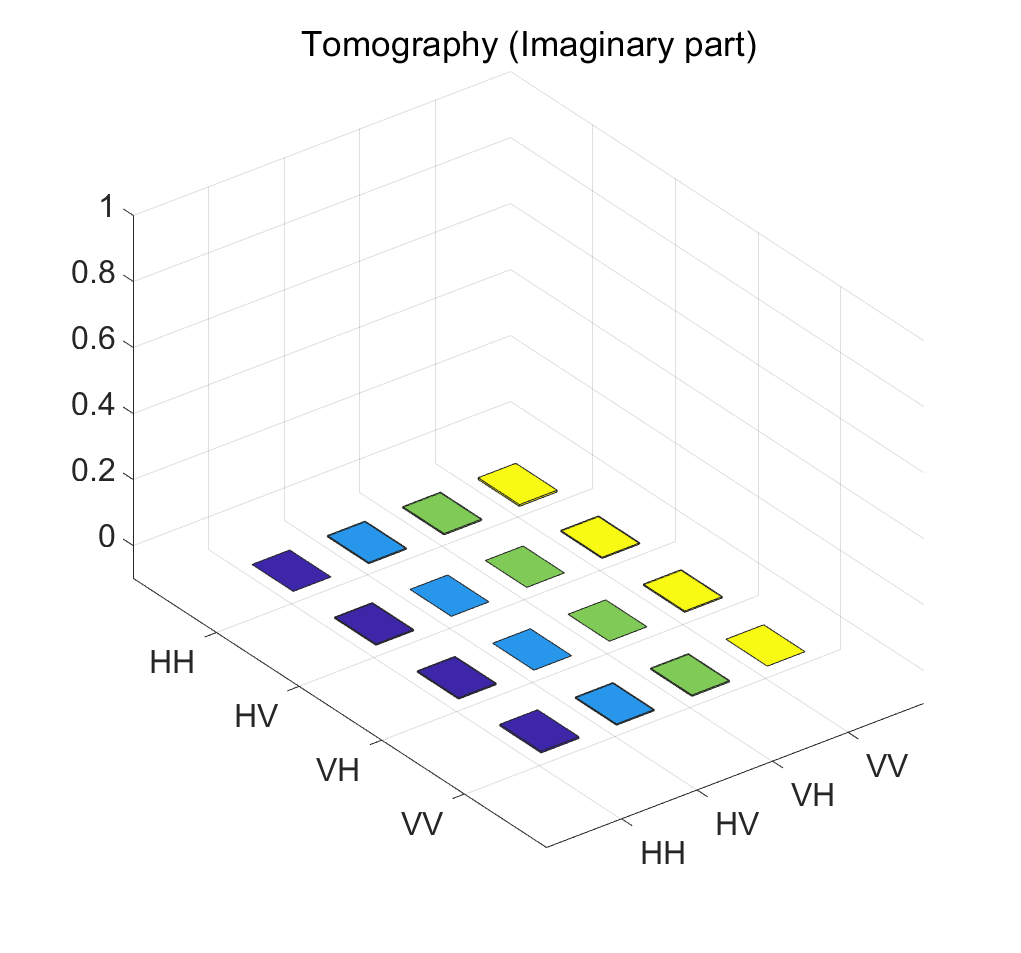}
	\caption{\footnotesize Two-photon quantum state tomography of the entanglement source, which generate a state $|\psi>=\frac{1}{\sqrt2}(\ket{HH}+\ket{VV})$. Realpart on the left, and imaginary part on the right. }
	\label{figure:tomo}
\end{figure}
We developed a polarization-entangled photon source to generate high-fidelity four-photon W states and cluster states using the following configuration: A 775 nm pulsed pump laser (80 MHz repetition rate) with polarization state $\ket{A}$ was injected into an isosceles triangular Sagnac interferometer, where a dPBS (775/1550 nm) split the beam into counter-propagating components that pumped a ppKTP crystal to produce orthogonally polarized $\ket{HV}$ photon pairs at 1550 nm. When the clockwise and counter-clockwise paths were precisely overlapped, the parametric photons became path-indistinguishable at the PBS output ports, generating polarization-entangled state $|\psi>=\frac{1}{\sqrt2}(\ket{HV}+\ket{VH})$, and converting to $|\psi>=\frac{1}{\sqrt2}(\ket{HH}+\ket{VV})$ by a 45° HWP in one path mode. The two output modes were directly characterized via quantum state tomography using a measurement apparatus comprising QWPs, HWPs, and PBS. Under conditions of 1550±10 nm spectral filtering and 800 mW pump power, we measured an average two-photon coincidence rate of 60000 Hz in the $\ket{H}/\ket{V}$ , $\ket{A}/\ket{D}$ and $\ket{R}/\ket{L}$ bases, achieving a two-photon fidelity \cite{2001Measurement} of 0.996±0.002. The quantum state tomography results for the two-photon entangled source are presented in Fig.\ref{figure:tomo}.

\FloatBarrier

\subsection{Four Photon Dicke State}
To generate multiphoton entangled states with controlled fidelity, we systematically varied experimental conditions through the following approaches: (1) spectral filtering of parametric photons using different bandwidths (1550±5nm, 1550±10nm, or no filter); (2) precise temperature tuning of the ppKTP crystal via an electronic oven to modify spectral indistinguishability; (3) adjustment of walk-off compensation (KDP crystal angle) and optical path difference (via 1D translation stage) to control spatial indistinguishability. 

The experimental configurations (measurement bases and conditions) for characterizing four-photon Dicke state the detailed configurations are shown in Tables \ref{tab:D4con}, and, and the results of fidelity and witness measurements are detailed in Tables \ref{tab:D4a}, \ref{tab:D4b}, \ref{tab:D4c} and \ref{tab:D4d}.

\begin{table}[htbp]
	\centering
	\renewcommand\arraystretch{1.4}	
	\renewcommand\tabcolsep{2pt}
	\caption{Experimental configurations of four-photon Dicke state.} 
	\begin{tabular}{|c|c|c|c|c|}
		\hline
		Fidelity & $0.974 \pm 0.004$ & $0.951 \pm 0.003$ & $0.791 \pm 0.002$ & $0.510 \pm 0.003$  \\ \hline
		Spectral filtering & $1550\pm5$ nm & $1550\pm10$ nm & unfiltered & unfiltered  \\ \hline
		Pump Laser Power & 300 mW & 500 mW & 2350 mW & 2350 mW  \\ \hline
		PPKTP crystal temperature & 25$^\circ$C & 25$^\circ$C & 92$^\circ$C & 92$^\circ$C  \\ \hline
		Walk-off compensation & No offset & No offset & No offset & Offset of 9$^\circ$  \\ \hline
		Rate of counting & 0.20/s & 1.60/s & 270/s & 270/s \\ \hline
	\end{tabular}
	\label{tab:D4con}
\end{table}

\begin{table}[htbp]
	\centering
	\renewcommand\arraystretch{1.4}	
	\renewcommand\tabcolsep{2pt}
	\caption{Experimental results of four-photon Dicke state fidelity and witness measurements under fidelity F = 0.974 ± 0.004. Including the required 27 operator expectation values.Under experimental configuration with parametric photon filtering at 1550±5 nm, 300 mW pump power, ppKTP crystal temperature stabilized at 25℃, and zero walk-off compensation offset. Each measurement basis accumulated approximately 3000 events over a 15000-second integration time. Error bars are derived from raw data analysis incorporating Poissonian counting statistics.} 
	\begin{tabular}{|c|c|c|c|c|c|}
		\hline
		${}$ & Operators & Expectation value & ${}$ & Operators & Expectation value \\
		\hline
		$1$ & $X_{1}X_{2}$ & 0.656 ± 0.014 & $15$ & $Z_{1}Z_{2}$ & -0.368 ± 0.017 \\
		\hline
		$2$ & $X_{1}X_{3}$ & 0.659 ± 0.014 & $16$ & $Z_{1}Z_{3}$ & -0.298 ± 0.017 \\
		\hline
		$3$ & $X_{1}X_{4}$ & 0.660 ± 0.014 & $17$ & $Z_{1}Z_{4}$ & -0.328 ± 0.017 \\
		\hline
		$4$ & $X_{2}X_{3}$ & 0.662 ± 0.014 & $18$ & $Z_{2}Z_{3}$ & -0.326 ± 0.017 \\
		\hline
		$5$ & $X_{2}X_{4}$ & 0.656 ± 0.014 & $19$ & $Z_{2}Z_{4}$ & -0.296 ± 0.017 \\
		\hline
		$6$ & $X_{3}X_{4}$ & 0.650 ± 0.014 & $20$ & $Z_{3}Z_{4}$ & -0.367 ± 0.017 \\
		\hline
		$7$ & $X_{1}X_{2}X_{3}X_{4}$ & 0.985 ± 0.003 & $21$ & $Z_{1}Z_{2}Z_{3}Z_{4}$ & 0.982 ± 0.003 \\
		\hline
		$8$ & $Y_{1}Y_{2}$ & 0.657 ± 0.014 & $22$ & $\left[\frac{1}{\sqrt{2}}(X + Z)\right]^{\otimes 4}$ & -0.475 ± 0.016 \\
		\hline
		$9$ & $Y_{1}Y_{3}$ & 0.665 ± 0.014 & $23$ & $\left[\frac{1}{\sqrt{2}}(X - Z)\right]^{\otimes 4}$ & -0.457 ± 0.016 \\
		\hline
		$10$ & $Y_{1}Y_{4}$ & 0.659 ± 0.014 & $24$ & $\left[\frac{1}{\sqrt{2}}(Y + Z)\right]^{\otimes 4}$ & -0.419 ± 0.016 \\
		\hline
		$11$ & $Y_{2}Y_{3}$ & 0.660 ± 0.014 & $25$ & $\left[\frac{1}{\sqrt{2}}(Y - Z)\right]^{\otimes 4}$ & -0.490 ± 0.016 \\
		\hline
		$12$ & $Y_{2}Y_{4}$ & 0.662 ± 0.014 & $26$ & $\left[\frac{1}{\sqrt{2}}(X + Y)\right]^{\otimes 4}$ & 0.979 ± 0.004 \\
		\hline
		$13$ & $Y_{3}Y_{4}$ & 0.647 ± 0.014 & $27$ & $\left[\frac{1}{\sqrt{2}}(X - Y)\right]^{\otimes 4}$ & 0.979 ± 0.004 \\
		\hline
		$14$ & $Y_{1}Y_{2}Y_{3}Y_{4}$ & 0.970 ± 0.005 & & & \\
		\hline
		\multicolumn{2}{|c|}{$F_{\text{D}_{4}}$} & \multicolumn{4}{c|}{0.974 ± 0.004} \\
		\hline
		\multicolumn{2}{|c|}{Witnesses} & \multicolumn{2}{c|}{Experimental results} & \multicolumn{2}{c|}{Theoretical expectation(F=1)} \\
		\hline
		\multicolumn{2}{|c|}{$\cW_{1}$} & \multicolumn{2}{c|}{-0.00582 ± 0.00069} & \multicolumn{2}{c|}{-0.0065} \\
		\hline
		\multicolumn{2}{|c|}{$\cW_{2}$} & \multicolumn{2}{c|}{-0.00850 ± 0.00059} & \multicolumn{2}{c|}{-0.0093} \\
		\hline
		\multicolumn{2}{|c|}{$\cW_{3}$} & \multicolumn{2}{c|}{-0.0107 ± 0.0005} & \multicolumn{2}{c|}{-0.0117} \\
		\hline
		\multicolumn{2}{|c|}{$\cW_{4}$} & \multicolumn{2}{c|}{-0.0192 ± 0.0007} & \multicolumn{2}{c|}{-0.0199} \\
		\hline
		\multicolumn{2}{|c|}{$\cW_{5}$} & \multicolumn{2}{c|}{-0.0274 ± 0.0007} & \multicolumn{2}{c|}{-0.0285} \\
		\hline
	\end{tabular}
	\label{tab:D4a}
\end{table}

\begin{table}[htbp]
	\centering
	\renewcommand\arraystretch{1.4}	
	\renewcommand\tabcolsep{2pt}
	\caption{Experimental results of four-photon Dicke state fidelity and witness measurements under fidelity F = 0.951 ± 0.003. Including the required 27 operator expectation values. Under experimental configuration with parametric photon filtering at 1550±10 nm, 500 mW pump power, ppKTP crystal temperature stabilized at 25℃, and zero walk-off compensation offset. Each measurement basis accumulated approximately 4000 events over a 2500-second integration time. Error bars are derived from raw data analysis incorporating Poissonian counting statistics.}
	\begin{tabular}{|c|c|c|c|c|c|}
		\hline
		${}$ & Operators & Expectation value & ${}$ & Operators & Expectation value \\
		\hline
		$1$ & $X_{1}X_{2}$ & 0.684 ± 0.011 & $15$ & $Z_{1}Z_{2}$ & -0.349 ± 0.015 \\
		\hline
		$2$ & $X_{1}X_{3}$ & 0.611 ± 0.012 & $16$ & $Z_{1}Z_{3}$ & -0.347 ± 0.015 \\
		\hline
		$3$ & $X_{1}X_{4}$ & 0.626 ± 0.012 & $17$ & $Z_{1}Z_{4}$ & -0.275 ± 0.015 \\
		\hline
		$4$ & $X_{2}X_{3}$ & 0.621 ± 0.012 & $18$ & $Z_{2}Z_{3}$ & -0.278 ± 0.015 \\
		\hline
		$5$ & $X_{2}X_{4}$ & 0.606 ± 0.012 & $19$ & $Z_{2}Z_{4}$ & -0.348 ± 0.015 \\
		\hline
		$6$ & $X_{3}X_{4}$ & 0.682 ± 0.011 & $20$ & $Z_{3}Z_{4}$ & -0.347 ± 0.015 \\
		\hline
		$7$ & $X_{1}X_{2}X_{3}X_{4}$ & 0.952 ± 0.005 & $21$ & $Z_{1}Z_{2}Z_{3}Z_{4}$ & 0.945 ± 0.005 \\
		\hline
		$8$ & $Y_{1}Y_{2}$ & 0.676 ± 0.012 & $22$ & $\left[\frac{1}{\sqrt{2}}(X + Z)\right]^{\otimes 4}$ & -0.430 ± 0.015 \\
		\hline
		$9$ & $Y_{1}Y_{3}$ & 0.636 ± 0.012 & $23$ & $\left[\frac{1}{\sqrt{2}}(X - Z)\right]^{\otimes 4}$ & -0.453 ± 0.015 \\
		\hline
		$10$ & $Y_{1}Y_{4}$ & 0.615 ± 0.013 & $24$ & $\left[\frac{1}{\sqrt{2}}(Y + Z)\right]^{\otimes 4}$ & -0.447 ± 0.014 \\
		\hline
		$11$ & $Y_{2}Y_{3}$ & 0.618 ± 0.013 & $25$ & $\left[\frac{1}{\sqrt{2}}(Y - Z)\right]^{\otimes 4}$ & -0.461 ± 0.014 \\
		\hline
		$12$ & $Y_{2}Y_{4}$ & 0.635 ± 0.012 & $26$ & $\left[\frac{1}{\sqrt{2}}(X + Y)\right]^{\otimes 4}$ & 0.977 ± 0.003 \\
		\hline
		$13$ & $Y_{3}Y_{4}$ & 0.678 ± 0.012 & $27$ & $\left[\frac{1}{\sqrt{2}}(X - Y)\right]^{\otimes 4}$ & 0.960 ± 0.004 \\
		\hline
		$14$ & $Y_{1}Y_{2}Y_{3}Y_{4}$ & 0.966 ± 0.004 & & & \\
		\hline
		\multicolumn{2}{|c|}{$F_{\text{D}_{4}}$} & \multicolumn{4}{c|}{0.951 ± 0.003} \\
		\hline
		\multicolumn{2}{|c|}{Witnesses} & \multicolumn{2}{c|}{Experimental results} & \multicolumn{2}{c|}{Theoretical expectation(F=1)} \\
		\hline
		\multicolumn{2}{|c|}{$\cW_{1}$} & \multicolumn{2}{c|}{-0.00517 ± 0.00076} & \multicolumn{2}{c|}{-0.0065} \\
		\hline
		\multicolumn{2}{|c|}{$\cW_{2}$} & \multicolumn{2}{c|}{-0.00659 ± 0.00051} & \multicolumn{2}{c|}{-0.0093} \\
		\hline
		\multicolumn{2}{|c|}{$\cW_{3}$} & \multicolumn{2}{c|}{-0.00977 ± 0.00046} & \multicolumn{2}{c|}{-0.0117} \\
		\hline
		\multicolumn{2}{|c|}{$\cW_{4}$} & \multicolumn{2}{c|}{-0.0169 ± 0.0006} & \multicolumn{2}{c|}{-0.0199} \\
		\hline
		\multicolumn{2}{|c|}{$\cW_{5}$} & \multicolumn{2}{c|}{-0.0253 ± 0.0006} & \multicolumn{2}{c|}{-0.0285} \\
		\hline
	\end{tabular}
	\label{tab:D4b}
\end{table}

\begin{table}[htbp]
	\centering
	\caption{Experimental results of four-photon Dicke state fidelity and witness under fidelity F = 0.791 ± 0.002. Including the required 27 operator expectation values. With no spectral filtering of parametric light, a pump laser power of 2350 mW, a ppKTP crystal temperature of 92℃, and no walk-off compensation offset, approximately 11,000 events were collected per measurement basis over a 40-second integration time. Error bars are derived from raw data analysis incorporating Poissonian counting statistics.} 
	\label{tab:operator_expectations_c}
	\renewcommand\arraystretch{1.4}	
	\renewcommand\tabcolsep{2pt}
	\begin{tabular}{|c|c|c|c|c|c|}
		\hline
		${}$ & Operators & Expectation value & ${}$ & Operators & Expectation value \\
		\hline
		$1$ & $X_{1}X_{2}$ & 0.572 ± 0.008 & $15$ & $Z_{1}Z_{2}$ & -0.351 ± 0.009 \\
		\hline
		$2$ & $X_{1}X_{3}$ & 0.542 ± 0.008 & $16$ & $Z_{1}Z_{3}$ & -0.263 ± 0.009 \\
		\hline
		$3$ & $X_{1}X_{4}$ & 0.429 ± 0.009 & $17$ & $Z_{1}Z_{4}$ & -0.372 ± 0.009 \\
		\hline
		$4$ & $X_{2}X_{3}$ & 0.555 ± 0.007 & $18$ & $Z_{2}Z_{3}$ & -0.366 ± 0.008 \\
		\hline
		$5$ & $X_{2}X_{4}$ & 0.434 ± 0.009 & $19$ & $Z_{2}Z_{4}$ & -0.262 ± 0.009 \\
		\hline
		$6$ & $X_{3}X_{4}$ & 0.396 ± 0.009 & $20$ & $Z_{3}Z_{4}$ & -0.344 ± 0.009 \\
		\hline
		$7$ & $X_{1}X_{2}X_{3}X_{4}$ & 0.520 ± 0.008 & $21$ & $Z_{1}Z_{2}Z_{3}Z_{4}$ & 0.957 ± 0.003 \\
		\hline
		$8$ & $Y_{1}Y_{2}$ & 0.559 ± 0.008 & $22$ & $\left[\frac{1}{\sqrt{2}}(X + Z)\right]^{\otimes 4}$ & -0.371 ± 0.009 \\
		\hline
		$9$ & $Y_{1}Y_{3}$ & 0.563 ± 0.008 & $23$ & $\left[\frac{1}{\sqrt{2}}(X - Z)\right]^{\otimes 4}$ & -0.371 ± 0.009 \\
		\hline
		$10$ & $Y_{1}Y_{4}$ & 0.572 ± 0.007 & $24$ & $\left[\frac{1}{\sqrt{2}}(Y + Z)\right]^{\otimes 4}$ & -0.377 ± 0.009 \\
		\hline
		$11$ & $Y_{2}Y_{3}$ & 0.567 ± 0.008 & $25$ & $\left[\frac{1}{\sqrt{2}}(Y - Z)\right]^{\otimes 4}$ & -0.371 ± 0.009 \\
		\hline
		$12$ & $Y_{2}Y_{4}$ & 0.565 ± 0.008 & $26$ & $\left[\frac{1}{\sqrt{2}}(X + Y)\right]^{\otimes 4}$ & 0.704 ± 0.007 \\
		\hline
		$13$ & $Y_{3}Y_{4}$ & 0.560 ± 0.008 & $27$ & $\left[\frac{1}{\sqrt{2}}(X - Y)\right]^{\otimes 4}$ & 0.714 ± 0.007 \\
		\hline
		$14$ & $Y_{1}Y_{2}Y_{3}Y_{4}$ & 0.709 ± 0.007 & & & \\
		\hline
		\multicolumn{2}{|c|}{$F_{\text{D}_{4}}$} & \multicolumn{4}{c|}{0.791 ± 0.002} \\
		\hline
		\multicolumn{2}{|c|}{Witnesses} & \multicolumn{2}{c|}{Experimental results} & \multicolumn{2}{c|}{Theoretical expectation(F=1)} \\
		\hline
		\multicolumn{2}{|c|}{$\cW_{1}$} & \multicolumn{2}{c|}{0.00610 ± 0.00048} & \multicolumn{2}{c|}{-0.0065} \\
		\hline
		\multicolumn{2}{|c|}{$\cW_{2}$} & \multicolumn{2}{c|}{0.00371 ± 0.00034} & \multicolumn{2}{c|}{-0.0093} \\
		\hline
		\multicolumn{2}{|c|}{$\cW_{3}$} & \multicolumn{2}{c|}{0.00330 ± 0.00031} & \multicolumn{2}{c|}{-0.0117} \\
		\hline
		\multicolumn{2}{|c|}{$\cW_{4}$} & \multicolumn{2}{c|}{-0.00767 ± 0.00039} & \multicolumn{2}{c|}{-0.0199} \\
		\hline
		\multicolumn{2}{|c|}{$\cW_{5}$} & \multicolumn{2}{c|}{-0.0149 ± 0.0004} & \multicolumn{2}{c|}{-0.0285} \\
		\hline
	\end{tabular}
	\label{tab:D4c}
\end{table}

\begin{table}[htbp]
	\centering
	\renewcommand\arraystretch{1.4}	
	\renewcommand\tabcolsep{2pt}
	\caption{Experimental results of four-photon Dicke state fidelity and witness under fidelity F = 0.510 ± 0.003. Including the required 27 operator expectation values. With no spectral filtering of parametric light, a pump laser power of 2350 mW, a ppKTP crystal temperature of 92℃, and walk-off compensation offset of 7°. Approximately 11,000 events were collected per measurement basis over a 40-second integration time. Error bars are derived from raw data analysis incorporating Poissonian counting statistics.}
	\begin{tabular}{|c|c|c|c|c|c|}
		\hline
		${}$ & Operators & Expectation value & ${}$ & Operators & Expectation value \\
		\hline
		$1$ & $X_{1}X_{2}$ & 0.418 ± 0.009 & $15$ & $Z_{1}Z_{2}$ & -0.267 ± 0.009 \\
		\hline
		$2$ & $X_{1}X_{3}$ & 0.403 ± 0.009 & $16$ & $Z_{1}Z_{3}$ & -0.213 ± 0.009 \\
		\hline
		$3$ & $X_{1}X_{4}$ & 0.419 ± 0.009 & $17$ & $Z_{1}Z_{4}$ & -0.275 ± 0.009 \\
		\hline
		$4$ & $X_{2}X_{3}$ & 0.412 ± 0.009 & $18$ & $Z_{2}Z_{3}$ & -0.274 ± 0.009 \\
		\hline
		$5$ & $X_{2}X_{4}$ & 0.442 ± 0.009 & $19$ & $Z_{2}Z_{4}$ & -0.206 ± 0.009 \\
		\hline
		$6$ & $X_{3}X_{4}$ & 0.417 ± 0.009 & $20$ & $Z_{3}Z_{4}$ & -0.257 ± 0.009 \\
		\hline
		$7$ & $X_{1}X_{2}X_{3}X_{4}$ & 0.336 ± 0.009 & $21$ & $Z_{1}Z_{2}Z_{3}Z_{4}$ & 0.542 ± 0.008 \\
		\hline
		$8$ & $Y_{1}Y_{2}$ & 0.413 ± 0.009 & $22$ & $\left[\frac{1}{\sqrt{2}}(X + Z)\right]^{\otimes 4}$ & -0.100 ± 0.010 \\
		\hline
		$9$ & $Y_{1}Y_{3}$ & 0.398 ± 0.009 & $23$ & $\left[\frac{1}{\sqrt{2}}(X - Z)\right]^{\otimes 4}$ & -0.173 ± 0.010 \\
		\hline
		$10$ & $Y_{1}Y_{4}$ & 0.418 ± 0.009 & $24$ & $\left[\frac{1}{\sqrt{2}}(Y + Z)\right]^{\otimes 4}$ & -0.062 ± 0.010 \\
		\hline
		$11$ & $Y_{2}Y_{3}$ & 0.408 ± 0.009 & $25$ & $\left[\frac{1}{\sqrt{2}}(Y - Z)\right]^{\otimes 4}$ & -0.173 ± 0.010 \\
		\hline
		$12$ & $Y_{2}Y_{4}$ & 0.431 ± 0.009 & $26$ & $\left[\frac{1}{\sqrt{2}}(X + Y)\right]^{\otimes 4}$ & 0.256 ± 0.009 \\
		\hline
		$13$ & $Y_{3}Y_{4}$ & 0.405 ± 0.009 & $27$ & $\left[\frac{1}{\sqrt{2}}(X - Y)\right]^{\otimes 4}$ & 0.469 ± 0.009 \\
		\hline
		$14$ & $Y_{1}Y_{2}Y_{3}Y_{4}$ & 0.350 ± 0.009 & & & \\
		\hline
		\multicolumn{2}{|c|}{$F_{\text{D}_{4}}$} & \multicolumn{4}{c|}{0.510 ± 0.003} \\
		\hline
		\multicolumn{2}{|c|}{Witnesses} & \multicolumn{2}{c|}{Experimental results} & \multicolumn{2}{c|}{Theoretical expectation(F=1)} \\
		\hline
		\multicolumn{2}{|c|}{$\cW_{1}$} & \multicolumn{2}{c|}{0.0202 ± 0.0005} & \multicolumn{2}{c|}{-0.0065} \\
		\hline
		\multicolumn{2}{|c|}{$\cW_{2}$} & \multicolumn{2}{c|}{0.0176 ± 0.0004} & \multicolumn{2}{c|}{-0.0093} \\
		\hline
		\multicolumn{2}{|c|}{$\cW_{3}$} & \multicolumn{2}{c|}{0.0160 ± 0.0003} & \multicolumn{2}{c|}{-0.0117} \\
		\hline
		\multicolumn{2}{|c|}{$\cW_{4}$} & \multicolumn{2}{c|}{0.00729 ± 0.00040} & \multicolumn{2}{c|}{-0.0199} \\
		\hline
		\multicolumn{2}{|c|}{$\cW_{5}$} & \multicolumn{2}{c|}{0.00219 ± 0.00041} & \multicolumn{2}{c|}{-0.0285} \\
		\hline
	\end{tabular}
	\label{tab:D4d}
\end{table}

\FloatBarrier
\subsection{Three Photon W State}

The detailed configurations are shown in Tables \ref{tab:W3con}, and results for measuring the fidelity and witness of the three-photon W state are presented in Tables \ref{tab:W3a}, \ref{tab:W3b} and \ref{tab:W3c}.
\begin{table}[htbp]
	\centering
	\normalsize
	\setlength{\tabcolsep}{6pt} 
	\caption{Experimental configurations of three-photon W state.}
	\begin{tabular}{|c|c|c|c|}
		\hline
		Fidelity & $0.982 \pm 0.009$ & $0.777 \pm 0.004$ & $0.337 \pm 0.009$ \\ \hline
		Spectral filtering & $1550\pm10$ nm & unfiltered & unfiltered \\ \hline
		Pump Laser Power & 600 mW & 2350 mW & 2350 mW \\ \hline
		PPKTP crystal temperature & 25$^\circ$C & 93$^\circ$C & 93$^\circ$C \\ \hline
		Walk-off compensation & No offset & No offset & offset of 15$^\circ$ \\ \hline
		Rate of counting & 1.10/s & 140/s & 135/s \\ \hline
	\end{tabular}
	\label{tab:W3con}
\end{table}

\begin{table}[htbp]
	\centering
	\fontsize{8}{13}\selectfont
	\setlength{\tabcolsep}{6pt} 
	\caption{Experimental results of three-photon W state fidelity and witness under fidelity F = 0.982 ± 0.009. Including the required 41 operator expectation values. Under the conditions of a parametric light filtering wavelength of 1550±10nm, a pump laser power of 600mw, a PPKTP crystal temperature of 25℃, and walk-off compensation without offset, the data acquisition time for each basis vector is approximately 3600s, with 3800 events collected. Error bars are derived from raw data analysis incorporating Poissonian counting statistics.}
	\begin{tabular}{|c|c|c|c|c|c|}
		\hline
		${}$ & Operators & Expectation value & ${}$ & Operators & Expectation value \\
		\hline
		$1$ & $Z_{1}$ & -0.270 ± 0.015 & $22$ & $\left[\frac{1}{\sqrt{2}}(Z + Y)\right]_{1}$ & -0.211 ± 0.016 \\
		\hline
		$2$ & $Z_{2}$ & -0.365 ± 0.014 & $23$ & $\left[\frac{1}{\sqrt{2}}(Z + Y)\right]_{2}$ & -0.293 ± 0.016 \\
		\hline
		$3$ & $Z_{3}$ & -0.341 ± 0.015 & $24$ & $\left[\frac{1}{\sqrt{2}}(Z + Y)\right]_{3}$ & -0.217 ± 0.016 \\
		\hline
		$4$ & $Z_{1}Z_{2}$ & -0.333 ± 0.015 & $25$ & $\left[\frac{1}{\sqrt{2}}(Z + Y)\right]_{1}\left[\frac{1}{\sqrt{2}}(Z + Y)\right]_{2}$ & 0.159 ± 0.016 \\
		\hline
		$5$ & $Z_{1}Z_{3}$ & -0.370 ± 0.014 & $26$ & $\left[\frac{1}{\sqrt{2}}(Z + Y)\right]_{1}\left[\frac{1}{\sqrt{2}}(Z + Y)\right]_{3}$ & 0.149 ± 0.016 \\
		\hline
		$6$ & $Z_{2}Z_{3}$ & -0.270 ± 0.015 & $27$ & $\left[\frac{1}{\sqrt{2}}(Z + Y)\right]_{2}\left[\frac{1}{\sqrt{2}}(Z + Y)\right]_{3}$ & 0.146 ± 0.016 \\
		\hline
		$7$ & $Z_{1}Z_{2}Z_{3}$ & 0.950 ± 0.005 & $28$ & $\left[\frac{1}{\sqrt{2}}(Z - Y)\right]^{\otimes 3}$ & -0.309 ± 0.016 \\
		\hline
		$8$ & $\left[\frac{1}{\sqrt{2}}(Z + X)\right]_{1}$ & -0.208 ± 0.016 & $29$ & $\left[\frac{1}{\sqrt{2}}(Z - Y)\right]_{1}$ & -0.236 ± 0.016 \\
		\hline
		$9$ & $\left[\frac{1}{\sqrt{2}}(Z + X)\right]_{2}$ & -0.236 ± 0.015 & $30$ & $\left[\frac{1}{\sqrt{2}}(Z - Y)\right]_{2}$ & -0.264 ± 0.016 \\
		\hline
		$10$ & $\left[\frac{1}{\sqrt{2}}(Z + X)\right]_{3}$ & -0.218 ± 0.016 & $31$ & $\left[\frac{1}{\sqrt{2}}(Z - Y)\right]_{3}$ & -0.241 ± 0.016 \\
		\hline
		$11$ & $\left[\frac{1}{\sqrt{2}}(Z + X)\right]_{1}\left[\frac{1}{\sqrt{2}}(Z + X)\right]_{2}$ & 0.203 ± 0.016 & $32$ & $\left[\frac{1}{\sqrt{2}}(Z - Y)\right]_{1}\left[\frac{1}{\sqrt{2}}(Z - Y)\right]_{2}$ & 0.192 ± 0.016 \\
		\hline
		$12$ & $\left[\frac{1}{\sqrt{2}}(Z + X)\right]_{1}\left[\frac{1}{\sqrt{2}}(Z + X)\right]_{3}$ & 0.193 ± 0.016 & $33$ & $\left[\frac{1}{\sqrt{2}}(Z - Y)\right]_{1}\left[\frac{1}{\sqrt{2}}(Z - Y)\right]_{3}$ & 0.169 ± 0.016 \\
		\hline
		$13$ & $\left[\frac{1}{\sqrt{2}}(Z + X)\right]_{2}\left[\frac{1}{\sqrt{2}}(Z + X)\right]_{3}$ & 0.196 ± 0.016 & $34$ & $\left[\frac{1}{\sqrt{2}}(Z - Y)\right]_{2}\left[\frac{1}{\sqrt{2}}(Z - Y)\right]_{3}$ & 0.168 ± 0.016 \\
		\hline
		$14$ & $\left[\frac{1}{\sqrt{2}}(Z + X)\right]^{\otimes 3}$ & -0.422 ± 0.014 & $35$ & $\left[\frac{1}{\sqrt{2}}(Z - Y)\right]^{\otimes 3}$ & -0.335 ± 0.015 \\
		\hline
		$15$ & $\left[\frac{1}{\sqrt{2}}(Z - X)\right]_{1}$ & -0.249 ± 0.016 & $36$ & $X_{1}X_{2}$ & 0.677 ± 0.011 \\
		\hline
		$16$ & $\left[\frac{1}{\sqrt{2}}(Z - X)\right]_{2}$ & -0.260 ± 0.016 & $37$ & $X_{1}X_{3}$ & 0.614 ± 0.012 \\
		\hline
		$17$ & $\left[\frac{1}{\sqrt{2}}(Z - X)\right]_{3}$ & -0.254 ± 0.016 & $38$ & $X_{2}X_{3}$ & 0.602 ± 0.012 \\
		\hline
		$18$ & $\left[\frac{1}{\sqrt{2}}(Z - X)\right]_{1}\left[\frac{1}{\sqrt{2}}(Z - X)\right]_{2}$ & 0.170 ± 0.016 & $39$ & $Y_{1}Y_{2}$ & 0.676 ± 0.013 \\
		\hline
		$19$ & $\left[\frac{1}{\sqrt{2}}(Z - X)\right]_{1}\left[\frac{1}{\sqrt{2}}(Z - X)\right]_{3}$ & 0.154 ± 0.016 & $40$ & $Y_{1}Y_{3}$ & 0.687 ± 0.012 \\
		\hline
		$20$ & $\left[\frac{1}{\sqrt{2}}(Z - X)\right]_{2}\left[\frac{1}{\sqrt{2}}(Z - X)\right]_{3}$ & 0.127 ± 0.016 & $41$ & $Y_{2}Y_{3}$ & 0.676 ± 0.013 \\
		\hline
		$21$ & $\left[\frac{1}{\sqrt{2}}(Z - X)\right]^{\otimes 3}$ & -0.293 ± 0.016 & $42$ & $I$ & 1 \\
		\hline
		\multicolumn{2}{|c|}{$F_{W_{3}}$} & \multicolumn{4}{c|}{0.982 ± 0.009} \\
		\hline
		\multicolumn{2}{|c|}{Witnesses} & \multicolumn{2}{c|}{Experimental results} & \multicolumn{2}{c|}{Theoretical expectation(F=1)} \\
		\hline
		\multicolumn{2}{|c|}{$\cW_{1}$} & \multicolumn{2}{c|}{-0.027 ± 0.001} & \multicolumn{2}{c|}{-0.029} \\
		\hline
		\multicolumn{2}{|c|}{$\cW_{2}$} & \multicolumn{2}{c|}{-0.051 ± 0.001} & \multicolumn{2}{c|}{-0.055} \\
		\hline
	\end{tabular}
	\label{tab:W3a}
\end{table}

\begin{table}[htbp]
	\centering
	\renewcommand\arraystretch{1.4}	
	\renewcommand\tabcolsep{2pt}
	\caption{Experimental results of three-photon W state fidelity and witness under fidelity F = 0.777 ± 0.004. Including the required 41 operator expectation values. Under the conditions of unfiltered parametric light, a pump laser power of 2350mw, a PPKTP crystal temperature of 93℃, and no walk-off compensation offset, data were collected for each basis vector over approximately 180s with 25,000 events. Error bars are derived from raw data analysis incorporating Poissonian counting statistics.}
	\begin{tabular}{|c|c|c|c|c|c|}
		\hline
		${}$ & Operators & Expectation value & ${}$ & Operators & Expectation value \\
		\hline
		$1$ & $Z_{1}$ & -0.315 ± 0.006 & $22$ & $\left[\frac{1}{\sqrt{2}}(Z + Y)\right]_{1}$ & -0.233 ± 0.006 \\
		\hline
		$2$ & $Z_{2}$ & -0.349 ± 0.006 & $23$ & $\left[\frac{1}{\sqrt{2}}(Z + Y)\right]_{2}$ & -0.248 ± 0.006 \\
		\hline
		$3$ & $Z_{3}$ & -0.314 ± 0.006 & $24$ & $\left[\frac{1}{\sqrt{2}}(Z + Y)\right]_{3}$ & -0.241 ± 0.006 \\
		\hline
		$4$ & $Z_{1}Z_{2}$ & -0.318 ± 0.006 & $25$ & $\left[\frac{1}{\sqrt{2}}(Z + Y)\right]_{1}\left[\frac{1}{\sqrt{2}}(Z + Y)\right]_{2}$ & 0.078 ± 0.006 \\
		\hline
		$5$ & $Z_{1}Z_{3}$ & -0.350 ± 0.006 & $26$ & $\left[\frac{1}{\sqrt{2}}(Z + Y)\right]_{1}\left[\frac{1}{\sqrt{2}}(Z + Y)\right]_{3}$ & 0.081 ± 0.006 \\
		\hline
		$6$ & $Z_{2}Z_{3}$ & -0.311 ± 0.006 & $27$ & $\left[\frac{1}{\sqrt{2}}(Z + Y)\right]_{2}\left[\frac{1}{\sqrt{2}}(Z + Y)\right]_{3}$ & 0.074 ± 0.006 \\
		\hline
		$7$ & $Z_{1}Z_{2}Z_{3}$ & 0.956 ± 0.002 & $28$ & $\left[\frac{1}{\sqrt{2}}(Z - Y)\right]^{\otimes 3}$ & -0.139 ± 0.006 \\
		\hline
		$8$ & $\left[\frac{1}{\sqrt{2}}(Z + X)\right]_{1}$ & -0.233 ± 0.006 & $29$ & $\left[\frac{1}{\sqrt{2}}(Z - Y)\right]_{1}$ & -0.254 ± 0.006 \\
		\hline
		$9$ & $\left[\frac{1}{\sqrt{2}}(Z + X)\right]_{2}$ & -0.251 ± 0.006 & $30$ & $\left[\frac{1}{\sqrt{2}}(Z - Y)\right]_{2}$ & -0.243 ± 0.006 \\
		\hline
		$10$ & $\left[\frac{1}{\sqrt{2}}(Z + X)\right]_{3}$ & -0.236 ± 0.006 & $31$ & $\left[\frac{1}{\sqrt{2}}(Z - Y)\right]_{3}$ & -0.229 ± 0.006 \\
		\hline
		$11$ & $\left[\frac{1}{\sqrt{2}}(Z + X)\right]_{1}\left[\frac{1}{\sqrt{2}}(Z + X)\right]_{2}$ & 0.093 ± 0.006 & $32$ & $\left[\frac{1}{\sqrt{2}}(Z - Y)\right]_{1}\left[\frac{1}{\sqrt{2}}(Z - Y)\right]_{2}$ & 0.052 ± 0.006 \\
		\hline
		$12$ & $\left[\frac{1}{\sqrt{2}}(Z + X)\right]_{1}\left[\frac{1}{\sqrt{2}}(Z + X)\right]_{3}$ & 0.089 ± 0.006 & $33$ & $\left[\frac{1}{\sqrt{2}}(Z - Y)\right]_{1}\left[\frac{1}{\sqrt{2}}(Z - Y)\right]_{3}$ & 0.048 ± 0.006 \\
		\hline
		$13$ & $\left[\frac{1}{\sqrt{2}}(Z + X)\right]_{2}\left[\frac{1}{\sqrt{2}}(Z + X)\right]_{3}$ & 0.087 ± 0.006 & $34$ & $\left[\frac{1}{\sqrt{2}}(Z - Y)\right]_{2}\left[\frac{1}{\sqrt{2}}(Z - Y)\right]_{3}$ & 0.060 ± 0.006 \\
		\hline
		$14$ & $\left[\frac{1}{\sqrt{2}}(Z + X)\right]^{\otimes 3}$ & -0.156 ± 0.006 & $35$ & $\left[\frac{1}{\sqrt{2}}(Z - Y)\right]^{\otimes 3}$ & -0.092 ± 0.006 \\
		\hline
		$15$ & $\left[\frac{1}{\sqrt{2}}(Z - X)\right]_{1}$ & -0.236 ± 0.006 & $36$ & $X_{1}X_{2}$ & 0.437 ± 0.006 \\
		\hline
		$16$ & $\left[\frac{1}{\sqrt{2}}(Z - X)\right]_{2}$ & -0.259 ± 0.006 & $37$ & $X_{1}X_{3}$ & 0.440 ± 0.006 \\
		\hline
		$17$ & $\left[\frac{1}{\sqrt{2}}(Z - X)\right]_{3}$ & -0.238 ± 0.006 & $38$ & $X_{2}X_{3}$ & 0.438 ± 0.006 \\
		\hline
		$18$ & $\left[\frac{1}{\sqrt{2}}(Z - X)\right]_{1}\left[\frac{1}{\sqrt{2}}(Z - X)\right]_{2}$ & 0.059 ± 0.006 & $39$ & $Y_{1}Y_{2}$ & 0.449 ± 0.006 \\
		\hline
		$19$ & $\left[\frac{1}{\sqrt{2}}(Z - X)\right]_{1}\left[\frac{1}{\sqrt{2}}(Z - X)\right]_{3}$ & 0.035 ± 0.006 & $40$ & $Y_{1}Y_{3}$ & 0.452 ± 0.006 \\
		\hline
		$20$ & $\left[\frac{1}{\sqrt{2}}(Z - X)\right]_{2}\left[\frac{1}{\sqrt{2}}(Z - X)\right]_{3}$ & 0.035 ± 0.006 & $41$ & $Y_{2}Y_{3}$ & 0.438 ± 0.006 \\
		\hline
		$21$ & $\left[\frac{1}{\sqrt{2}}(Z - X)\right]^{\otimes 3}$ & -0.081 ± 0.006 & $42$ & $I$ & 1 \\
		\hline
		\multicolumn{2}{|c|}{$F_{\text{W}_{3}}$} & \multicolumn{4}{c|}{0.777 ± 0.004} \\
		\hline
		\multicolumn{2}{|c|}{Witnesses} & \multicolumn{2}{c|}{Experimental results} & \multicolumn{2}{c|}{Theoretical expectation(F=1)} \\
		\hline
		\multicolumn{2}{|c|}{$\cW_{1}$} & \multicolumn{2}{c|}{0.011 ± 0.001} & \multicolumn{2}{c|}{-0.029} \\
		\hline
		\multicolumn{2}{|c|}{$\cW_{2}$} & \multicolumn{2}{c|}{-0.013 ± 0.001} & \multicolumn{2}{c|}{-0.055} \\
		\hline
	\end{tabular}
	\label{tab:W3b}
\end{table}

\begin{table}[htbp]
	\centering
	\renewcommand\arraystretch{1.4}	
	\renewcommand\tabcolsep{2pt}
	\caption{Experimental results of three-photon W state fidelity and witness under fidelity F = 0.337 ± 0.009. Including the required 41 operator expectation values. Under the conditions of unfiltered parametric light, a pump laser power of 2350mw, a PPKTP crystal temperature of 93℃, and no walk-off compensation offset of 15°, data were collected for each basis vector over approximately 180s with 25,000 events. Error bars are derived from raw data analysis incorporating Poissonian counting statistics.}
	\begin{tabular}{|c|c|c|c|c|c|}
		\hline
		${}$ & Operators & Expectation value & ${}$ & Operators & Expectation value \\
		\hline
		$1$ & $Z_{1}$ & -0.033 ± 0.014 & $22$ & $\left[\frac{1}{\sqrt{2}}(Z + Y)\right]_{1}$ & -0.228 ± 0.013 \\
		\hline
		$2$ & $Z_{2}$ & -0.077 ± 0.014 & $23$ & $\left[\frac{1}{\sqrt{2}}(Z + Y)\right]_{2}$ & -0.239 ± 0.013 \\
		\hline
		$3$ & $Z_{3}$ & -0.040 ± 0.014 & $24$ & $\left[\frac{1}{\sqrt{2}}(Z + Y)\right]_{3}$ & -0.218 ± 0.013 \\
		\hline
		$4$ & $Z_{1}Z_{2}$ & -0.041 ± 0.014 & $25$ & $\left[\frac{1}{\sqrt{2}}(Z + Y)\right]_{1}\left[\frac{1}{\sqrt{2}}(Z + Y)\right]_{2}$ & -0.076 ± 0.014 \\
		\hline
		$5$ & $Z_{1}Z_{3}$ & -0.053 ± 0.014 & $26$ & $\left[\frac{1}{\sqrt{2}}(Z + Y)\right]_{1}\left[\frac{1}{\sqrt{2}}(Z + Y)\right]_{3}$ & -0.069 ± 0.014 \\
		\hline
		$6$ & $Z_{2}Z_{3}$ & -0.039 ± 0.014 & $27$ & $\left[\frac{1}{\sqrt{2}}(Z + Y)\right]_{2}\left[\frac{1}{\sqrt{2}}(Z + Y)\right]_{3}$ & -0.067 ± 0.014 \\
		\hline
		$7$ & $Z_{1}Z_{2}Z_{3}$ & -0.152 ± 0.014 & $28$ & $\left[\frac{1}{\sqrt{2}}(Z - Y)\right]^{\otimes 3}$ & 0.152 ± 0.014 \\
		\hline
		$8$ & $\left[\frac{1}{\sqrt{2}}(Z + X)\right]_{1}$ & -0.219 ± 0.013 & $29$ & $\left[\frac{1}{\sqrt{2}}(Z - Y)\right]_{1}$ & 0.156 ± 0.014 \\
		\hline
		$9$ & $\left[\frac{1}{\sqrt{2}}(Z + X)\right]_{2}$ & -0.222 ± 0.013 & $30$ & $\left[\frac{1}{\sqrt{2}}(Z - Y)\right]_{2}$ & 0.135 ± 0.014 \\
		\hline
		$10$ & $\left[\frac{1}{\sqrt{2}}(Z + X)\right]_{3}$ & -0.201 ± 0.013 & $31$ & $\left[\frac{1}{\sqrt{2}}(Z - Y)\right]_{3}$ & 0.152 ± 0.014 \\
		\hline
		$11$ & $\left[\frac{1}{\sqrt{2}}(Z + X)\right]_{1}\left[\frac{1}{\sqrt{2}}(Z + X)\right]_{2}$ & -0.033 ± 0.014 & $32$ & $\left[\frac{1}{\sqrt{2}}(Z - Y)\right]_{1}\left[\frac{1}{\sqrt{2}}(Z - Y)\right]_{2}$ & 0.475 ± 0.012 \\
		\hline
		$12$ & $\left[\frac{1}{\sqrt{2}}(Z + X)\right]_{1}\left[\frac{1}{\sqrt{2}}(Z + X)\right]_{3}$ & -0.022 ± 0.014 & $33$ & $\left[\frac{1}{\sqrt{2}}(Z - Y)\right]_{1}\left[\frac{1}{\sqrt{2}}(Z - Y)\right]_{3}$ & 0.483 ± 0.012 \\
		\hline
		$13$ & $\left[\frac{1}{\sqrt{2}}(Z + X)\right]_{2}\left[\frac{1}{\sqrt{2}}(Z + X)\right]_{3}$ & -0.016 ± 0.014 & $34$ & $\left[\frac{1}{\sqrt{2}}(Z - Y)\right]_{2}\left[\frac{1}{\sqrt{2}}(Z - Y)\right]_{3}$ & 0.486 ± 0.012 \\
		\hline
		$14$ & $\left[\frac{1}{\sqrt{2}}(Z + X)\right]^{\otimes 3}$ & 0.069 ± 0.014 & $35$ & $\left[\frac{1}{\sqrt{2}}(Z - Y)\right]^{\otimes 3}$ & -0.111 ± 0.014 \\
		\hline
		$15$ & $\left[\frac{1}{\sqrt{2}}(Z - X)\right]_{1}$ & 0.118 ± 0.014 & $36$ & $X_{1}X_{2}$ & 0.483 ± 0.012 \\
		\hline
		$16$ & $\left[\frac{1}{\sqrt{2}}(Z - X)\right]_{2}$ & 0.120 ± 0.014 & $37$ & $X_{1}X_{3}$ & 0.483 ± 0.012 \\
		\hline
		$17$ & $\left[\frac{1}{\sqrt{2}}(Z - X)\right]_{3}$ & 0.144 ± 0.013 & $38$ & $X_{2}X_{3}$ & 0.470 ± 0.012 \\
		\hline
		$18$ & $\left[\frac{1}{\sqrt{2}}(Z - X)\right]_{1}\left[\frac{1}{\sqrt{2}}(Z - X)\right]_{2}$ & 0.471 ± 0.012 & $39$ & $Y_{1}Y_{2}$ & 0.422 ± 0.013 \\
		\hline
		$19$ & $\left[\frac{1}{\sqrt{2}}(Z - X)\right]_{1}\left[\frac{1}{\sqrt{2}}(Z - X)\right]_{3}$ & 0.470 ± 0.012 & $40$ & $Y_{1}Y_{3}$ & 0.440 ± 0.012 \\
		\hline
		$20$ & $\left[\frac{1}{\sqrt{2}}(Z - X)\right]_{2}\left[\frac{1}{\sqrt{2}}(Z - X)\right]_{3}$ & 0.472 ± 0.012 & $41$ & $Y_{2}Y_{3}$ & 0.447 ± 0.012 \\
		\hline
		$21$ & $\left[\frac{1}{\sqrt{2}}(Z - X)\right]^{\otimes 3}$ & -0.143 ± 0.013 & $42$ & $I$ & 1 \\
		\hline
		\multicolumn{2}{|c|}{$F_{\text{W}_{3}}$} & \multicolumn{4}{c|}{0.337 ± 0.009} \\
		\hline
		\multicolumn{2}{|c|}{Witnesses} & \multicolumn{2}{c|}{Experimental results} & \multicolumn{2}{c|}{Theoretical expectation(F=1)} \\
		\hline
		\multicolumn{2}{|c|}{$\cW_{1}$} & \multicolumn{2}{c|}{0.040 ± 0.001} & \multicolumn{2}{c|}{-0.029} \\
		\hline
		\multicolumn{2}{|c|}{$\cW_{2}$} & \multicolumn{2}{c|}{0.035 ± 0.001} & \multicolumn{2}{c|}{-0.055} \\
		\hline
	\end{tabular}
	\label{tab:W3c}
\end{table}

\FloatBarrier
\subsection{Four Photon W State}

In the preparation of the four-photon W state, the photon pairs $\vert HH \rangle$ in spatial mode 3 and the photons $\frac{1}{\sqrt2}(\vert H \rangle+\vert V \rangle)$ in spatial mode 2 need to simultaneously arrive at the beam splitter (BS) in a superposition mode. The half-wave plate (HWP) before the entanglement source dPBS and the HWP before the interference BS are adjusted so that both sources contribute an H photon at the BS. We performed Hong-Ou-Mandel quantum interference \cite{1987Measurement} between the entanglement source and the directly pumped spontaneous parametric down-conversion (SPDC) source to test the indistinguishability of the two photons in spatial and frequency modes at the BS, thereby ensuring the consistency of the optical paths of the two spatial modes during state preparation. As shown in Figure \ref{figure:HOM}, at zero delay, a dip appears in the quadruple coincidence count, and the measured interference visibility is 0.989 ± 0.004. 
The detailed configurations are shown in Tables \ref{tab:W4con}, and the measurement results of the fidelity and witness of the four-photon W state are shown in Tables \ref{tab:W4a}, \ref{tab:W4b}, and \ref{tab:W4c}.
\begin{figure}[t]
	\centering
	\includegraphics[scale=0.3]{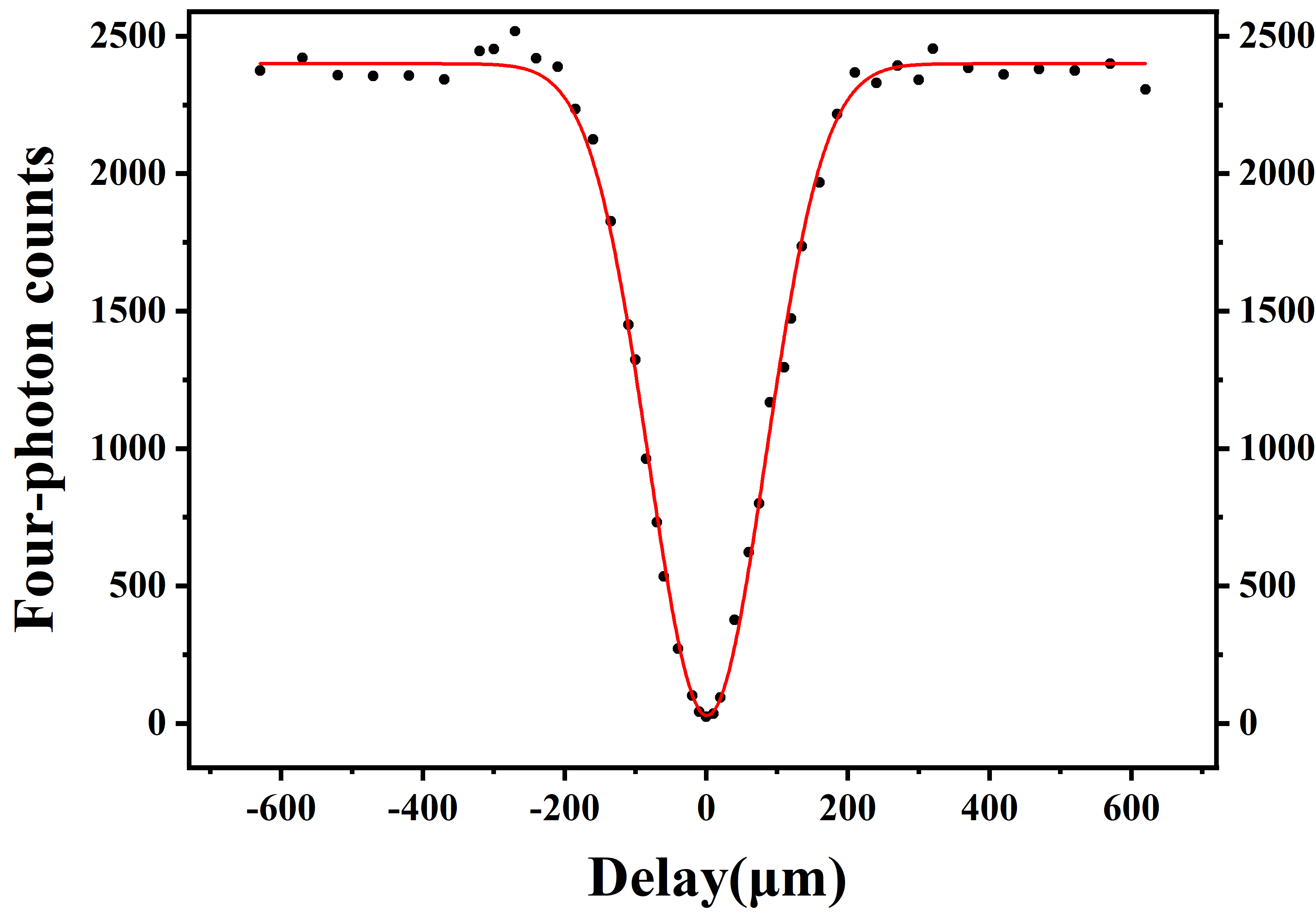}
	\caption{\footnotesize HOM interference curve. By adjusting the one-dimensional displacement stage, the delay amount at the offset center of the displacement stage is plotted as the horizontal coordinate, and the four-body coincidence count within 1200s is plotted as the vertical coordinate; the entanglement source laser pump power is 300mw, and the directly pumped SPDC source power is 319mw. During measurement, 1550±10nm filtering is performed.}
	\label{figure:HOM}
\end{figure}
\begin{table}[h]
	\centering
	\renewcommand\arraystretch{1.4}	
	\renewcommand\tabcolsep{2pt} 
	\caption{Experimental configurations of four-photon W state.}
	\begin{tabular}{|c|c|c|c|}
		\hline
		Fidelity & $0.971 \pm 0.006$ & $0.919 \pm 0.004$ & $0.720 \pm 0.004$ \\ \hline
		Spectral filtering & $1550\pm10$ nm & unfiltered & unfiltered \\ \hline
		Pump Laser Power of entangled source & 200 mW & 1000 mW & 1000 mW \\ \hline
		Pump Laser Power of direct SPDC & 400 mW & 1200 mW & 1200 mW \\ \hline
		PPKTP crystal temperature (both) & 25$^\circ$C & 70$^\circ$C & 70$^\circ$C \\ \hline
		Walk-off compensation & No offset & No offset & offset of 15$^\circ$ \\ \hline
		interference optical range difference & 0 & 0 & 40 $\mu$m \\ \hline
		Rate of counting & 0.20/s & 21/s & 25/s \\ \hline
	\end{tabular}
	\label{tab:W4con}
\end{table}

\begin{table}[h]
	\centering
	\renewcommand\arraystretch{1.4}	
	\renewcommand\tabcolsep{2pt}
	\caption{Experimental results of four-photon W state fidelity and witness under fidelity F = 0.971 ± 0.006. Including the required 90 operator expectation values. Under the conditions of 1550±10nm filtered parametric light, pump power of entangled source in 200mw, and direct SPDC source in 400mw, PPKTP crystal temperature of 25℃, and no walk-off compensation offset, no path difference in BS interference. X/Y/Z bases: 30,000 s (~6,600 events). Other bases: 10,000 s (~2,300 events). Error bars are derived from raw data analysis incorporating Poissonian counting statistics.}
	\begin{tabular}{|c|c|c|c|c|c|c|c|c|}
		\hline
		${}$ & Operators & Expectation value & ${}$ & Operators & Expectation value & ${}$ & Operators & Expectation value \\
		\hline
		$1$ & $Z_{1}$ & 0.546 ± 0.010 & $31$ & $\left[\frac{1}{\sqrt{2}}(Z + X)\right]_{2,3,4}$ & 0.345 ± 0.019 & $61$ & $\left[\frac{1}{\sqrt{2}}(Z + Y)\right]_{2,3,4}$ & 0.310 ± 0.020 \\
		\hline
		$2$ & $Z_{2}$ & 0.473 ± 0.011 & $32$ & $\left[\frac{1}{\sqrt{2}}(Z + X)\right]^{\otimes 4}$ & 0.455 ± 0.018 & $62$ & $\left[\frac{1}{\sqrt{2}}(Z + Y)\right]^{\otimes 4}$ & 0.500 ± 0.018 \\
		\hline
		$3$ & $Z_{3}$ & 0.520 ± 0.011 & $33$ & $\left[\frac{1}{\sqrt{2}}(Z - X)\right]_{1}$ & 0.377 ± 0.018 & $63$ & $\left[\frac{1}{\sqrt{2}}(Z - Y)\right]_{1}$ & 0.365 ± 0.019 \\
		\hline
		$4$ & $Z_{4}$ & 0.449 ± 0.011 & $34$ & $\left[\frac{1}{\sqrt{2}}(Z - X)\right]_{2}$ & 0.353 ± 0.019 & $64$ & $\left[\frac{1}{\sqrt{2}}(Z - Y)\right]_{2}$ & 0.371 ± 0.019 \\
		\hline
		$5$ & $Z_{1}Z_{2}$ & 0.031 ± 0.012 & $35$ & $\left[\frac{1}{\sqrt{2}}(Z - X)\right]_{3}$ & 0.353 ± 0.019 & $65$ & $\left[\frac{1}{\sqrt{2}}(Z - Y)\right]_{3}$ & 0.386 ± 0.019 \\
		\hline
		$6$ & $Z_{1}Z_{3}$ & 0.076 ± 0.012 & $36$ & $\left[\frac{1}{\sqrt{2}}(Z - X)\right]_{4}$ & 0.386 ± 0.018 & $66$ & $\left[\frac{1}{\sqrt{2}}(Z - Y)\right]_{4}$ & 0.320 ± 0.019 \\
		\hline
		$7$ & $Z_{1}Z_{4}$ & 0.000 ± 0.012 & $37$ & $\left[\frac{1}{\sqrt{2}}(Z - X)\right]_{1,2}$ & 0.266 ± 0.019 & $67$ & $\left[\frac{1}{\sqrt{2}}(Z - Y)\right]_{1,2}$ & 0.262 ± 0.019 \\
		\hline
		$8$ & $Z_{2}Z_{3}$ & 0.004 ± 0.012 & $38$ & $\left[\frac{1}{\sqrt{2}}(Z - X)\right]_{1,3}$ & 0.277 ± 0.019 & $68$ & $\left[\frac{1}{\sqrt{2}}(Z - Y)\right]_{1,3}$ & 0.264 ± 0.019 \\
		\hline
		$9$ & $Z_{2}Z_{4}$ & -0.069 ± 0.012 & $39$ & $\left[\frac{1}{\sqrt{2}}(Z - X)\right]_{1,4}$ & 0.206 ± 0.020 & $69$ & $\left[\frac{1}{\sqrt{2}}(Z - Y)\right]_{1,4}$ & 0.216 ± 0.020 \\
		\hline
		$10$ & $Z_{3}Z_{4}$ & -0.027 ± 0.012 & $40$ & $\left[\frac{1}{\sqrt{2}}(Z - X)\right]_{2,3}$ & 0.249 ± 0.020 & $70$ & $\left[\frac{1}{\sqrt{2}}(Z - Y)\right]_{2,3}$ & 0.251 ± 0.020 \\
		\hline
		$11$ & $Z_{1}Z_{2}Z_{3}$ & -0.444 ± 0.011 & $41$ & $\left[\frac{1}{\sqrt{2}}(Z - X)\right]_{2,4}$ & 0.180 ± 0.020 & $71$ & $\left[\frac{1}{\sqrt{2}}(Z - Y)\right]_{2,4}$ & 0.226 ± 0.020 \\
		\hline
		$12$ & $Z_{1}Z_{2}Z_{4}$ & -0.509 ± 0.011 & $42$ & $\left[\frac{1}{\sqrt{2}}(Z - X)\right]_{3,4}$ & 0.238 ± 0.019 & $72$ & $\left[\frac{1}{\sqrt{2}}(Z - Y)\right]_{3,4}$ & 0.211 ± 0.020 \\
		\hline
		$13$ & $Z_{1}Z_{3}Z_{4}$ & -0.467 ± 0.011 & $43$ & $\left[\frac{1}{\sqrt{2}}(Z - X)\right]_{1,2,3}$ & 0.348 ± 0.019 & $73$ & $\left[\frac{1}{\sqrt{2}}(Z - Y)\right]_{1,2,3}$ & 0.326 ± 0.019 \\
		\hline
		$14$ & $Z_{2}Z_{3}Z_{4}$ & -0.535 ± 0.011 & $44$ & $\left[\frac{1}{\sqrt{2}}(Z - X)\right]_{1,2,4}$ & 0.311 ± 0.019 & $74$ & $\left[\frac{1}{\sqrt{2}}(Z - Y)\right]_{1,2,4}$ & 0.328 ± 0.019 \\
		\hline
		$15$ & $Z^{\otimes 4}$ & -0.981 ± 0.002 & $45$ & $\left[\frac{1}{\sqrt{2}}(Z - X)\right]_{1,3,4}$ & 0.305 ± 0.019 & $75$ & $\left[\frac{1}{\sqrt{2}}(Z - Y)\right]_{1,3,4}$ & 0.340 ± 0.019 \\
		\hline
		$16$ & $X^{\otimes 4}$ & 0.007 ± 0.012 & $46$ & $\left[\frac{1}{\sqrt{2}}(Z - X)\right]_{2,3,4}$ & 0.303 ± 0.019 & $76$ & $\left[\frac{1}{\sqrt{2}}(Z - Y)\right]_{2,3,4}$ & 0.288 ± 0.019 \\
		\hline
		$17$ & $Y^{\otimes 4}$ & 0.004 ± 0.012 & $47$ & $\left[\frac{1}{\sqrt{2}}(Z - X)\right]^{\otimes 4}$ & 0.472 ± 0.018 & $77$ & $\left[\frac{1}{\sqrt{2}}(Z - Y)\right]^{\otimes 4}$ & 0.474 ± 0.018 \\
		\hline
		$18$ & $\left[\frac{1}{\sqrt{2}}(Z + X)\right]_{1}$ & 0.366 ± 0.019 & $48$ & $\left[\frac{1}{\sqrt{2}}(Z + Y)\right]_{1}$ & 0.324 ± 0.020 & $78$ & $X_{1}X_{2}$ & 0.469 ± 0.011 \\
		\hline
		$19$ & $\left[\frac{1}{\sqrt{2}}(Z + X)\right]_{2}$ & 0.330 ± 0.019 & $49$ & $\left[\frac{1}{\sqrt{2}}(Z + Y)\right]_{2}$ & 0.346 ± 0.020 & $79$ & $Y_{1}Y_{2}$ & 0.474 ± 0.011 \\
		\hline
		$20$ & $\left[\frac{1}{\sqrt{2}}(Z + X)\right]_{3}$ & 0.349 ± 0.019 & $50$ & $\left[\frac{1}{\sqrt{2}}(Z + Y)\right]_{3}$ & 0.341 ± 0.020 & $80$ & $X_{2}X_{3}$ & 0.486 ± 0.011 \\
		\hline
		$21$ & $\left[\frac{1}{\sqrt{2}}(Z + X)\right]_{4}$ & 0.309 ± 0.020 & $51$ & $\left[\frac{1}{\sqrt{2}}(Z + Y)\right]_{4}$ & 0.353 ± 0.020 & $81$ & $Y_{2}Y_{3}$ & 0.491 ± 0.011 \\
		\hline
		$22$ & $\left[\frac{1}{\sqrt{2}}(Z + X)\right]_{1,2}$ & 0.252 ± 0.020 & $52$ & $\left[\frac{1}{\sqrt{2}}(Z + Y)\right]_{1,2}$ & 0.263 ± 0.020 & $82$ & $X_{3}X_{4}$ & 0.506 ± 0.010 \\
		\hline
		$23$ & $\left[\frac{1}{\sqrt{2}}(Z + X)\right]_{1,3}$ & 0.262 ± 0.020 & $53$ & $\left[\frac{1}{\sqrt{2}}(Z + Y)\right]_{1,3}$ & 0.277 ± 0.020 & $83$ & $Y_{3}Y_{4}$ & 0.500 ± 0.011 \\
		\hline
		$24$ & $\left[\frac{1}{\sqrt{2}}(Z + X)\right]_{1,4}$ & 0.221 ± 0.020 & $54$ & $\left[\frac{1}{\sqrt{2}}(Z + Y)\right]_{1,4}$ & 0.242 ± 0.020 & $84$ & $X_{1}X_{4}$ & 0.473 ± 0.011 \\
		\hline
		$25$ & $\left[\frac{1}{\sqrt{2}}(Z + X)\right]_{2,3}$ & 0.249 ± 0.020 & $55$ & $\left[\frac{1}{\sqrt{2}}(Z + Y)\right]_{2,3}$ & 0.242 ± 0.020 & $85$ & $Y_{1}Y_{4}$ & 0.470 ± 0.011 \\
		\hline
		$26$ & $\left[\frac{1}{\sqrt{2}}(Z + X)\right]_{2,4}$ & 0.303 ± 0.020 & $56$ & $\left[\frac{1}{\sqrt{2}}(Z + Y)\right]_{2,4}$ & 0.208 ± 0.020 & $86$ & $X_{1}X_{3}$ & 0.483 ± 0.011 \\
		\hline
		$27$ & $\left[\frac{1}{\sqrt{2}}(Z + X)\right]_{3,4}$ & 0.253 ± 0.020 & $57$ & $\left[\frac{1}{\sqrt{2}}(Z + Y)\right]_{3,4}$ & 0.240 ± 0.020 & $87$ & $Y_{1}Y_{3}$ & 0.478 ± 0.011 \\
		\hline
		$28$ & $\left[\frac{1}{\sqrt{2}}(Z + X)\right]_{1,2,3}$ & 0.385 ± 0.019 & $58$ & $\left[\frac{1}{\sqrt{2}}(Z + Y)\right]_{1,2,3}$ & 0.420 ± 0.019 & $88$ & $X_{2}X_{4}$ & 0.487 ± 0.011 \\
		\hline
		$29$ & $\left[\frac{1}{\sqrt{2}}(Z + X)\right]_{1,2,4}$ & 0.357 ± 0.019 & $59$ & $\left[\frac{1}{\sqrt{2}}(Z + Y)\right]_{1,2,4}$ & 0.350 ± 0.020 & $89$ & $Y_{2}Y_{4}$ & 0.492 ± 0.011 \\
		\hline
		$30$ & $\left[\frac{1}{\sqrt{2}}(Z + X)\right]_{1,3,4}$ & 0.339 ± 0.019 & $60$ & $\left[\frac{1}{\sqrt{2}}(Z + Y)\right]_{1,3,4}$ & 0.373 ± 0.019 & $90$ & $I$ & 1 \\
		\hline
		\multicolumn{3}{|c|}{$F_{\text{W}_{4}}$} & \multicolumn{6}{c|}{0.971 ± 0.006} \\
		\hline
		\multicolumn{3}{|c|}{Witnesses} & \multicolumn{3}{c|}{Experimental result} & \multicolumn{3}{c|}{Theoretical expectation(F=1)} \\
		\hline
		\multicolumn{3}{|c|}{$\cW_{1}$} & \multicolumn{3}{c|}{-0.00349 ± 0.00047} & \multicolumn{3}{c|}{-0.0047} \\
		\hline
		\multicolumn{3}{|c|}{$\cW_{2}$} & \multicolumn{3}{c|}{-0.00522 ± 0.00049} & \multicolumn{3}{c|}{-0.0070} \\
		\hline
		\multicolumn{3}{|c|}{$\cW_{3}$} & \multicolumn{3}{c|}{-0.00751 ± 0.00047} & \multicolumn{3}{c|}{-0.0090} \\
		\hline
		\multicolumn{3}{|c|}{$\cW_{4}$} & \multicolumn{3}{c|}{-0.00823 ± 0.00049} & \multicolumn{3}{c|}{-0.0095} \\
		\hline
		\multicolumn{3}{|c|}{$\cW_{5}$} & \multicolumn{3}{c|}{-0.00987 ± 0.00048} & \multicolumn{3}{c|}{-0.0114} \\
		\hline
	\end{tabular}
	\label{tab:W4a}
\end{table}

\begin{table}[h]
	\centering
	\renewcommand\arraystretch{1.4}	
	\renewcommand\tabcolsep{2pt}
	\caption{Experimental results of four-photon W state fidelity and witness under fidelity F = 0.919 ± 0.004. Including the required 90 operator expectation values. Under the conditions of unfiltered parametric light, pump power of entangled source in 1000mw, and direct SPDC source in 1200mw, PPKTP crystal temperature of 70℃, and no walk-off compensation offset, no path difference in BS interference. Data were collected for each basis in 300 s (about 7000 events). Error bars are derived from raw data analysis incorporating Poissonian counting statistics.}
	\begin{tabular}{|c|c|c|c|c|c|c|c|c|}
		\hline
		{} & Operators & Expectation value & {} & Operators & Expectation value & {} & Operators & Expectation value \\
		\hline
		\(1\) & \(Z_{1}\) & 0.528 ± 0.010 & \(31\) & \(\left[\frac{1}{\sqrt{2}}(Z + X)\right]_{2,3,4}\) & 0.298 ± 0.011 & \(61\) & \(\left[\frac{1}{\sqrt{2}}(Z + Y)\right]_{2,3,4}\) & 0.297 ± 0.011 \\
		\hline
		\(2\) & \(Z_{2}\) & 0.500 ± 0.010 & \(32\) & \(\left[\frac{1}{\sqrt{2}}(Z + X)\right]^{\otimes 4}\) & 0.395 ± 0.011 & \(62\) & \(\left[\frac{1}{\sqrt{2}}(Z + Y)\right]^{\otimes 4}\) & 0.412 ± 0.011 \\
		\hline
		\(3\) & \(Z_{3}\) & 0.528 ± 0.010 & \(33\) & \(\left[\frac{1}{\sqrt{2}}(Z - X)\right]_{1}\) & 0.392 ± 0.011 & \(63\) & \(\left[\frac{1}{\sqrt{2}}(Z - Y)\right]_{1}\) & 0.393 ± 0.011 \\
		\hline
		\(4\) & \(Z_{4}\) & 0.416 ± 0.010 & \(34\) & \(\left[\frac{1}{\sqrt{2}}(Z - X)\right]_{2}\) & 0.396 ± 0.011 & \(64\) & \(\left[\frac{1}{\sqrt{2}}(Z - Y)\right]_{2}\) & 0.359 ± 0.011 \\
		\hline
		\(5\) & \(Z_{1}Z_{2}\) & 0.057 ± 0.011 & \(35\) & \(\left[\frac{1}{\sqrt{2}}(Z - X)\right]_{3}\) & 0.379 ± 0.011 & \(65\) & \(\left[\frac{1}{\sqrt{2}}(Z - Y)\right]_{3}\) & 0.351 ± 0.011 \\
		\hline
		\(6\) & \(Z_{1}Z_{3}\) & 0.084 ± 0.011 & \(36\) & \(\left[\frac{1}{\sqrt{2}}(Z - X)\right]_{4}\) & 0.299 ± 0.011 & \(66\) & \(\left[\frac{1}{\sqrt{2}}(Z - Y)\right]_{4}\) & 0.337 ± 0.011 \\
		\hline
		\(7\) & \(Z_{1}Z_{4}\) & -0.044 ± 0.011 & \(37\) & \(\left[\frac{1}{\sqrt{2}}(Z - X)\right]_{1,2}\) & 0.263 ± 0.011 & \(67\) & \(\left[\frac{1}{\sqrt{2}}(Z - Y)\right]_{1,2}\) & 0.259 ± 0.011 \\
		\hline
		\(8\) & \(Z_{2}Z_{3}\) & 0.055 ± 0.011 & \(38\) & \(\left[\frac{1}{\sqrt{2}}(Z - X)\right]_{1,3}\) & 0.270 ± 0.011 & \(68\) & \(\left[\frac{1}{\sqrt{2}}(Z - Y)\right]_{1,3}\) & 0.257 ± 0.011 \\
		\hline
		\(9\) & \(Z_{2}Z_{4}\) & -0.077 ± 0.011 & \(39\) & \(\left[\frac{1}{\sqrt{2}}(Z - X)\right]_{1,4}\) & 0.204 ± 0.012 & \(69\) & \(\left[\frac{1}{\sqrt{2}}(Z - Y)\right]_{1,4}\) & 0.199 ± 0.011 \\
		\hline
		\(10\) & \(Z_{3}Z_{4}\) & -0.051 ± 0.011 & \(40\) & \(\left[\frac{1}{\sqrt{2}}(Z - X)\right]_{2,3}\) & 0.246 ± 0.011 & \(70\) & \(\left[\frac{1}{\sqrt{2}}(Z - Y)\right]_{2,3}\) & 0.248 ± 0.011 \\
		\hline
		\(11\) & \(Z_{1}Z_{2}Z_{3}\) & -0.406 ± 0.010 & \(41\) & \(\left[\frac{1}{\sqrt{2}}(Z - X)\right]_{2,4}\) & 0.218 ± 0.012 & \(71\) & \(\left[\frac{1}{\sqrt{2}}(Z - Y)\right]_{2,4}\) & 0.217 ± 0.011 \\
		\hline
		\(12\) & \(Z_{1}Z_{2}Z_{4}\) & -0.512 ± 0.010 & \(42\) & \(\left[\frac{1}{\sqrt{2}}(Z - X)\right]_{3,4}\) & 0.219 ± 0.012 & \(72\) & \(\left[\frac{1}{\sqrt{2}}(Z - Y)\right]_{3,4}\) & 0.223 ± 0.011 \\
		\hline
		\(13\) & \(Z_{1}Z_{3}Z_{4}\) & -0.486 ± 0.010 & \(43\) & \(\left[\frac{1}{\sqrt{2}}(Z - X)\right]_{1,2,3}\) & 0.327 ± 0.011 & \(73\) & \(\left[\frac{1}{\sqrt{2}}(Z - Y)\right]_{1,2,3}\) & 0.332 ± 0.011 \\
		\hline
		\(14\) & \(Z_{2}Z_{3}Z_{4}\) & -0.519 ± 0.010 & \(44\) & \(\left[\frac{1}{\sqrt{2}}(Z - X)\right]_{1,2,4}\) & 0.297 ± 0.011 & \(74\) & \(\left[\frac{1}{\sqrt{2}}(Z - Y)\right]_{1,2,4}\) & 0.302 ± 0.011 \\
		\hline
		\(15\) & \(Z^{\otimes 4}\) & -0.975 ± 0.003 & \(45\) & \(\left[\frac{1}{\sqrt{2}}(Z - X)\right]_{1,3,4}\) & 0.311 ± 0.011 & \(75\) & \(\left[\frac{1}{\sqrt{2}}(Z - Y)\right]_{1,3,4}\) & 0.295 ± 0.011 \\
		\hline
		\(16\) & \(X^{\otimes 4}\) & 0.007 ± 0.012 & \(46\) & \(\left[\frac{1}{\sqrt{2}}(Z - X)\right]_{2,3,4}\) & 0.296 ± 0.011 & \(76\) & \(\left[\frac{1}{\sqrt{2}}(Z - Y)\right]_{2,3,4}\) & 0.306 ± 0.011 \\
		\hline
		\(17\) & \(Y^{\otimes 4}\) & -0.008 ± 0.012 & \(47\) & \(\left[\frac{1}{\sqrt{2}}(Z - X)\right]^{\otimes 4}\) & 0.413 ± 0.011 & \(77\) & \(\left[\frac{1}{\sqrt{2}}(Z - Y)\right]^{\otimes 4}\) & 0.429 ± 0.011 \\
		\hline
		\(18\) & \(\left[\frac{1}{\sqrt{2}}(Z + X)\right]_{1}\) & 0.363 ± 0.011 & \(48\) & \(\left[\frac{1}{\sqrt{2}}(Z + Y)\right]_{1}\) & 0.390 ± 0.011 & \(78\) & \(X_{1}X_{2}\) & 0.407 ± 0.011 \\
		\hline
		\(19\) & \(\left[\frac{1}{\sqrt{2}}(Z + X)\right]_{2}\) & 0.367 ± 0.011 & \(49\) & \(\left[\frac{1}{\sqrt{2}}(Z + Y)\right]_{2}\) & 0.363 ± 0.011 & \(79\) & \(Y_{1}Y_{2}\) & 0.392 ± 0.011 \\
		\hline
		\(20\) & \(\left[\frac{1}{\sqrt{2}}(Z + X)\right]_{3}\) & 0.372 ± 0.011 & \(50\) & \(\left[\frac{1}{\sqrt{2}}(Z + Y)\right]_{3}\) & 0.395 ± 0.011 & \(80\) & \(X_{2}X_{3}\) & 0.437 ± 0.011 \\
		\hline
		\(21\) & \(\left[\frac{1}{\sqrt{2}}(Z + X)\right]_{4}\) & 0.294 ± 0.011 & \(51\) & \(\left[\frac{1}{\sqrt{2}}(Z + Y)\right]_{4}\) & 0.321 ± 0.011 & \(81\) & \(Y_{2}Y_{3}\) & 0.416 ± 0.011 \\
		\hline
		\(22\) & \(\left[\frac{1}{\sqrt{2}}(Z + X)\right]_{1,2}\) & 0.258 ± 0.011 & \(52\) & \(\left[\frac{1}{\sqrt{2}}(Z + Y)\right]_{1,2}\) & 0.250 ± 0.011 & \(82\) & \(X_{3}X_{4}\) & 0.497 ± 0.010 \\
		\hline
		\(23\) & \(\left[\frac{1}{\sqrt{2}}(Z + X)\right]_{1,3}\) & 0.249 ± 0.011 & \(53\) & \(\left[\frac{1}{\sqrt{2}}(Z + Y)\right]_{1,3}\) & 0.257 ± 0.011 & \(83\) & \(Y_{3}Y_{4}\) & 0.498 ± 0.010 \\
		\hline
		\(24\) & \(\left[\frac{1}{\sqrt{2}}(Z + X)\right]_{1,4}\) & 0.179 ± 0.011 & \(54\) & \(\left[\frac{1}{\sqrt{2}}(Z + Y)\right]_{1,4}\) & 0.227 ± 0.011 & \(84\) & \(X_{1}X_{4}\) & 0.432 ± 0.011 \\
		\hline
		\(25\) & \(\left[\frac{1}{\sqrt{2}}(Z + X)\right]_{2,3}\) & 0.246 ± 0.011 & \(55\) & \(\left[\frac{1}{\sqrt{2}}(Z + Y)\right]_{2,3}\) & 0.247 ± 0.011 & \(85\) & \(Y_{1}Y_{4}\) & 0.409 ± 0.011 \\
		\hline
		\(26\) & \(\left[\frac{1}{\sqrt{2}}(Z + X)\right]_{2,4}\) & 0.207 ± 0.011 & \(56\) & \(\left[\frac{1}{\sqrt{2}}(Z + Y)\right]_{2,4}\) & 0.220 ± 0.011 & \(86\) & \(X_{1}X_{3}\) & 0.430 ± 0.011 \\
		\hline
		\(27\) & \(\left[\frac{1}{\sqrt{2}}(Z + X)\right]_{3,4}\) & 0.185 ± 0.011 & \(57\) & \(\left[\frac{1}{\sqrt{2}}(Z + Y)\right]_{3,4}\) & 0.212 ± 0.011 & \(87\) & \(Y_{1}Y_{3}\) & 0.417 ± 0.011 \\
		\hline
		\(28\) & \(\left[\frac{1}{\sqrt{2}}(Z + X)\right]_{1,2,3}\) & 0.304 ± 0.011 & \(58\) & \(\left[\frac{1}{\sqrt{2}}(Z + Y)\right]_{1,2,3}\) & 0.312 ± 0.011 & \(88\) & \(X_{2}X_{4}\) & 0.488 ± 0.010 \\
		\hline
		\(29\) & \(\left[\frac{1}{\sqrt{2}}(Z + X)\right]_{1,2,4}\) & 0.288 ± 0.011 & \(59\) & \(\left[\frac{1}{\sqrt{2}}(Z + Y)\right]_{1,2,4}\) & 0.307 ± 0.011 & \(89\) & \(Y_{2}Y_{4}\) & 0.474 ± 0.010 \\
		\hline
		\(30\) & \(\left[\frac{1}{\sqrt{2}}(Z + X)\right]_{1,3,4}\) & 0.259 ± 0.011 & \(60\) & \(\left[\frac{1}{\sqrt{2}}(Z + Y)\right]_{1,3,4}\) & 0.304 ± 0.011 & \(90\) & \(I\) & 1 \\
		\hline
		\multicolumn{3}{|c|}{\(F_{\text{W}_{4}}\)} & \multicolumn{6}{c|}{0.919 ± 0.004} \\
		\hline
		\multicolumn{3}{|c|}{Witnesses} & \multicolumn{3}{c|}{Experimental result} & \multicolumn{3}{c|}{Theoretical expectation(F=1)} \\
		\hline
		\multicolumn{3}{|c|}{$\cW_{1}$} & \multicolumn{3}{c|}{0.000184 ± 0.000451} & \multicolumn{3}{c|}{-0.0047} \\
		\hline
		\multicolumn{3}{|c|}{$\cW_{2}$} & \multicolumn{3}{c|}{-0.00153 ± 0.00047} & \multicolumn{3}{c|}{-0.0070} \\
		\hline
		\multicolumn{3}{|c|}{$\cW_{3}$} & \multicolumn{3}{c|}{-0.00351 ± 0.00045} & \multicolumn{3}{c|}{-0.0090} \\
		\hline
		\multicolumn{3}{|c|}{$\cW_{4}$} & \multicolumn{3}{c|}{-0.00416 ± 0.00047} & \multicolumn{3}{c|}{-0.0095} \\
		\hline
		\multicolumn{3}{|c|}{$\cW_{5}$} & \multicolumn{3}{c|}{-0.00619 ± 0.00045} & \multicolumn{3}{c|}{-0.0114} \\
		\hline
	\end{tabular}
	\label{tab:W4b}
\end{table}

\begin{table}[h]
	\centering
	\renewcommand\arraystretch{1.4}	
	\renewcommand\tabcolsep{2pt}
	\caption{Experimental results of four-photon W state fidelity and witness under fidelity F = 0.720 ± 0.004. Including the required 90 operator expectation values. Under the conditions of unfiltered parametric light, pump power of entangled source in 1000mw, and direct SPDC source in 1000mw, PPKTP crystal temperature of 70℃, and walk-off compensation no offset, 40$\mu$m path difference in BS interference. Data were collected for each basis in 300 s (about 8000 events). Error bars are derived from raw data analysis incorporating Poissonian counting statistics.}
	\begin{tabular}{|c|c|c|c|c|c|c|c|c|}
		\hline
		{} & Operators & Expectation value & {} & Operators & Expectation value & {} & Operators & Expectation value \\
		\hline
		\(1\) & \(Z_{1}\) & 0.587 ± 0.009 & \(31\) & \(\left[\frac{1}{\sqrt{2}}(Z + X)\right]_{2,3,4}\) & 0.211 ± 0.011 & \(61\) & \(\left[\frac{1}{\sqrt{2}}(Z + Y)\right]_{2,3,4}\) & 0.194 ± 0.011 \\
		\hline
		\(2\) & \(Z_{2}\) & 0.526 ± 0.009 & \(32\) & \(\left[\frac{1}{\sqrt{2}}(Z + X)\right]^{\otimes 4}\) & 0.261 ± 0.011 & \(62\) & \(\left[\frac{1}{\sqrt{2}}(Z + Y)\right]^{\otimes 4}\) & 0.227 ± 0.011 \\
		\hline
		\(3\) & \(Z_{3}\) & 0.557 ± 0.009 & \(33\) & \(\left[\frac{1}{\sqrt{2}}(Z - X)\right]_{1}\) & 0.438 ± 0.010 & \(63\) & \(\left[\frac{1}{\sqrt{2}}(Z - Y)\right]_{1}\) & 0.405 ± 0.010 \\
		\hline
		\(4\) & \(Z_{4}\) & 0.306 ± 0.010 & \(34\) & \(\left[\frac{1}{\sqrt{2}}(Z - X)\right]_{2}\) & 0.432 ± 0.010 & \(64\) & \(\left[\frac{1}{\sqrt{2}}(Z - Y)\right]_{2}\) & 0.408 ± 0.010 \\
		\hline
		\(5\) & \(Z_{1}Z_{2}\) & 0.132 ± 0.011 & \(35\) & \(\left[\frac{1}{\sqrt{2}}(Z - X)\right]_{3}\) & 0.404 ± 0.010 & \(65\) & \(\left[\frac{1}{\sqrt{2}}(Z - Y)\right]_{3}\) & 0.404 ± 0.010 \\
		\hline
		\(6\) & \(Z_{1}Z_{3}\) & 0.166 ± 0.011 & \(36\) & \(\left[\frac{1}{\sqrt{2}}(Z - X)\right]_{4}\) & 0.218 ± 0.011 & \(66\) & \(\left[\frac{1}{\sqrt{2}}(Z - Y)\right]_{4}\) & 0.253 ± 0.011 \\
		\hline
		\(7\) & \(Z_{1}Z_{4}\) & -0.100 ± 0.011 & \(37\) & \(\left[\frac{1}{\sqrt{2}}(Z - X)\right]_{1,2}\) & 0.235 ± 0.011 & \(67\) & \(\left[\frac{1}{\sqrt{2}}(Z - Y)\right]_{1,2}\) & 0.220 ± 0.011 \\
		\hline
		\(8\) & \(Z_{2}Z_{3}\) & 0.105 ± 0.011 & \(38\) & \(\left[\frac{1}{\sqrt{2}}(Z - X)\right]_{1,3}\) & 0.228 ± 0.011 & \(68\) & \(\left[\frac{1}{\sqrt{2}}(Z - Y)\right]_{1,3}\) & 0.236 ± 0.011 \\
		\hline
		\(9\) & \(Z_{2}Z_{4}\) & -0.162 ± 0.011 & \(39\) & \(\left[\frac{1}{\sqrt{2}}(Z - X)\right]_{1,4}\) & -0.036 ± 0.011 & \(69\) & \(\left[\frac{1}{\sqrt{2}}(Z - Y)\right]_{1,4}\) & -0.026 ± 0.011 \\
		\hline
		\(10\) & \(Z_{3}Z_{4}\) & -0.126 ± 0.011 & \(40\) & \(\left[\frac{1}{\sqrt{2}}(Z - X)\right]_{2,3}\) & 0.249 ± 0.011 & \(70\) & \(\left[\frac{1}{\sqrt{2}}(Z - Y)\right]_{2,3}\) & 0.221 ± 0.011 \\
		\hline
		\(11\) & \(Z_{1}Z_{2}Z_{3}\) & -0.296 ± 0.010 & \(41\) & \(\left[\frac{1}{\sqrt{2}}(Z - X)\right]_{2,4}\) & 0.114 ± 0.011 & \(71\) & \(\left[\frac{1}{\sqrt{2}}(Z - Y)\right]_{2,4}\) & 0.131 ± 0.011 \\
		\hline
		\(12\) & \(Z_{1}Z_{2}Z_{4}\) & -0.551 ± 0.009 & \(42\) & \(\left[\frac{1}{\sqrt{2}}(Z - X)\right]_{3,4}\) & 0.135 ± 0.011 & \(72\) & \(\left[\frac{1}{\sqrt{2}}(Z - Y)\right]_{3,4}\) & 0.119 ± 0.011 \\
		\hline
		\(13\) & \(Z_{1}Z_{3}Z_{4}\) & -0.516 ± 0.009 & \(43\) & \(\left[\frac{1}{\sqrt{2}}(Z - X)\right]_{1,2,3}\) & 0.210 ± 0.011 & \(73\) & \(\left[\frac{1}{\sqrt{2}}(Z - Y)\right]_{1,2,3}\) & 0.210 ± 0.011 \\
		\hline
		\(14\) & \(Z_{2}Z_{3}Z_{4}\) & -0.574 ± 0.009 & \(44\) & \(\left[\frac{1}{\sqrt{2}}(Z - X)\right]_{1,2,4}\) & 0.050 ± 0.011 & \(74\) & \(\left[\frac{1}{\sqrt{2}}(Z - Y)\right]_{1,2,4}\) & 0.057 ± 0.011 \\
		\hline
		\(15\) & \(Z^{\otimes 4}\) & -0.976 ± 0.002 & \(45\) & \(\left[\frac{1}{\sqrt{2}}(Z - X)\right]_{1,3,4}\) & 0.076 ± 0.011 & \(75\) & \(\left[\frac{1}{\sqrt{2}}(Z - Y)\right]_{1,3,4}\) & 0.066 ± 0.011 \\
		\hline
		\(16\) & \(X^{\otimes 4}\) & 0.017 ± 0.011 & \(46\) & \(\left[\frac{1}{\sqrt{2}}(Z - X)\right]_{2,3,4}\) & 0.158 ± 0.011 & \(76\) & \(\left[\frac{1}{\sqrt{2}}(Z - Y)\right]_{2,3,4}\) & 0.155 ± 0.011 \\
		\hline
		\(17\) & \(Y^{\otimes 4}\) & 0.020 ± 0.011 & \(47\) & \(\left[\frac{1}{\sqrt{2}}(Z - X)\right]^{\otimes 4}\) & 0.162 ± 0.011 & \(77\) & \(\left[\frac{1}{\sqrt{2}}(Z - Y)\right]^{\otimes 4}\) & 0.164 ± 0.011 \\
		\hline
		\(18\) & \(\left[\frac{1}{\sqrt{2}}(Z + X)\right]_{1}\) & 0.398 ± 0.011 & \(48\) & \(\left[\frac{1}{\sqrt{2}}(Z + Y)\right]_{1}\) & 0.438 ± 0.010 & \(78\) & \(X_{1}X_{2}\) & 0.307 ± 0.011 \\
		\hline
		\(19\) & \(\left[\frac{1}{\sqrt{2}}(Z + X)\right]_{2}\) & 0.383 ± 0.011 & \(49\) & \(\left[\frac{1}{\sqrt{2}}(Z + Y)\right]_{2}\) & 0.404 ± 0.010 & \(79\) & \(Y_{1}Y_{2}\) & 0.302 ± 0.011 \\
		\hline
		\(20\) & \(\left[\frac{1}{\sqrt{2}}(Z + X)\right]_{3}\) & 0.405 ± 0.011 & \(50\) & \(\left[\frac{1}{\sqrt{2}}(Z + Y)\right]_{3}\) & 0.423 ± 0.010 & \(80\) & \(X_{2}X_{3}\) & 0.313 ± 0.011 \\
		\hline
		\(21\) & \(\left[\frac{1}{\sqrt{2}}(Z + X)\right]_{4}\) & 0.290 ± 0.011 & \(51\) & \(\left[\frac{1}{\sqrt{2}}(Z + Y)\right]_{4}\) & 0.250 ± 0.011 & \(81\) & \(Y_{2}Y_{3}\) & 0.323 ± 0.011 \\
		\hline
		\(22\) & \(\left[\frac{1}{\sqrt{2}}(Z + X)\right]_{1,2}\) & 0.248 ± 0.011 & \(52\) & \(\left[\frac{1}{\sqrt{2}}(Z + Y)\right]_{1,2}\) & 0.261 ± 0.011 & \(82\) & \(X_{3}X_{4}\) & 0.399 ± 0.010 \\
		\hline
		\(23\) & \(\left[\frac{1}{\sqrt{2}}(Z + X)\right]_{1,3}\) & 0.263 ± 0.011 & \(53\) & \(\left[\frac{1}{\sqrt{2}}(Z + Y)\right]_{1,3}\) & 0.257 ± 0.011 & \(83\) & \(Y_{3}Y_{4}\) & 0.389 ± 0.011 \\
		\hline
		\(24\) & \(\left[\frac{1}{\sqrt{2}}(Z + X)\right]_{1,4}\) & 0.072 ± 0.011 & \(54\) & \(\left[\frac{1}{\sqrt{2}}(Z + Y)\right]_{1,4}\) & 0.046 ± 0.011 & \(84\) & \(X_{1}X_{4}\) & 0.141 ± 0.011 \\
		\hline
		\(25\) & \(\left[\frac{1}{\sqrt{2}}(Z + X)\right]_{2,3}\) & 0.249 ± 0.011 & \(55\) & \(\left[\frac{1}{\sqrt{2}}(Z + Y)\right]_{2,3}\) & 0.263 ± 0.011 & \(85\) & \(Y_{1}Y_{4}\) & 0.155 ± 0.011 \\
		\hline
		\(26\) & \(\left[\frac{1}{\sqrt{2}}(Z + X)\right]_{2,4}\) & 0.151 ± 0.011 & \(56\) & \(\left[\frac{1}{\sqrt{2}}(Z + Y)\right]_{2,4}\) & 0.141 ± 0.011 & \(86\) & \(X_{1}X_{3}\) & 0.278 ± 0.011 \\
		\hline
		\(27\) & \(\left[\frac{1}{\sqrt{2}}(Z + X)\right]_{3,4}\) & 0.142 ± 0.011 & \(57\) & \(\left[\frac{1}{\sqrt{2}}(Z + Y)\right]_{3,4}\) & 0.127 ± 0.011 & \(87\) & \(Y_{1}Y_{3}\) & 0.319 ± 0.011 \\
		\hline
		\(28\) & \(\left[\frac{1}{\sqrt{2}}(Z + X)\right]_{1,2,3}\) & 0.271 ± 0.011 & \(58\) & \(\left[\frac{1}{\sqrt{2}}(Z + Y)\right]_{1,2,3}\) & 0.260 ± 0.011 & \(88\) & \(X_{2}X_{4}\) & 0.399 ± 0.010 \\
		\hline
		\(29\) & \(\left[\frac{1}{\sqrt{2}}(Z + X)\right]_{1,2,4}\) & 0.151 ± 0.011 & \(59\) & \(\left[\frac{1}{\sqrt{2}}(Z + Y)\right]_{1,2,4}\) & 0.147 ± 0.011 & \(89\) & \(Y_{2}Y_{4}\) & 0.408 ± 0.010 \\
		\hline
		\(30\) & \(\left[\frac{1}{\sqrt{2}}(Z + X)\right]_{1,3,4}\) & 0.152 ± 0.011 & \(60\) & \(\left[\frac{1}{\sqrt{2}}(Z + Y)\right]_{1,3,4}\) & 0.103 ± 0.011 & \(90\) & \(I\) & 1 \\
		\hline
		\multicolumn{3}{|c|}{\(F_{\text{W}_{4}}\)} & \multicolumn{6}{c|}{0.720 ± 0.004} \\
		\hline
		\multicolumn{3}{|c|}{Witnesses} & \multicolumn{3}{c|}{Experimental result} & \multicolumn{3}{c|}{Theoretical expectation(F=1)} \\
		\hline
		\multicolumn{3}{|c|}{$\cW_{1}$} & \multicolumn{3}{c|}{0.00830 ± 0.00045} & \multicolumn{3}{c|}{-0.0047} \\
		\hline
		\multicolumn{3}{|c|}{$\cW_{2}$} & \multicolumn{3}{c|}{0.00547 ± 0.00046} & \multicolumn{3}{c|}{-0.0070} \\
		\hline
		\multicolumn{3}{|c|}{$\cW_{3}$} & \multicolumn{3}{c|}{0.00759 ± 0.00044} & \multicolumn{3}{c|}{-0.0090} \\
		\hline
		\multicolumn{3}{|c|}{$\cW_{4}$} & \multicolumn{3}{c|}{0.00698 ± 0.00046} & \multicolumn{3}{c|}{-0.0095} \\
		\hline
		\multicolumn{3}{|c|}{$\cW_{5}$} & \multicolumn{3}{c|}{0.00379 ± 0.00043} & \multicolumn{3}{c|}{-0.0114} \\
		\hline
	\end{tabular}
	\label{tab:W4c}
\end{table}

\FloatBarrier
\subsection{Four Photon Cluster State}

The the detailed configurations are shown in Tables \ref{tab:C4con}, and results for measuring the fidelity and witness of the four-photon cluster state are presented in Tables \ref{tab:C4a} and \ref{tab:C4b}.
\begin{table}[h]
	\centering
	\renewcommand\arraystretch{1.4}	
	\renewcommand\tabcolsep{2pt}
	\caption{Experimental configurations of four-photon Cluster state.}
	\begin{tabular}{|c|c|c|}
		\hline
		Fidelity & $0.968 \pm 0.002$ & $0.880 \pm 0.003$ \\ \hline
		Spectral filtering & $1550\pm10$ nm & unfiltered \\ \hline
		Pump Laser Power of entangled source & 240 mW & 1100 mW \\ \hline
		Pump Laser Power of direct SPDC & 240 mW & 1100 mW \\ \hline
		PPKTP crystal temperature (both) & 25$^\circ$C & 25$^\circ$C \\ \hline
		interference optical range difference & 0 & 0 \\ \hline
		Rate of counting & 0.70/s & 68/s \\ \hline
	\end{tabular}
	\label{tab:C4con}
\end{table}

\begin{table}[h]
	\centering
	\renewcommand\arraystretch{1.4}	
	\renewcommand\tabcolsep{2pt}
	\caption{Experimental results of four-photon Cluster state fidelity and witness under fidelity F = 0.968 ± 0.002. Including the required 16 operator expectation values. Under the conditions of 1550±10nm filtered parametric light, pump power of the two entangled source in 240mw, PPKTP crystal temperature of 25℃, no path difference in PDBS interference. Data were collected for each basis in 2100 s (about 1500 events). Error bars are derived from raw data analysis incorporating Poissonian counting statistics.}
	\begin{tabular}{|c|c|c|c|c|c|}
		\hline
		\({}\) & Operators & Expectation value & \({}\) & Operators & Expectation \\
		\hline
		\(1\) & \(Z_{1}Z_{2}\) & 0.991 ± 0.003 & \(9\) & \(X_{1}X_{2}Z_{4}\) & 0.966 ± 0.007 \\
		\hline
		\(2\) & \(X_{1}X_{2}Z_{3}\) & 0.967 ± 0.007 & \(10\) & \(Z_{2}Y_{3}Y_{4}\) & 0.956 ± 0.008 \\
		\hline
		\(3\) & \(Z_{2}X_{3}X_{4}\) & 0.956 ± 0.008 & \(11\) & \(Y_{1}X_{2}Y_{3}X_{4}\) & 0.946 ± 0.009 \\
		\hline
		\(4\) & \(Z_{3}Z_{4}\) & 0.996 ± 0.002 & \(12\) & \(Y_{1}Y_{2}Z_{4}\) & 0.963 ± 0.007 \\
		\hline
		\(5\) & \(Y_{1}Y_{2}Z_{3}\) & 0.965 ± 0.007 & \(13\) & \(Z_{1}Y_{3}Y_{4}\) & 0.957 ± 0.008 \\
		\hline
		\(6\) & \(Z_{1}X_{3}X_{4}\) & 0.960 ± 0.007 & \(14\) & \(X_{1}Y_{2}X_{3}Y_{4}\) & 0.958 ± 0.008 \\
		\hline
		\(7\) & \(Z_{1}Z_{2}Z_{3}Z_{4}\) & 0.990 ± 0.004 & \(15\) & \(Y_{1}X_{2}X_{3}Y_{4}\) & 0.953 ± 0.008 \\
		\hline
		\(8\) & \(X_{1}Y_{2}Y_{3}X_{4}\) & 0.962 ± 0.007 & \(16\) & \(I\) & 1 \\
		\hline
		\multicolumn{2}{|c|}{\(F_{\text{C}_{4}}\)} & \multicolumn{4}{c|}{0.968 ± 0.002} \\
		\hline
		\multicolumn{2}{|c|}{Witnesses} & \multicolumn{2}{c|}{Experimental results} & \multicolumn{2}{c|}{Theoretical expectation(F=1)} \\
		\hline
		\multicolumn{2}{|c|}{\(\cW_{1}\)} & \multicolumn{2}{c|}{-0.0132 ± 0.0002} & \multicolumn{2}{c|}{-0.0156} \\
		\hline
		\multicolumn{2}{|c|}{\(\cW_{2}\)} & \multicolumn{2}{c|}{-0.0287 ± 0.0002} & \multicolumn{2}{c|}{-0.0312} \\
		\hline
		\multicolumn{2}{|c|}{\(\cW_{3}\)} & \multicolumn{2}{c|}{-0.0378 ± 0.0003} & \multicolumn{2}{c|}{-0.0417} \\
		\hline
		\multicolumn{2}{|c|}{\(\cW_{4}\)} & \multicolumn{2}{c|}{-0.0573 ± 0.0003} & \multicolumn{2}{c|}{-0.0625} \\
		\hline
	\end{tabular}
	\label{tab:C4a}
\end{table}

\begin{table}[h]
	\centering
	\renewcommand\arraystretch{1.4}	
	\renewcommand\tabcolsep{2pt}
	\caption{Experimental results of four-photon Cluster state fidelity and witness under fidelity F = 0.880 ± 0.003. Including the required 16 operator expectation values. Under the conditions of 1550±10nm filtered parametric light, pump power of the two entangled source in 240mw, PPKTP crystal temperature of 25℃, no path difference in PDBS interference. Data are collected for each basis in 2100 s (about 1500 events). Error bars are derived from raw data analysis incorporating Poissonian counting statistics.}
	\begin{tabular}{|c|c|c|c|c|c|}
		\hline
		\({}\) & Operators & Expectation value & \({}\) & Operators & Expectation \\
		\hline
		\(1\) & \(Z_{1}Z_{2}\) & 0.969 ± 0.006 & \(9\) & \(X_{1}X_{2}Z_{4}\) & 0.859 ± 0.013 \\
		\hline
		\(2\) & \(X_{1}X_{2}Z_{3}\) & 0.860 ± 0.013 & \(10\) & \(Z_{2}Y_{3}Y_{4}\) & 0.857 ± 0.013 \\
		\hline
		\(3\) & \(Z_{2}X_{3}X_{4}\) & 0.868 ± 0.012 & \(11\) & \(Y_{1}X_{2}Y_{3}X_{4}\) & 0.844 ± 0.013 \\
		\hline
		\(4\) & \(Z_{3}Z_{4}\) & 0.982 ± 0.005 & \(12\) & \(Y_{1}Y_{2}Z_{4}\) & 0.815 ± 0.014 \\
		\hline
		\(5\) & \(Y_{1}Y_{2}Z_{3}\) & 0.820 ± 0.014 & \(13\) & \(Z_{1}Y_{3}Y_{4}\) & 0.840 ± 0.013 \\
		\hline
		\(6\) & \(Z_{1}X_{3}X_{4}\) & 0.859 ± 0.013 & \(14\) & \(X_{1}Y_{2}X_{3}Y_{4}\) & 0.865 ± 0.012 \\
		\hline
		\(7\) & \(Z_{1}Z_{2}Z_{3}Z_{4}\) & 0.951 ± 0.007 & \(15\) & \(Y_{1}X_{2}X_{3}Y_{4}\) & 0.858 ± 0.013 \\
		\hline
		\(8\) & \(X_{1}Y_{2}Y_{3}X_{4}\) & 0.839 ± 0.013 & \(16\) & \(I\) & 1 \\
		\hline
		\multicolumn{2}{|c|}{\(F_{\text{C}_{4}}\)} & \multicolumn{4}{c|}{0.880 ± 0.003} \\
		\hline
		\multicolumn{2}{|c|}{Witnesses} & \multicolumn{2}{c|}{Experimental results} & \multicolumn{2}{c|}{Theoretical expectation(F=1)} \\
		\hline
		\multicolumn{2}{|c|}{\(\cW_{1}\)} & \multicolumn{2}{c|}{-0.0056 ± 0.0004} & \multicolumn{2}{c|}{-0.0156} \\
		\hline
		\multicolumn{2}{|c|}{\(\cW_{2}\)} & \multicolumn{2}{c|}{-0.0207 ± 0.0004} & \multicolumn{2}{c|}{-0.0312} \\
		\hline
		\multicolumn{2}{|c|}{\(\cW_{3}\)} & \multicolumn{2}{c|}{-0.0268 ± 0.0005} & \multicolumn{2}{c|}{-0.0417} \\
		\hline
		\multicolumn{2}{|c|}{\(\cW_{4}\)} & \multicolumn{2}{c|}{-0.0419 ± 0.0006} & \multicolumn{2}{c|}{-0.0625} \\
		\hline
	\end{tabular}
	\label{tab:C4b}
\end{table}

\FloatBarrier
\subsection{Robustness Verification}
To verify the robustness of the EDL framework against measurement noise, we demonstrate three deflection angles under two conditions: (a) measurement noise acts on the basis X, Y, and Z, and (b)  measurement noise acts solely on the basis Y. The positions of the points collected in our experiment within the theoretically predicted images are shown in Figures \ref{figure:p}(a) and (b).

\begin{figure}[t]
	\centering
	\includegraphics[scale=0.2]{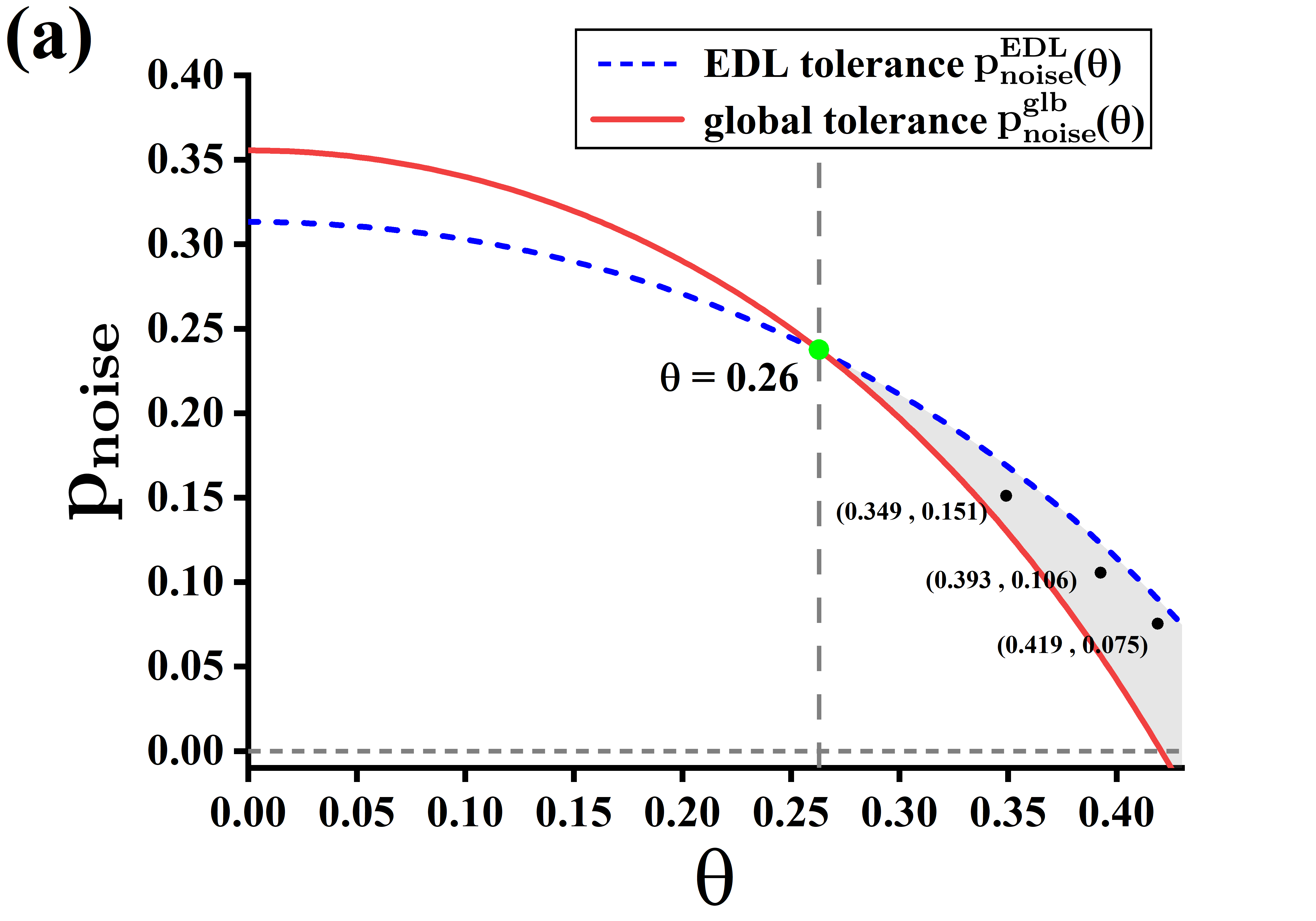}
	\includegraphics[scale=0.2]{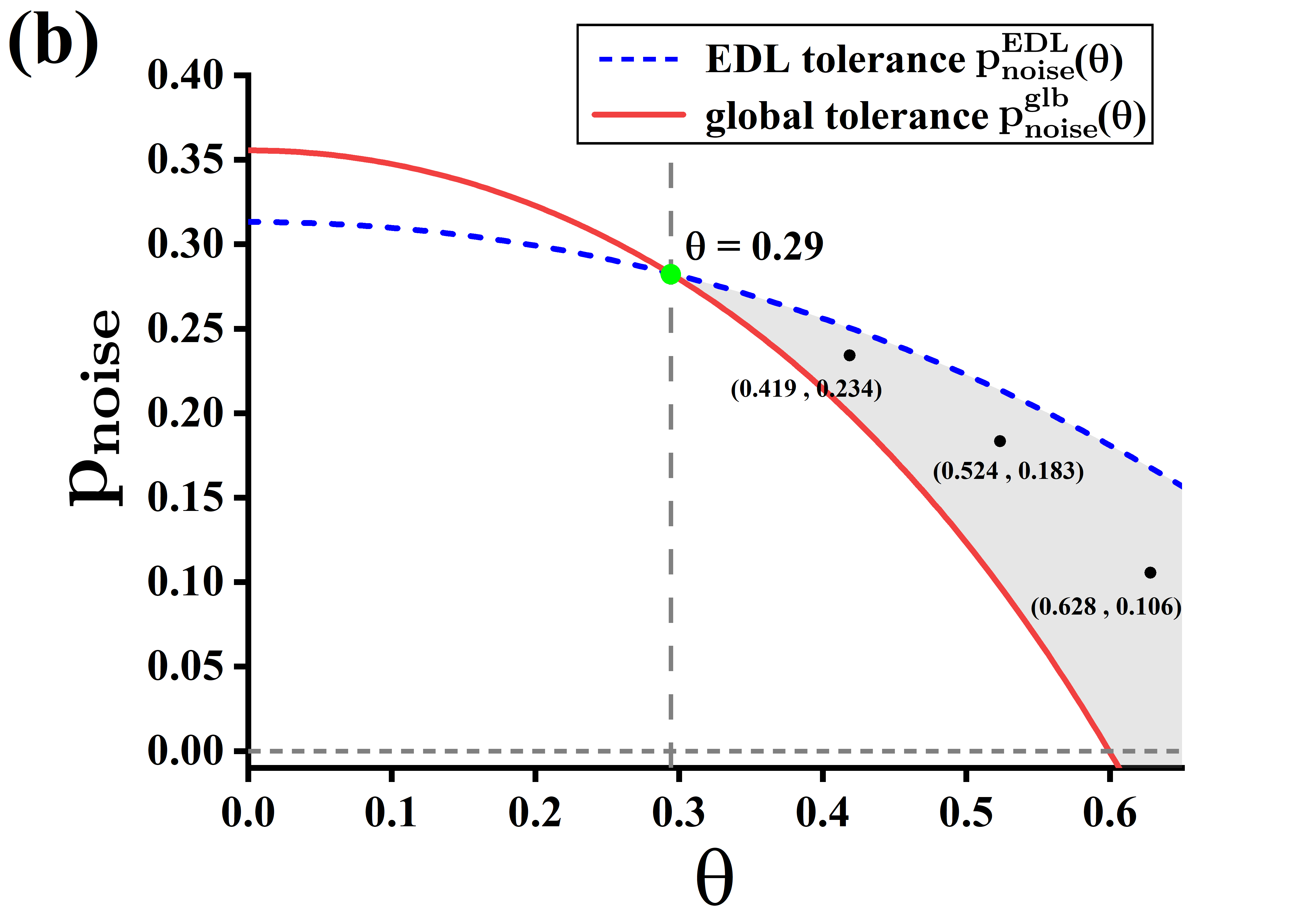}
	\caption{\footnotesize The position of experimental results on the theoretical image. The horizontal coordinate $\theta$ of the data points represents the deflection angle set in the experiment, while the vertical coordinate $p_{\text{noise}}=1-F_{\text{D}_4}$ is calculated from the fidelity of the initial state prepared in the experiment. (a) measurement noise acts on the basis X, Y, and Z, (b)  measurement noise acts solely on the basis Y.}
	\label{figure:p}
\end{figure}

The specific experimental configurations are shown in Tables \ref{tab:D4compare} and \ref{tab:D4comparey}. Detailed results including initial state and noise introdued state in different $\theta$ of condition (a) are shown in Tables \ref{tab:D4R11}, \ref{tab:D4R12}, \ref{tab:D4R21}, \ref{tab:D4R22}, \ref{tab:D4R31}, and \ref{tab:D4R32}, while condition (b) are shown in Tables \ref{tab:D4yR11}, \ref{tab:D4yR12}, \ref{tab:D4yR21}, \ref{tab:D4yR22}, \ref{tab:D4yR31}, and \ref{tab:D4yR32}.

\begin{table}[htbp]
	\centering
	\caption{Experimental configurations of robustness verification based on four-photon Dicke state (noise introduced on basis X, Y, and Z).} 
	\renewcommand\arraystretch{1.4}	
	\renewcommand\tabcolsep{2pt}
	\begin{tabular}{|c|c|c|c|}
		\hline
		\textbf{condition} & \textbf{$\theta=2\pi/15$} & \textbf{$\theta=\pi/8$} & \textbf{$\theta=\pi/9$} \\ \hline
		Fidelity(original) & $0.925 \pm 0.003$ & $0.894 \pm 0.003$ & $0.849 \pm 0.003$  \\ \hline
		Fidelity(noise introduced) & $0.611 \pm 0.004$ & $0.603 \pm 0.004$ & $0.645 \pm 0.004$  \\ \hline
		Spectral filtering & unfiltered & unfiltered & unfiltered  \\ \hline
		Pump Laser Power & 1350 mW & 3100 mW & 3100 mW   \\ \hline
		PPKTP crystal temperature & 17$^\circ$C & 17$^\circ$C & 50$^\circ$C  \\ \hline
		Walk-off compensation & No offset & No offset & No offset \\ \hline
		Rate of counting & 18.5/s & 98/s & 98/s \\ \hline
	\end{tabular}
	\label{tab:D4compare}
\end{table}

\begin{table}[htbp]
	\centering
	\renewcommand\arraystretch{1.4}	
	\renewcommand\tabcolsep{2pt}
	\caption{Results of fidelity and witness of the initial four-photon Dicke state prepared for robustness verification with noise $\theta=2\pi/15$. Including the required 27 operator expectation values. Under the conditions of no filtering of the parametric light, pump laser power of 1350mw, PPKTP crystal temperature of 17℃. Data are collected for each basis in 300 s (about 5600 events). Error bars are derived from raw data analysis incorporating Poissonian counting statistics.}
	\begin{tabular}{|c|c|c|c|c|c|}
		\hline
		\multicolumn{2}{|c|}{Operators} & Expectation value & \multicolumn{2}{c|}{Operators} & Expectation value \\ \hline
		1 & $X_{1}X_{2}$ & $0.655 \pm 0.010$ & 15 & $Z_{1}Z_{2}$ & $-0.357 \pm 0.012$ \\ \hline
		2 & $X_{1}X_{3}$ & $0.571 \pm 0.011$ & 16 & $Z_{1}Z_{3}$ & $-0.360 \pm 0.012$ \\ \hline
		3 & $X_{1}X_{4}$ & $0.572 \pm 0.011$ & 17 & $Z_{1}Z_{4}$ & $-0.253 \pm 0.013$ \\ \hline
		4 & $X_{2}X_{3}$ & $0.577 \pm 0.011$ & 18 & $Z_{2}Z_{3}$ & $-0.250 \pm 0.013$ \\ \hline
		5 & $X_{2}X_{4}$ & $0.574 \pm 0.011$ & 19 & $Z_{2}Z_{4}$ & $-0.354 \pm 0.012$ \\ \hline
		6 & $X_{3}X_{4}$ & $0.661 \pm 0.010$ & 20 & $Z_{3}Z_{4}$ & $-0.359 \pm 0.012$ \\ \hline
		7 & $X_{1}X_{2}X_{3}X_{4}$ & $0.945 \pm 0.004$ & 21 & $Z_{1}Z_{2}Z_{3}Z_{4}$ & $0.934 \pm 0.005$ \\ \hline
		8 & $Y_{1}Y_{2}$ & $0.682 \pm 0.010$ & 22 & $[\frac{1}{\sqrt{2}}(X + Z)]^{\otimes 4}$ & $-0.427 \pm 0.012$ \\ \hline
		9 & $Y_{1}Y_{3}$ & $0.586 \pm 0.011$ & 23 & $[\frac{1}{\sqrt{2}}(X - Z)]^{\otimes 4}$ & $-0.453 \pm 0.012$ \\ \hline
		10 & $Y_{1}Y_{4}$ & $0.592 \pm 0.010$ & 24 & $[\frac{1}{\sqrt{2}}(Y + Z)]^{\otimes 4}$ & $-0.434 \pm 0.012$ \\ \hline
		11 & $Y_{2}Y_{3}$ & $0.591 \pm 0.010$ & 25 & $[\frac{1}{\sqrt{2}}(Y - Z)]^{\otimes 4}$ & $-0.434 \pm 0.012$ \\ \hline
		12 & $Y_{2}Y_{4}$ & $0.586 \pm 0.011$ & 26 & $[\frac{1}{\sqrt{2}}(X + Y)]^{\otimes 4}$ & $0.936 \pm 0.005$ \\ \hline
		13 & $Y_{3}Y_{4}$ & $0.685 \pm 0.009$ & 27 & $[\frac{1}{\sqrt{2}}(X - Y)]^{\otimes 4}$ & $0.932 \pm 0.005$ \\ \hline
		14 & $Y_{1}Y_{2}Y_{3}Y_{4}$ & $0.937 \pm 0.005$ & \multicolumn{2}{c|}{$W_{\text{proj}}$} & $-0.258 \pm 0.003$ \\ \hline
		\multicolumn{2}{|c|}{$F_{\text{D}_{4}^{(2)}}$} & $0.925 \pm 0.003$ & \multicolumn{2}{c|}{$W_{5}$} & $-0.0224 \pm 0.0005$ \\ \hline
	\end{tabular}
	\label{tab:D4R11}
\end{table}

\begin{table}[htbp]
	\centering
	\renewcommand\arraystretch{1.4}	
	\renewcommand\tabcolsep{2pt}
	\caption{Results of fidelity and witness of four-photon Dicke state at noise $\theta=2\pi/15$. Including the required 27 operator expectation values. Under the conditions of no filtering of the parametric light, pump laser power of 1350mw, PPKTP crystal temperature of 17℃. Data are collected for each basis in 300 s (about 5600 events). Error bars are derived from raw data analysis incorporating Poissonian counting statistics.}
	\begin{tabular}{|c|c|c|c|c|c|}
		\hline
		\multicolumn{2}{|c|}{Operators} & Expectation value & \multicolumn{2}{c|}{Operators} & Expectation value \\ \hline
		1 & $X_{1}X_{2}$ & $0.661 \pm 0.010$ & 15 & $Z_{1}Z_{2}$ & $-0.227 \pm 0.013$ \\ \hline
		2 & $X_{1}X_{3}$ & $0.574 \pm 0.011$ & 16 & $Z_{1}Z_{3}$ & $-0.185 \pm 0.013$ \\ \hline
		3 & $X_{1}X_{4}$ & $0.570 \pm 0.011$ & 17 & $Z_{1}Z_{4}$ & $-0.121 \pm 0.013$ \\ \hline
		4 & $X_{2}X_{3}$ & $0.568 \pm 0.011$ & 18 & $Z_{2}Z_{3}$ & $-0.140 \pm 0.013$ \\ \hline
		5 & $X_{2}X_{4}$ & $0.588 \pm 0.011$ & 19 & $Z_{2}Z_{4}$ & $-0.194 \pm 0.013$ \\ \hline
		6 & $X_{3}X_{4}$ & $0.663 \pm 0.010$ & 20 & $Z_{3}Z_{4}$ & $-0.173 \pm 0.013$ \\ \hline
		7 & $X_{1}X_{2}X_{3}X_{4}$ & $0.937 \pm 0.005$ & 21 & $Z_{1}Z_{2}Z_{3}Z_{4}$ & $0.204 \pm 0.013$ \\ \hline
		8 & $Y_{1}Y_{2}$ & $0.506 \pm 0.011$ & 22 & $[\frac{1}{\sqrt{2}}(X + Z)]^{\otimes 4}$ & $-0.187 \pm 0.013$ \\ \hline
		9 & $Y_{1}Y_{3}$ & $0.431 \pm 0.012$ & 23 & $[\frac{1}{\sqrt{2}}(X - Z)]^{\otimes 4}$ & $-0.329 \pm 0.013$ \\ \hline
		10 & $Y_{1}Y_{4}$ & $0.465 \pm 0.012$ & 24 & $[\frac{1}{\sqrt{2}}(Y + Z)]^{\otimes 4}$ & $-0.381 \pm 0.012$ \\ \hline
		11 & $Y_{2}Y_{3}$ & $0.462 \pm 0.012$ & 25 & $[\frac{1}{\sqrt{2}}(Y - Z)]^{\otimes 4}$ & $-0.009 \pm 0.013$ \\ \hline
		12 & $Y_{2}Y_{4}$ & $0.486 \pm 0.012$ & 26 & $[\frac{1}{\sqrt{2}}(X + Y)]^{\otimes 4}$ & $0.682 \pm 0.010$ \\ \hline
		13 & $Y_{3}Y_{4}$ & $0.543 \pm 0.011$ & 27 & $[\frac{1}{\sqrt{2}}(X - Y)]^{\otimes 4}$ & $0.480 \pm 0.012$ \\ \hline
		14 & $Y_{1}Y_{2}Y_{3}Y_{4}$ & $0.280 \pm 0.013$ & \multicolumn{2}{c|}{$W_{\text{proj}}$} & $0.056 \pm 0.004$ \\ \hline
		\multicolumn{2}{|c|}{$F_{\text{D}_{4}^{(2)}}$} & $0.611 \pm 0.004$ & \multicolumn{2}{c|}{$W_{5}$} & $-0.0029 \pm 0.0006$ \\ \hline
	\end{tabular}
	\label{tab:D4R12}
\end{table}

\begin{table}[htbp]
	\centering
	\renewcommand\arraystretch{1.4}	
	\renewcommand\tabcolsep{2pt}
	\caption{Results of fidelity and witness of the initial four-photon Dicke state prepared for robustness verification with noise $\theta=\pi/8$. Including the required 27 operator expectation values. Under the conditions of no filtering of the parametric light, pump laser power of 3100mw, PPKTP crystal temperature of 17℃. Data are collected for each basis in 60 s (about 5600 events). Error bars are derived from raw data analysis incorporating Poissonian counting statistics.}
	\begin{tabular}{|c|c|c|c|c|c|}
		\hline
		\multicolumn{2}{|c|}{Operators} & Expectation value & \multicolumn{2}{c|}{Operators} & Expectation value \\ \hline
		1 & $X_{1}X_{2}$ & $0.651 \pm 0.010$ & 15 & $Z_{1}Z_{2}$ & $-0.346 \pm 0.012$ \\ \hline
		2 & $X_{1}X_{3}$ & $0.566 \pm 0.011$ & 16 & $Z_{1}Z_{3}$ & $-0.343 \pm 0.012$ \\ \hline
		3 & $X_{1}X_{4}$ & $0.564 \pm 0.011$ & 17 & $Z_{1}Z_{4}$ & $-0.266 \pm 0.012$ \\ \hline
		4 & $X_{2}X_{3}$ & $0.558 \pm 0.011$ & 18 & $Z_{2}Z_{3}$ & $-0.267 \pm 0.012$ \\ \hline
		5 & $X_{2}X_{4}$ & $0.584 \pm 0.011$ & 19 & $Z_{2}Z_{4}$ & $-0.344 \pm 0.012$ \\ \hline
		6 & $X_{3}X_{4}$ & $0.647 \pm 0.010$ & 20 & $Z_{3}Z_{4}$ & $-0.318 \pm 0.012$ \\ \hline
		7 & $X_{1}X_{2}X_{3}X_{4}$ & $0.892 \pm 0.006$ & 21 & $Z_{1}Z_{2}Z_{3}Z_{4}$ & $0.884 \pm 0.006$ \\ \hline
		8 & $Y_{1}Y_{2}$ & $0.664 \pm 0.010$ & 22 & $[\frac{1}{\sqrt{2}}(X + Z)]^{\otimes 4}$ & $-0.424 \pm 0.012$ \\ \hline
		9 & $Y_{1}Y_{3}$ & $0.596 \pm 0.011$ & 23 & $[\frac{1}{\sqrt{2}}(X - Z)]^{\otimes 4}$ & $-0.399 \pm 0.012$ \\ \hline
		10 & $Y_{1}Y_{4}$ & $0.599 \pm 0.011$ & 24 & $[\frac{1}{\sqrt{2}}(Y + Z)]^{\otimes 4}$ & $-0.401 \pm 0.012$ \\ \hline
		11 & $Y_{2}Y_{3}$ & $0.598 \pm 0.011$ & 25 & $[\frac{1}{\sqrt{2}}(Y - Z)]^{\otimes 4}$ & $-0.397 \pm 0.012$ \\ \hline
		12 & $Y_{2}Y_{4}$ & $0.595 \pm 0.011$ & 26 & $[\frac{1}{\sqrt{2}}(X + Y)]^{\otimes 4}$ & $0.888 \pm 0.006$ \\ \hline
		13 & $Y_{3}Y_{4}$ & $0.668 \pm 0.010$ & 27 & $[\frac{1}{\sqrt{2}}(X - Y)]^{\otimes 4}$ & $0.916 \pm 0.005$ \\ \hline
		14 & $Y_{1}Y_{2}Y_{3}Y_{4}$ & $0.906 \pm 0.006$ & \multicolumn{2}{c|}{$W_{\text{proj}}$} & $-0.228 \pm 0.003$ \\ \hline
		\multicolumn{2}{|c|}{$F_{\text{D}_{4}^{(2)}}$} & $0.894 \pm 0.003$ & \multicolumn{2}{c|}{$W_{5}$} & $-0.0213 \pm 0.0005$ \\ \hline
	\end{tabular}
	\label{tab:D4R21}
\end{table}

\begin{table}[htbp]
	\centering
	\renewcommand\arraystretch{1.4}	
	\renewcommand\tabcolsep{2pt}
	\caption{Results of verify the fidelity and witness of four-photon Dicke state at noise $\theta=\pi/8$. Including the required 27 operator expectation values. Under the conditions of no filtering of the parametric light, pump laser power of 3100mw, PPKTP crystal temperature of 17℃. Data are collected for each basis in 60 s (about 5600 events). Error bars are derived from raw data analysis incorporating Poissonian counting statistics.}
	\begin{tabular}{|c|c|c|c|c|c|}
		\hline
		\multicolumn{2}{|c|}{Operators} & Expectation value & \multicolumn{2}{c|}{Operators} & Expectation value \\ \hline
		1 & $X_{1}X_{2}$ & $0.670 \pm 0.010$ & 15 & $Z_{1}Z_{2}$ & $-0.197 \pm 0.013$ \\ \hline
		2 & $X_{1}X_{3}$ & $0.583 \pm 0.011$ & 16 & $Z_{1}Z_{3}$ & $-0.162 \pm 0.013$ \\ \hline
		3 & $X_{1}X_{4}$ & $0.592 \pm 0.011$ & 17 & $Z_{1}Z_{4}$ & $-0.149 \pm 0.013$ \\ \hline
		4 & $X_{2}X_{3}$ & $0.591 \pm 0.011$ & 18 & $Z_{2}Z_{3}$ & $-0.122 \pm 0.013$ \\ \hline
		5 & $X_{2}X_{4}$ & $0.609 \pm 0.010$ & 19 & $Z_{2}Z_{4}$ & $-0.207 \pm 0.013$ \\ \hline
		6 & $X_{3}X_{4}$ & $0.662 \pm 0.010$ & 20 & $Z_{3}Z_{4}$ & $-0.152 \pm 0.013$ \\ \hline
		7 & $X_{1}X_{2}X_{3}X_{4}$ & $0.897 \pm 0.006$ & 21 & $Z_{1}Z_{2}Z_{3}Z_{4}$ & $0.171 \pm 0.013$ \\ \hline
		8 & $Y_{1}Y_{2}$ & $0.540 \pm 0.011$ & 22 & $[\frac{1}{\sqrt{2}}(X + Z)]^{\otimes 4}$ & $-0.154 \pm 0.013$ \\ \hline
		9 & $Y_{1}Y_{3}$ & $0.473 \pm 0.011$ & 23 & $[\frac{1}{\sqrt{2}}(X - Z)]^{\otimes 4}$ & $-0.217 \pm 0.013$ \\ \hline
		10 & $Y_{1}Y_{4}$ & $0.478 \pm 0.011$ & 24 & $[\frac{1}{\sqrt{2}}(Y + Z)]^{\otimes 4}$ & $-0.362 \pm 0.012$ \\ \hline
		11 & $Y_{2}Y_{3}$ & $0.489 \pm 0.011$ & 25 & $[\frac{1}{\sqrt{2}}(Y - Z)]^{\otimes 4}$ & $-0.022 \pm 0.013$ \\ \hline
		12 & $Y_{2}Y_{4}$ & $0.506 \pm 0.011$ & 26 & $[\frac{1}{\sqrt{2}}(X + Y)]^{\otimes 4}$ & $0.747 \pm 0.009$ \\ \hline
		13 & $Y_{3}Y_{4}$ & $0.530 \pm 0.011$ & 27 & $[\frac{1}{\sqrt{2}}(X - Y)]^{\otimes 4}$ & $0.388 \pm 0.012$ \\ \hline
		14 & $Y_{1}Y_{2}Y_{3}Y_{4}$ & $0.360 \pm 0.012$ & \multicolumn{2}{c|}{$W_{\text{proj}}$} & $0.064 \pm 0.004$ \\ \hline
		\multicolumn{2}{|c|}{$F_{\text{D}_{4}^{(2)}}$} & $0.603 \pm 0.004$ & \multicolumn{2}{c|}{$W_{5}$} & $-0.0037 \pm 0.0006$ \\ \hline
	\end{tabular}
	\label{tab:D4R22}
\end{table}

\begin{table}[htbp]
	\centering
	\renewcommand\arraystretch{1.4}	
	\renewcommand\tabcolsep{2pt}
	\caption{Results of fidelity and witness of the initial four-photon Dicke state prepared for robustness verification with noise $\theta=\pi/9$. Including the required 27 operator expectation values. Under the conditions of no filtering of the parametric light, pump laser power of 3100mw, PPKTP crystal temperature of 50℃. Data are collected for each basis in 60 s (about 5600 events). Error bars are derived from raw data analysis incorporating Poissonian counting statistics.}
	\begin{tabular}{|c|c|c|c|c|c|}
		\hline
		\multicolumn{2}{|c|}{Operators} & Expectation value & \multicolumn{2}{c|}{Operators} & Expectation value \\ \hline
		1 & $X_{1}X_{2}$ & $0.618 \pm 0.010$ & 15 & $Z_{1}Z_{2}$ & $-0.342 \pm 0.012$ \\ \hline
		2 & $X_{1}X_{3}$ & $0.554 \pm 0.011$ & 16 & $Z_{1}Z_{3}$ & $-0.349 \pm 0.012$ \\ \hline
		3 & $X_{1}X_{4}$ & $0.543 \pm 0.011$ & 17 & $Z_{1}Z_{4}$ & $-0.256 \pm 0.012$ \\ \hline
		4 & $X_{2}X_{3}$ & $0.548 \pm 0.011$ & 18 & $Z_{2}Z_{3}$ & $-0.255 \pm 0.012$ \\ \hline
		5 & $X_{2}X_{4}$ & $0.560 \pm 0.011$ & 19 & $Z_{2}Z_{4}$ & $-0.349 \pm 0.012$ \\ \hline
		6 & $X_{3}X_{4}$ & $0.612 \pm 0.010$ & 20 & $Z_{3}Z_{4}$ & $-0.333 \pm 0.012$ \\ \hline
		7 & $X_{1}X_{2}X_{3}X_{4}$ & $0.805 \pm 0.008$ & 21 & $Z_{1}Z_{2}Z_{3}Z_{4}$ & $0.886 \pm 0.006$ \\ \hline
		8 & $Y_{1}Y_{2}$ & $0.643 \pm 0.010$ & 22 & $[\frac{1}{\sqrt{2}}(X + Z)]^{\otimes 4}$ & $-0.385 \pm 0.012$ \\ \hline
		9 & $Y_{1}Y_{3}$ & $0.560 \pm 0.011$ & 23 & $[\frac{1}{\sqrt{2}}(X - Z)]^{\otimes 4}$ & $-0.373 \pm 0.012$ \\ \hline
		10 & $Y_{1}Y_{4}$ & $0.551 \pm 0.011$ & 24 & $[\frac{1}{\sqrt{2}}(Y + Z)]^{\otimes 4}$ & $-0.377 \pm 0.012$ \\ \hline
		11 & $Y_{2}Y_{3}$ & $0.553 \pm 0.011$ & 25 & $[\frac{1}{\sqrt{2}}(Y - Z)]^{\otimes 4}$ & $-0.381 \pm 0.012$ \\ \hline
		12 & $Y_{2}Y_{4}$ & $0.565 \pm 0.011$ & 26 & $[\frac{1}{\sqrt{2}}(X + Y)]^{\otimes 4}$ & $0.801 \pm 0.008$ \\ \hline
		13 & $Y_{3}Y_{4}$ & $0.649 \pm 0.010$ & 27 & $[\frac{1}{\sqrt{2}}(X - Y)]^{\otimes 4}$ & $0.803 \pm 0.008$ \\ \hline
		14 & $Y_{1}Y_{2}Y_{3}Y_{4}$ & $0.814 \pm 0.008$ & \multicolumn{2}{c|}{$W_{\text{proj}}$} & $-0.182 \pm 0.003$ \\ \hline
		\multicolumn{2}{|c|}{$F_{\text{D}_{4}^{(2)}}$} & $0.849 \pm 0.003$ & \multicolumn{2}{c|}{$W_{5}$} & $-0.0188 \pm 0.0005$ \\ \hline
	\end{tabular}
	\label{tab:D4R31}
\end{table}

\begin{table}[htbp]
	\centering
	\renewcommand\arraystretch{1.4}	
	\renewcommand\tabcolsep{2pt}
	\caption{Results of verify the fidelity and witness of four-photon Dicke state at noise $\theta=\pi/9$. Including the required 27 operator expectation values. Under the conditions of no filtering of the parametric light, pump laser power of 3100mw, PPKTP crystal temperature of 50℃. Data are collected for each basis in 60 s (about 5600 events). Error bars are derived from raw data analysis incorporating Poissonian counting statistics.}
	\begin{tabular}{|c|c|c|c|c|c|}
		\hline
		\multicolumn{2}{|c|}{Operators} & Expectation value & \multicolumn{2}{c|}{Operators} & Expectation value \\ \hline
		1 & $X_{1}X_{2}$ & $0.630 \pm 0.010$ & 15 & $Z_{1}Z_{2}$ & $-0.253 \pm 0.013$ \\ \hline
		2 & $X_{1}X_{3}$ & $0.552 \pm 0.011$ & 16 & $Z_{1}Z_{3}$ & $-0.218 \pm 0.013$ \\ \hline
		3 & $X_{1}X_{4}$ & $0.547 \pm 0.011$ & 17 & $Z_{1}Z_{4}$ & $-0.160 \pm 0.013$ \\ \hline
		4 & $X_{2}X_{3}$ & $0.549 \pm 0.011$ & 18 & $Z_{2}Z_{3}$ & $-0.160 \pm 0.013$ \\ \hline
		5 & $X_{2}X_{4}$ & $0.559 \pm 0.011$ & 19 & $Z_{2}Z_{4}$ & $-0.251 \pm 0.013$ \\ \hline
		6 & $X_{3}X_{4}$ & $0.631 \pm 0.010$ & 20 & $Z_{3}Z_{4}$ & $-0.222 \pm 0.013$ \\ \hline
		7 & $X_{1}X_{2}X_{3}X_{4}$ & $0.804 \pm 0.008$ & 21 & $Z_{1}Z_{2}Z_{3}Z_{4}$ & $0.364 \pm 0.012$ \\ \hline
		8 & $Y_{1}Y_{2}$ & $0.552 \pm 0.011$ & 22 & $[\frac{1}{\sqrt{2}}(X + Z)]^{\otimes 4}$ & $-0.246 \pm 0.013$ \\ \hline
		9 & $Y_{1}Y_{3}$ & $0.473 \pm 0.012$ & 23 & $[\frac{1}{\sqrt{2}}(X - Z)]^{\otimes 4}$ & $-0.255 \pm 0.013$ \\ \hline
		10 & $Y_{1}Y_{4}$ & $0.479 \pm 0.012$ & 24 & $[\frac{1}{\sqrt{2}}(Y + Z)]^{\otimes 4}$ & $-0.339 \pm 0.012$ \\ \hline
		11 & $Y_{2}Y_{3}$ & $0.497 \pm 0.011$ & 25 & $[\frac{1}{\sqrt{2}}(Y - Z)]^{\otimes 4}$ & $-0.171 \pm 0.013$ \\ \hline
		12 & $Y_{2}Y_{4}$ & $0.505 \pm 0.011$ & 26 & $[\frac{1}{\sqrt{2}}(X + Y)]^{\otimes 4}$ & $0.665 \pm 0.010$ \\ \hline
		13 & $Y_{3}Y_{4}$ & $0.556 \pm 0.011$ & 27 & $[\frac{1}{\sqrt{2}}(X - Y)]^{\otimes 4}$ & $0.462 \pm 0.011$ \\ \hline
		14 & $Y_{1}Y_{2}Y_{3}Y_{4}$ & $0.391 \pm 0.012$ & \multicolumn{2}{c|}{$W_{\text{proj}}$} & $0.022 \pm 0.004$ \\ \hline
		\multicolumn{2}{|c|}{$F_{\text{D}_{4}^{(2)}}$} & $0.645 \pm 0.004$ & \multicolumn{2}{c|}{$W_{5}$} & $-0.0063 \pm 0.0006$ \\ \hline
	\end{tabular}
	\label{tab:D4R32}
\end{table}

\begin{table}[htbp]
	\centering
	\caption{Experimental configurations of robustness verification based on four-photon Dicke state (noise introduced only on basis Y).} 
	\renewcommand\arraystretch{1.4}	
	\renewcommand\tabcolsep{2pt}
	\begin{tabular}{|c|c|c|c|}
		\hline
		\textbf{condition} & \textbf{$\theta=\pi/5$} & \textbf{$\theta=\pi/6$} & \textbf{$2\pi/15$} \\ \hline
		Fidelity(original) & $0.894 \pm 0.003$ & $0.817 \pm 0.003$ & $0.766 \pm 0.003$  \\ \hline
		Fidelity(noise introduced) & $0.613 \pm 0.003$ & $0.620 \pm 0.003$ & $0.650 \pm 0.003$  \\ \hline
		spectral filtering & unfiltered & unfiltered & unfiltered  \\ \hline
		Pump Laser Power & 3100 mW & 3100 mW & 3100 mW   \\ \hline
		PPKTP crystal temperature & 17$^\circ$C & 60$^\circ$C & 70$^\circ$C  \\ \hline
		Walk-off compensation & No offset & No offset & No offset \\ \hline
		Rate of counting & 98/s & 98/s & 101/s \\ \hline
	\end{tabular}
	\label{tab:D4comparey}
\end{table}

\begin{table}[htbp]
	\centering
	\renewcommand\arraystretch{1.4}	
	\renewcommand\tabcolsep{2pt}
	\caption{Results of fidelity and witness of the initial four-photon Dicke state prepared for robustness verification with noise $\theta=\pi/5$. Including the required 27 operator expectation values. Under the conditions of no filtering of the parametric light, pump laser power of 3100mw, PPKTP crystal temperature of 17℃(the same condition as \ref{tab:D4R21}). Data are collected for each basis in 60 s (about 5600 events). Error bars are derived from raw data analysis incorporating Poissonian counting statistics.}
	\begin{tabular}{|c|c|c|c|c|c|}
		\hline
		\multicolumn{2}{|c|}{Operators} & Expectation value & \multicolumn{2}{c|}{Operators} & Expectation value \\ \hline
		1 & $X_{1}X_{2}$ & $0.651 \pm 0.010$ & 15 & $Z_{1}Z_{2}$ & $-0.346 \pm 0.012$ \\ \hline
		2 & $X_{1}X_{3}$ & $0.566 \pm 0.011$ & 16 & $Z_{1}Z_{3}$ & $-0.343 \pm 0.012$ \\ \hline
		3 & $X_{1}X_{4}$ & $0.564 \pm 0.011$ & 17 & $Z_{1}Z_{4}$ & $-0.266 \pm 0.012$ \\ \hline
		4 & $X_{2}X_{3}$ & $0.558 \pm 0.011$ & 18 & $Z_{2}Z_{3}$ & $-0.267 \pm 0.012$ \\ \hline
		5 & $X_{2}X_{4}$ & $0.584 \pm 0.011$ & 19 & $Z_{2}Z_{4}$ & $-0.344 \pm 0.012$ \\ \hline
		6 & $X_{3}X_{4}$ & $0.647 \pm 0.010$ & 20 & $Z_{3}Z_{4}$ & $-0.318 \pm 0.012$ \\ \hline
		7 & $X_{1}X_{2}X_{3}X_{4}$ & $0.892 \pm 0.006$ & 21 & $Z_{1}Z_{2}Z_{3}Z_{4}$ & $0.884 \pm 0.006$ \\ \hline
		8 & $Y_{1}Y_{2}$ & $0.664 \pm 0.010$ & 22 & $[\frac{1}{\sqrt{2}}(X + Z)]^{\otimes 4}$ & $-0.424 \pm 0.012$ \\ \hline
		9 & $Y_{1}Y_{3}$ & $0.596 \pm 0.011$ & 23 & $[\frac{1}{\sqrt{2}}(X - Z)]^{\otimes 4}$ & $-0.399 \pm 0.012$ \\ \hline
		10 & $Y_{1}Y_{4}$ & $0.599 \pm 0.011$ & 24 & $[\frac{1}{\sqrt{2}}(Y + Z)]^{\otimes 4}$ & $-0.401 \pm 0.012$ \\ \hline
		11 & $Y_{2}Y_{3}$ & $0.598 \pm 0.011$ & 25 & $[\frac{1}{\sqrt{2}}(Y - Z)]^{\otimes 4}$ & $-0.397 \pm 0.012$ \\ \hline
		12 & $Y_{2}Y_{4}$ & $0.595 \pm 0.011$ & 26 & $[\frac{1}{\sqrt{2}}(X + Y)]^{\otimes 4}$ & $0.888 \pm 0.006$ \\ \hline
		13 & $Y_{3}Y_{4}$ & $0.668 \pm 0.010$ & 27 & $[\frac{1}{\sqrt{2}}(X - Y)]^{\otimes 4}$ & $0.916 \pm 0.005$ \\ \hline
		14 & $Y_{1}Y_{2}Y_{3}Y_{4}$ & $0.906 \pm 0.006$ & \multicolumn{2}{c|}{$W_{\text{proj}}$} & $-0.228 \pm 0.003$ \\ \hline
		\multicolumn{2}{|c|}{$F_{\text{D}_{4}^{(2)}}$} & $0.894 \pm 0.003$ & \multicolumn{2}{c|}{$W_{5}$} & $-0.0213 \pm 0.0005$ \\ \hline
	\end{tabular}
	\label{tab:D4yR11}
\end{table}

\begin{table}[htbp]
	\centering
	\renewcommand\arraystretch{1.4}	
	\renewcommand\tabcolsep{2pt}
	\caption{Results of fidelity and witness of four-photon Dicke state at noise $\theta=2\pi/15$. Including the required 27 operator expectation values. Under the conditions of no filtering of the parametric light, pump laser power of 3100mw, PPKTP crystal temperature of 17℃. Data are collected for each basis in 60 s (about 5600 events). Error bars are derived from raw data analysis incorporating Poissonian counting statistics.}
	\begin{tabular}{|c|c|c|c|c|c|}
		\hline
		\multicolumn{2}{|c|}{Operators} & Expectation value & \multicolumn{2}{c|}{Operators} & Expectation value \\ \hline
		1 & $X_{1}X_{2}$ & $0.652 \pm 0.010$ & 15 & $Z_{1}Z_{2}$ & $-0.342 \pm 0.012$ \\ \hline
		2 & $X_{1}X_{3}$ & $0.574 \pm 0.011$ & 16 & $Z_{1}Z_{3}$ & $-0.349 \pm 0.012$ \\ \hline
		3 & $X_{1}X_{4}$ & $0.592 \pm 0.011$ & 17 & $Z_{1}Z_{4}$ & $-0.254 \pm 0.013$ \\ \hline
		4 & $X_{2}X_{3}$ & $0.595 \pm 0.011$ & 18 & $Z_{2}Z_{3}$ & $-0.248 \pm 0.013$ \\ \hline
		5 & $X_{2}X_{4}$ & $0.592 \pm 0.011$ & 19 & $Z_{2}Z_{4}$ & $-0.356 \pm 0.012$ \\ \hline
		6 & $X_{3}X_{4}$ & $0.654 \pm 0.010$ & 20 & $Z_{3}Z_{4}$ & $-0.331 \pm 0.012$ \\ \hline
		7 & $X_{1}X_{2}X_{3}X_{4}$ & $0.902 \pm 0.006$ & 21 & $Z_{1}Z_{2}Z_{3}Z_{4}$ & $0.886 \pm 0.006$ \\ \hline
		8 & $Y_{1}Y_{2}$ & $0.340 \pm 0.012$ & 22 & $[\frac{1}{\sqrt{2}}(X + Z)]^{\otimes 4}$ & $-0.398 \pm 0.012$ \\ \hline
		9 & $Y_{1}Y_{3}$ & $0.303 \pm 0.012$ & 23 & $[\frac{1}{\sqrt{2}}(X - Z)]^{\otimes 4}$ & $-0.418 \pm 0.012$ \\ \hline
		10 & $Y_{1}Y_{4}$ & $0.322 \pm 0.012$ & 24 & $[\frac{1}{\sqrt{2}}(Y + Z)]^{\otimes 4}$ & $-0.041 \pm 0.013$ \\ \hline
		11 & $Y_{2}Y_{3}$ & $0.323 \pm 0.012$ & 25 & $[\frac{1}{\sqrt{2}}(Y - Z)]^{\otimes 4}$ & $0.009 \pm 0.013$ \\ \hline
		12 & $Y_{2}Y_{4}$ & $0.321 \pm 0.012$ & 26 & $[\frac{1}{\sqrt{2}}(X + Y)]^{\otimes 4}$ & $0.276 \pm 0.013$ \\ \hline
		13 & $Y_{3}Y_{4}$ & $0.364 \pm 0.012$ & 27 & $[\frac{1}{\sqrt{2}}(X - Y)]^{\otimes 4}$ & $0.189 \pm 0.013$ \\ \hline
		14 & $Y_{1}Y_{2}Y_{3}Y_{4}$ & $-0.207 \pm 0.013$ & \multicolumn{2}{c|}{$W_{proj}$} & $0.053 \pm 0.003$ \\ \hline
		\multicolumn{2}{|c|}{$F_{D_{4}^{(2)}}$} & $0.613 \pm 0.003$ & \multicolumn{2}{c|}{$W_{5}$} & $-0.0085 \pm 0.0005$ \\ \hline
	\end{tabular}
	\label{tab:D4yR12}
\end{table}

\begin{table}[htbp]
	\centering
	\renewcommand\arraystretch{1.4}	
	\renewcommand\tabcolsep{2pt}
	\caption{Results of fidelity and witness of the initial four-photon Dicke state prepared for robustness verification with noise $\theta=\pi/6$. Including the required 27 operator expectation values. Under the conditions of no filtering of the parametric light, pump laser power of 3100mw, PPKTP crystal temperature of 60℃. Data are collected for each basis in 60 s (about 5900 events). Error bars are derived from raw data analysis incorporating Poissonian counting statistics.}
	\begin{tabular}{|c|c|c|c|c|c|}
		\hline
		\multicolumn{2}{|c|}{Operators} & Expectation value & \multicolumn{2}{c|}{Operators} & Expectation value \\ \hline
		1 & $X_{1}X_{2}$ & $0.613 \pm 0.010$ & 15 & $Z_{1}Z_{2}$ & $-0.337 \pm 0.012$ \\ \hline
		2 & $X_{1}X_{3}$ & $0.530 \pm 0.011$ & 16 & $Z_{1}Z_{3}$ & $-0.356 \pm 0.012$ \\ \hline
		3 & $X_{1}X_{4}$ & $0.539 \pm 0.011$ & 17 & $Z_{1}Z_{4}$ & $-0.266 \pm 0.012$ \\ \hline
		4 & $X_{2}X_{3}$ & $0.543 \pm 0.011$ & 18 & $Z_{2}Z_{3}$ & $-0.265 \pm 0.012$ \\ \hline
		5 & $X_{2}X_{4}$ & $0.538 \pm 0.011$ & 19 & $Z_{2}Z_{4}$ & $-0.346 \pm 0.012$ \\ \hline
		6 & $X_{3}X_{4}$ & $0.602 \pm 0.010$ & 20 & $Z_{3}Z_{4}$ & $-0.328 \pm 0.012$ \\ \hline
		7 & $X_{1}X_{2}X_{3}X_{4}$ & $0.739 \pm 0.009$ & 21 & $Z_{1}Z_{2}Z_{3}Z_{4}$ & $0.898 \pm 0.006$ \\ \hline
		8 & $Y_{1}Y_{2}$ & $0.594 \pm 0.010$ & 22 & $[\frac{1}{\sqrt{2}}(X + Z)]^{\otimes 4}$ & $-0.379 \pm 0.012$ \\ \hline
		9 & $Y_{1}Y_{3}$ & $0.531 \pm 0.011$ & 23 & $[\frac{1}{\sqrt{2}}(X - Z)]^{\otimes 4}$ & $-0.372 \pm 0.012$ \\ \hline
		10 & $Y_{1}Y_{4}$ & $0.540 \pm 0.011$ & 24 & $[\frac{1}{\sqrt{2}}(Y + Z)]^{\otimes 4}$ & $-0.359 \pm 0.012$ \\ \hline
		11 & $Y_{2}Y_{3}$ & $0.540 \pm 0.011$ & 25 & $[\frac{1}{\sqrt{2}}(Y - Z)]^{\otimes 4}$ & $-0.342 \pm 0.012$ \\ \hline
		12 & $Y_{2}Y_{4}$ & $0.539 \pm 0.011$ & 26 & $[\frac{1}{\sqrt{2}}(X + Y)]^{\otimes 4}$ & $0.734 \pm 0.009$ \\ \hline
		13 & $Y_{3}Y_{4}$ & $0.591 \pm 0.010$ & 27 & $[\frac{1}{\sqrt{2}}(X - Y)]^{\otimes 4}$ & $0.717 \pm 0.009$ \\ \hline
		14 & $Y_{1}Y_{2}Y_{3}Y_{4}$ & $0.737 \pm 0.009$ & \multicolumn{2}{c|}{$W_{\text{proj}}$} & $-0.150 \pm 0.003$ \\ \hline
		\multicolumn{2}{|c|}{$F_{\text{D}_{4}^{(2)}}$} & $0.817 \pm 0.003$ & \multicolumn{2}{c|}{$W_{5}$} & $-0.0170 \pm 0.0005$ \\ \hline
	\end{tabular}
	\label{tab:D4yR21}
\end{table}

\begin{table}[htbp]
	\centering
	\renewcommand\arraystretch{1.4}	
	\renewcommand\tabcolsep{2pt}
	\caption{Results of fidelity and witness of four-photon Dicke state at noise $\theta=\pi/6$. Including the required 27 operator expectation values. Under the conditions of no filtering of the parametric light, pump laser power of 3100mw, PPKTP crystal temperature of 60℃. Data are collected for each basis in 60 s (about 5900 events). Error bars are derived from raw data analysis incorporating Poissonian counting statistics.}
	\begin{tabular}{|c|c|c|c|c|c|}
		\hline
		\multicolumn{2}{|c|}{Operators} & Expectation value & \multicolumn{2}{c|}{Operators} & Expectation value \\ \hline
		1 & $X_{1}X_{2}$ & $0.594 \pm 0.010$ & 15 & $Z_{1}Z_{2}$ & $-0.340 \pm 0.012$ \\ \hline
		2 & $X_{1}X_{3}$ & $0.523 \pm 0.011$ & 16 & $Z_{1}Z_{3}$ & $-0.338 \pm 0.012$ \\ \hline
		3 & $X_{1}X_{4}$ & $0.548 \pm 0.011$ & 17 & $Z_{1}Z_{4}$ & $-0.271 \pm 0.013$ \\ \hline
		4 & $X_{2}X_{3}$ & $0.544 \pm 0.011$ & 18 & $Z_{2}Z_{3}$ & $-0.267 \pm 0.013$ \\ \hline
		5 & $X_{2}X_{4}$ & $0.538 \pm 0.011$ & 19 & $Z_{2}Z_{4}$ & $-0.334 \pm 0.012$ \\ \hline
		6 & $X_{3}X_{4}$ & $0.592 \pm 0.010$ & 20 & $Z_{3}Z_{4}$ & $-0.335 \pm 0.012$ \\ \hline
		7 & $X_{1}X_{2}X_{3}X_{4}$ & $0.725 \pm 0.009$ & 21 & $Z_{1}Z_{2}Z_{3}Z_{4}$ & $0.888 \pm 0.006$ \\ \hline
		8 & $Y_{1}Y_{2}$ & $0.382 \pm 0.012$ & 22 & $[\frac{1}{\sqrt{2}}(X + Z)]^{\otimes 4}$ & $-0.366 \pm 0.012$ \\ \hline
		9 & $Y_{1}Y_{3}$ & $0.340 \pm 0.012$ & 23 & $[\frac{1}{\sqrt{2}}(X - Z)]^{\otimes 4}$ & $-0.366 \pm 0.012$ \\ \hline
		10 & $Y_{1}Y_{4}$ & $0.341 \pm 0.012$ & 24 & $[\frac{1}{\sqrt{2}}(Y + Z)]^{\otimes 4}$ & $-0.088 \pm 0.013$ \\ \hline
		11 & $Y_{2}Y_{3}$ & $0.368 \pm 0.012$ & 25 & $[\frac{1}{\sqrt{2}}(Y - Z)]^{\otimes 4}$ & $-0.132 \pm 0.013$ \\ \hline
		12 & $Y_{2}Y_{4}$ & $0.358 \pm 0.012$ & 26 & $[\frac{1}{\sqrt{2}}(X + Y)]^{\otimes 4}$ & $0.267 \pm 0.012$ \\ \hline
		13 & $Y_{3}Y_{4}$ & $0.388 \pm 0.012$ & 27 & $[\frac{1}{\sqrt{2}}(X - Y)]^{\otimes 4}$ & $0.300 \pm 0.012$ \\ \hline
		14 & $Y_{1}Y_{2}Y_{3}Y_{4}$ & $-0.051 \pm 0.013$ & \multicolumn{2}{c|}{$W_{\text{proj}}$} & $0.046 \pm 0.003$ \\ \hline
		\multicolumn{2}{|c|}{$F_{\text{D}_{4}^{(2)}}$} & $0.620 \pm 0.003$ & \multicolumn{2}{c|}{$W_{5}$} & $-0.0077 \pm 0.0005$ \\ \hline
	\end{tabular}
	\label{tab:D4yR22}
\end{table}

\begin{table}[htbp]
	\centering
	\renewcommand\arraystretch{1.4}	
	\renewcommand\tabcolsep{2pt}
	\caption{Results of fidelity and witness of the initial four-photon Dicke state prepared for robustness verification with noise $\theta=2\pi/15$. Including the required 27 operator expectation values. Under the conditions of no filtering of the parametric light, pump laser power of 3100mw, PPKTP crystal temperature of 70℃. Data are collected for each basis in 60 s (about 5600 events). Error bars are derived from raw data analysis incorporating Poissonian counting statistics.}
	\begin{tabular}{|c|c|c|c|c|c|}
		\hline
		\multicolumn{2}{|c|}{Operators} & Expectation value & \multicolumn{2}{c|}{Operators} & Expectation value \\ \hline
		1 & $X_{1}X_{2}$ & $0.566 \pm 0.011$ & 15 & $Z_{1}Z_{2}$ & $-0.355 \pm 0.012$ \\ \hline
		2 & $X_{1}X_{3}$ & $0.495 \pm 0.011$ & 16 & $Z_{1}Z_{3}$ & $-0.363 \pm 0.012$ \\ \hline
		3 & $X_{1}X_{4}$ & $0.501 \pm 0.011$ & 17 & $Z_{1}Z_{4}$ & $-0.240 \pm 0.012$ \\ \hline
		4 & $X_{2}X_{3}$ & $0.494 \pm 0.011$ & 18 & $Z_{2}Z_{3}$ & $-0.238 \pm 0.012$ \\ \hline
		5 & $X_{2}X_{4}$ & $0.500 \pm 0.011$ & 19 & $Z_{2}Z_{4}$ & $-0.352 \pm 0.012$ \\ \hline
		6 & $X_{3}X_{4}$ & $0.558 \pm 0.011$ & 20 & $Z_{3}Z_{4}$ & $-0.340 \pm 0.012$ \\ \hline
		7 & $X_{1}X_{2}X_{3}X_{4}$ & $0.632 \pm 0.010$ & 21 & $Z_{1}Z_{2}Z_{3}Z_{4}$ & $0.891 \pm 0.006$ \\ \hline
		8 & $Y_{1}Y_{2}$ & $0.553 \pm 0.011$ & 22 & $[\frac{1}{\sqrt{2}}(X + Z)]^{\otimes 4}$ & $-0.345 \pm 0.012$ \\ \hline
		9 & $Y_{1}Y_{3}$ & $0.499 \pm 0.011$ & 23 & $[\frac{1}{\sqrt{2}}(X - Z)]^{\otimes 4}$ & $-0.352 \pm 0.012$ \\ \hline
		10 & $Y_{1}Y_{4}$ & $0.491 \pm 0.011$ & 24 & $[\frac{1}{\sqrt{2}}(Y + Z)]^{\otimes 4}$ & $-0.364 \pm 0.012$ \\ \hline
		11 & $Y_{2}Y_{3}$ & $0.499 \pm 0.011$ & 25 & $[\frac{1}{\sqrt{2}}(Y - Z)]^{\otimes 4}$ & $-0.323 \pm 0.012$ \\ \hline
		12 & $Y_{2}Y_{4}$ & $0.508 \pm 0.011$ & 26 & $[\frac{1}{\sqrt{2}}(X + Y)]^{\otimes 4}$ & $0.645 \pm 0.010$ \\ \hline
		13 & $Y_{3}Y_{4}$ & $0.560 \pm 0.011$ & 27 & $[\frac{1}{\sqrt{2}}(X - Y)]^{\otimes 4}$ & $0.627 \pm 0.010$ \\ \hline
		14 & $Y_{1}Y_{2}Y_{3}Y_{4}$ & $0.643 \pm 0.010$ & \multicolumn{2}{c|}{$W_{\text{proj}}$} & $-0.099 \pm 0.003$ \\ \hline
		\multicolumn{2}{|c|}{$F_{\text{D}_{4}^{(2)}}$} & $0.766 \pm 0.003$ & \multicolumn{2}{c|}{$W_{5}$} & $-0.0132 \pm 0.0005$ \\ \hline
	\end{tabular}
	\label{tab:D4yR31}
\end{table}

\begin{table}[htbp]
	\centering
	\renewcommand\arraystretch{1.4}	
	\renewcommand\tabcolsep{2pt}
	\caption{Results of fidelity and witness of four-photon Dicke state at noise $\theta=2\pi/15$. Including the required 27 operator expectation values. Under the conditions of no filtering of the parametric light, pump laser power of 3100mw, PPKTP crystal temperature of 70℃. Data are collected for each basis in 60 s (about 5600 events). Error bars are derived from raw data analysis incorporating Poissonian counting statistics.}
	\begin{tabular}{|c|c|c|c|c|c|}
		\hline
		\multicolumn{2}{|c|}{Operators} & Expectation value & \multicolumn{2}{c|}{Operators} & Expectation value \\ \hline
		1 & $X_{1}X_{2}$ & $0.586 \pm 0.010$ & 15 & $Z_{1}Z_{2}$ & $-0.351 \pm 0.012$ \\ \hline
		2 & $X_{1}X_{3}$ & $0.491 \pm 0.011$ & 16 & $Z_{1}Z_{3}$ & $-0.332 \pm 0.012$ \\ \hline
		3 & $X_{1}X_{4}$ & $0.512 \pm 0.011$ & 17 & $Z_{1}Z_{4}$ & $-0.273 \pm 0.012$ \\ \hline
		4 & $X_{2}X_{3}$ & $0.508 \pm 0.011$ & 18 & $Z_{2}Z_{3}$ & $-0.274 \pm 0.012$ \\ \hline
		5 & $X_{2}X_{4}$ & $0.521 \pm 0.011$ & 19 & $Z_{2}Z_{4}$ & $-0.331 \pm 0.012$ \\ \hline
		6 & $X_{3}X_{4}$ & $0.579 \pm 0.010$ & 20 & $Z_{3}Z_{4}$ & $-0.338 \pm 0.012$ \\ \hline
		7 & $X_{1}X_{2}X_{3}X_{4}$ & $0.645 \pm 0.010$ & 21 & $Z_{1}Z_{2}Z_{3}Z_{4}$ & $0.901 \pm 0.005$ \\ \hline
		8 & $Y_{1}Y_{2}$ & $0.419 \pm 0.012$ & 22 & $[\frac{1}{\sqrt{2}}(X + Z)]^{\otimes 4}$ & $-0.338 \pm 0.012$ \\ \hline
		9 & $Y_{1}Y_{3}$ & $0.364 \pm 0.012$ & 23 & $[\frac{1}{\sqrt{2}}(X - Z)]^{\otimes 4}$ & $-0.348 \pm 0.012$ \\ \hline
		10 & $Y_{1}Y_{4}$ & $0.384 \pm 0.012$ & 24 & $[\frac{1}{\sqrt{2}}(Y + Z)]^{\otimes 4}$ & $-0.154 \pm 0.013$ \\ \hline
		11 & $Y_{2}Y_{3}$ & $0.401 \pm 0.012$ & 25 & $[\frac{1}{\sqrt{2}}(Y - Z)]^{\otimes 4}$ & $-0.241 \pm 0.012$ \\ \hline
		12 & $Y_{2}Y_{4}$ & $0.408 \pm 0.012$ & 26 & $[\frac{1}{\sqrt{2}}(X + Y)]^{\otimes 4}$ & $0.353 \pm 0.012$ \\ \hline
		13 & $Y_{3}Y_{4}$ & $0.417 \pm 0.012$ & 27 & $[\frac{1}{\sqrt{2}}(X - Y)]^{\otimes 4}$ & $0.361 \pm 0.012$ \\ \hline
		14 & $Y_{1}Y_{2}Y_{3}Y_{4}$ & $0.115 \pm 0.013$ & \multicolumn{2}{c|}{$W_{\text{proj}}$} & $0.017 \pm 0.003$ \\ \hline
		\multicolumn{2}{|c|}{$F_{\text{D}_{4}^{(2)}}$} & $0.650 \pm 0.003$ & \multicolumn{2}{c|}{$W_{5}$} & $-0.0085 \pm 0.0005$ \\ \hline
	\end{tabular}
	\label{tab:D4yR32}
\end{table}

\end{widetext}


\bibliographystyle{apsrev4-2}
\clearpage
\bibliography{reference}

 \end{document}